\documentstyle[aps,epsfig,floats,amssymb,graphicx,psfrag,amsfonts,amsmath,bbm]{revtex}
\draft
\newcommand{\nwc}{\newcommand}

\nwc{\bdm} {\begin{displaymath}}
\nwc{\edm} {\end{displaymath}}
\nwc{\bea} {\be\ba{rcl}}
\nwc{\Tr} {\textrm{Tr}}
\nwc{\LL} {\cal L}
\begin{document}
\twocolumn[\hsize\textwidth\columnwidth\hsize\csname
@twocolumnfalse\endcsname

\title{Bosonic effective action for interacting fermions}

\author{C. Wetterich}

\address{
Institut f{\"u}r Theoretische Physik,
Philosophenweg 16, 69120 Heidelberg, Germany}

\maketitle

\begin{abstract}
We compare different versions of a bosonic description for
systems of interacting fermions, with
particular emphasis on the free energy functional.
The bosonic effective action makes the
issue of symmetries particularly transparent and we
present for the Hubbard model an exact mapping between repulsive
and attractive interactions. A systematic
expansion for the bosonic effective action
starts with a solution to the  lowest order Schwinger-Dyson
or gap equation.
We propose a two particle irreducible formulation of an exact functional renormalization
group equation for computations beyond leading order. On this basis we suggest a
renormalized gap equation. This approach is compared with functional
renormalization in a partially bosonized setting.
\end{abstract}
\pacs{PACS numbers: 11.10.-z, 11.10.Hi, 11.10.St  \hfill }

 ]


\section {Introduction}
\label{introduction}
Systems of strongly correlated electrons often show the phenomenon of spontaneous symmetry
breaking or a behavior close to it. Examples are antiferromagnetism or superconductivity
in the Hubbard model \cite{HM}. Spontaneous symmetry breaking (SSB) is
also common to many other
fermionic systems, like effective quark-models for the strong interactions
\cite{NJL} or ultracold fermionic atoms \cite{FA}.
The spontaneous breaking of symmetry is introduced by a bosonic order
parameter or vacuum expectation value (vev), typically corresponding to the condensation
of fermion-fermion or fermion-antifermion (electron-hole) pairs. Examples are Cooper
pairs $\sim \psi\psi$ for superconductors or the spin vector $\vec{m}\sim \overline{\psi}
\vec{\tau}\psi$ for ferromagnets. A reliable quantitative description of SSB - for example
the computation of the order parameter and the mass gap in the low temperature phase - is
not easy in a purely fermionic language. The simplest possible effective potential for the
magnetization in a ferromagnet involves already four powers of $\vec{m}$, e.g.
\begin{equation}\label{eq:fourpowers}
U =-\frac{\mu^2}{2}\vec{m}^2+\frac{\lambda}{8}\left(\vec{m}^2\right)^2
\end{equation}
The term $\sim m^4$ corresponds to an eight-fermion-interaction which is hard to compute in
a purely fermionic picture. Moreover, the effective potential (or free energy) often does
not take a simple polynomial form as in the example (\ref{eq:fourpowers}).

Standard approaches to this problem are the mean field theory (MFT) or the solution of a
gap equation based on the Schwinger-Dyson (SD) equation  \cite{SD}.
In both methods one treats effectively the fermionic fluctuations in a
bosonic ``background field'' or ``mean field''. This is manifest in MFT where the fermionic
fluctuations are truncated in quadratic order. The remaining fermionic functional integral is
Gaussian and the free energy for a given mean field can be computed in this approximation.
Minimization of the free energy yields a self-consistency equation for the mean field. The
shortcoming is the complete omission of the effective bosonic fluctuations or, equivalently,
the fermionic fluctuations beyond the quadratic approximation. For strongly interacting
systems this approximation is not expected to be quantitatively accurate.

This issue is most easily addressed by partial bosonization via a Hubbard-Stratonovich-
transformation \cite{HS}. In the partially bosonized language the partition function is written
as a functional integral over fermion and (composite) boson fields
and MFT precisely corresponds to the neglection of the bosonic
fluctuations. Partial bosonization is, however, not unique and the results on SSB depend
strongly on the choice of the mean field. This is a well known problem for the Hubbard
model where certain implementations of ``partial bosonization'' are consistent with the
spin rotation symmetry but fail to reproduce the Hartree-Fock result (i.e. lowest order
Schwinger-Dyson equation) whereas others are consistent with Hartree-Fock but do not exhibit
the correct symmetries \footnote{This situation was summarized by \cite{HJS} where also a cure
based on nonlinear fields is proposed.}. The problem can be traced back to the possibility
of Fierz-reordering of a local interaction. For the Hubbard model
a systematic discussion of the strong
quantitative importance of the ``Fierz ambiguity'' can be found in \cite{BBW}. In a more
general context the origin and the cure of
this ``Fierz ambiguity'' are discussed in detail in \cite{IW}.
There it is shown that even a rough inclusion of the
bosonic fluctuations greatly reduces the
ambiguity. The remaining ambiguity can be used for an error estimate of a given
approximation - physical results must, of course, be
independent of the choice of a specific partial bosonization. All this clearly demonstrates
the importance of the bosonic fluctuations neglected in MFT.

The SD-approach is formally a fermionic formulation and the lowest order gap equation is one
loop exact.\footnote{This contrasts with MFT \cite{IW}.}
However, an apparent problem is the reconstruction
of the free energy from the solution of the gap equation. This becomes crucial when the gap
equation admits solutions with different order parameters and the free energy corresponding
to the different solutions has to be compared.\footnote{See \cite{BR} for
a recent discussion of this issue in color superconductivity.
Their attempt to reconstruct the free energy may not always work.} Furthermore, the lowest
order SD-equation for a given order parameter generically takes the form of a MFT
computation for a particular choice of partial bosonization. This shows that the gap equation
leaves out the effective bosonic fluctuations\footnote{In contrast to MFT the bosonic
fluctuations do not contribute in lowest order in the coupling constant.},
just as MFT. This shortcoming becomes
particularly important when the composite bosons are correlated on large length scales
(or have a small renormalized mass), as characteristic for critical behavior in the
vicinity of a second order phase transition. \footnote{The gap equation only gives the
MFT values for critical exponents.}

Going beyond the MFT and SD approaches, which are established since many years, requires a
method which should (a) allow for a computation of the free energy, (b) be one loop exact
and (c) include the effects of the bosonic fluctuations. One possible approach relies on the
exact renormalization group equation for the effective average action \cite{ERGE}. It is based on
partial bosonization, but the Fierz ambiguity of MFT can now be cured \cite{IW} by the inclusion
of the bosonic fluctuations and an appropriate ``rebosonization'' of multi-fermion
interactions at every scale \cite{GW}. Partial bosonization
brings a redundancy of the description: a four fermion interaction amplitude
can be composed from a direct fermionic vertex and a piece from the exchange of bosons.
Rebosonization avoids this redundancy by eliminating (parts of) the direct vertex in terms of
interactions involving bosons. Partially bosonized functional renormalization
has been employed successfully
\cite{JuW} to the issue of chiral symmetry breaking within effective interacting quark
models for the strong interactions. Recently, it has also led to a description of the
low temperature antiferromagnetic phase for the two dimensional Hubbard model for small
doping \cite{BBWN}.

In this paper we compare partial bosonization with
a method based on the ``bosonic effective action'' (BEA). The latter approach was pioneered
in solid state physics by Luttinger and Ward \cite{LW}. (See \cite{Hau} for a review.)
As an alternative to partial bosonization one introduces bosonic sources for fermion
bilinears, but no explicit bosonic fluctuating fields.
As usual, the effective action obtains from an appropriate Legendre transform and
is a functional of bosonic fields. We argue that there is no need to include fermionic
sources - as a result the BEA depends {\em only} on bosonic variables.
The original fermionic problem is then completely translated into a purely bosonic language.
In particular, this avoids the complications of redundancy by construction. Indeed,
all operators with possible nonzero expectation values are of bosonic nature - as,
for example, the fermion propagator or any order parameter. The general possibility of a
purely bosonic description should therefore be of no surprise. It is also clear that a
purely bosonic description will, in general, not only involve local fields. The bosonic
fields correspond to fermion bilinears $\psi_\alpha\psi_\beta$ where $\psi_\alpha$ and
$\psi_\beta$ may be fermion fields at different locations.

The BEA is based on the two particle irreducible (2PI) effective action \cite{LW,Baym,CIT}.
Indeed, the loop expansion of the BEA involves 2PI graphs in the fermionic language.
However, we choose here a
{\em complete} translation to a bosonic picture, such that the BEA depends {\em only} on
composite bosonic fields.
Some of the formulae in the first sections
of this paper are well known from earlier work on the
$2PI$-effective action. The present derivation highlights the close link between the
$SD$-equation and the bosonic effective action and makes several aspects particularly
simple and transparent.

The new aspects of this work concern several issues. We put emphasis
on the free energy functional which is needed for a comparison of different solutions of the
gap equation. In particular, we present an explicit local formula in the case of a local
gap in sect. \ref{freeenergy}. This can be generalized to multi-fermion interactions
(beyond four fermion interactions) in a straightforward way \cite{JW}.
(Previous discussions of the effective potential or free energy can be found in
\cite{GO}). We also construct a systematic mapping between the fermionic effective action
$(1PI)$ and the bosonic effective action $(2PI)$. This helps for an understanding
of the issue of redundancy. In particular, an exact $SD$-identity for the
BEA permits simple systematic expansions.

Furthermore, we provide for a general
discussion of the symmetries of the BEA. They go far beyond the symmetries of the fermionic
effective action since the BEA respects all symmetries of the interaction term.
In particular, two models with the same interaction but different fermion kinetic or
mass terms lead to the same BEA. Their difference only results in a different source
term in the field equations. We show how the symmetry transformations can be used to map
models with different fermionic quadratic terms into each other. The large degree
of symmetry is particularly apparent in a formulation where particles and antiparticles
(electrons and holes) are treated as different components of a fermionic Grassmann field.
This formulation leads also to simple mappings between models with different interactions.
As an example, we discuss an exact mapping between the Hubbard models with repulsive
and attractive interaction. Results from numerical simulations in the attractive case
can therefore be used as an information about appropriate quantities in the model with
a repulsive interaction.

One of the main tasks of this paper is a detailed comparison between the BEA and
partial bosonization
based on the Hubbard-Stratonovich-transformation. We concentrate on the
correlation functions that are computed in both methods. This highlights the
advantages and problematic points of the two approaches.

For a first time, we derive exact functional renormalization group equations
within the $2PI$ formalism. We restrict our aim here to a formal establishment of these
equations and a first qualitative discussion, while postponing practical applications to
future work. In particular, we propose a simple renormalized gap equation. It has the
same structure as the lowest order gap equation, but with ``microscopic'' couplings and
propagators now replaced by renormalized ones. Even for local microscopic interactions the
renormalized coupling will not remain local, in general. This can lead to new structures
in the space of possible solutions. For example, one may find a gap equation for a
nonlocal superconducting order parameter in the Hubbard model which is not present
in lowest order. We also propose an improvement of the ``lowest order'' flow equation
for the four fermion vertex by including the effects of a gap. This can avoid the
previously observed divergence of the renormalized coupling in the Hubbard model
\cite{MEZ,16A,16B}. Furthermore, we develop
``bosonic renormalization group equations'' which permit to establish a close
contact with the renormalization group for bosonic systems. This allows for a description
of critical exponents at second order phase transitions beyond the lowest
order ``mean field values''. We compare the different versions of the
functional renormalization group
for the BEA with the one for partial bosonization.

This paper is organized as follows: After defining the BEA in
sect. \ref{bosoniceffective} we translate in
sect. \ref{schwingerdyson} the fermionic SD-equation into a
bosonic formulation. The resulting identity for the BEA can be used for the construction
of a systematic loop expansion for the BEA and the free energy. The final results of
these sections are well-known \cite{Hau}. It is obvious that the
BEA is a rather simple object which can explicitely be written down in rather high orders
of the loop expansion. The price to pay is the comparatively high complexity of the field
equation which needs to be solved in order to find the minima of the free energy. The BEA
also has a very high degree of symmetry
(sect. \ref{symmetries}). It is invariant under all anomaly
free transformations
leaving the ``classical'' fermionic interaction invariant, irrespective of the form of the
classical fermion propagator.

In sects. \ref{localinteractions}-\ref{antiferromagnetism} we turn to
models with local four-fermion-
interactions. Based on the well-known local gap equation
(sect. \ref{gapequation}) and a simple local formula for the free
energy (sect. \ref{freeenergy}) we discuss antiferromagnetism in
the two dimensional Hubbard model in sect. \ref{antiferromagnetism}.
Already the lowest order computation of the
free energy realizes all symmetries in a simple linear way (including spin rotations) and is
nevertheless consistent with Hartree-Fock.
We compute the free energy for the antiferromagnetic state and sketch how it can be compared
with other phases.

In sects. \ref{symmetrieshubbard}, \ref{mappingbetween} we turn to the symmetries of the
Hubbard model and describe  the exact map between the models with repulsive and
attractive interaction.

Sect. \ref{partialbosonization} is devoted to partial bosonization. We first describe a
mixed effective action which supplements the BEA by a redundant but essentially
local piece for the propagation and interaction of fermions. Subsequently, we turn to a
``local fermion boson model'' as defined by a Hubbard-Stratonovich transformation.
The identities relating it to the mixed effective action or to the BEA are exact.
However, the different versions of bosonization will typically use different
approximation schemes.

The powerful tool of the exact renormalization group equation for the effective average
action \cite{ERGE} can be directly implemented for the BEA
(sect. \ref{exactfermionic}). There is no need for
rebosonization \cite{GW} in this formulation. In particular, we advocate the use of a
``renormalized gap equation'' in sect. \ref{renormalizedgapequation}
and propose a ``gap-improved'' flow equation for the evolution of an appropriate
running ``four fermion coupling''. We derive the bosonic functional renormalization
group equation in sect. \ref{bosonicrenormalizationgroup}.
After comparison with the functional
renormalization group for partial bosonization in \ref{functionalrenormalization} our
conclusions are drawn in sect. \ref{conclusions}.

\section {Bosonic effective action}
\label{bosoniceffective}
We start by collecting all fermionic degrees of freedom in a set of Grassmann variables
$\tilde{\psi}_\alpha$, where the collective index $\alpha$ labels momenta (or locations) as
well as possible internal degrees of freedom including spin. It also differentiates between
what is usually called $\psi$ and $\overline{\psi}$, e.g. electrons and holes or
particles and antiparticles. For a model with quartic fermionic
interaction the partition function reads
\begin{eqnarray}
Z[\eta,j]&=&\int  D \tilde{\psi} \exp( \eta_{\alpha} \tilde{\psi}_{\alpha}
+\frac{1}{2}j_{\alpha \beta} \tilde{\psi}_{\alpha} \tilde{\psi}_\beta\nonumber\\
&&-\frac{1}{24} \lambda_{\alpha \beta \gamma \delta} \tilde{\psi}_\alpha \tilde{\psi}_ \beta
\tilde{\psi}_\gamma \tilde{\psi}_\delta )
\end{eqnarray}

\noindent
Due to the anticommutation property of the Grassmann variables
$\lambda_{\alpha\beta\gamma\delta}$ is totally antisymmetric in all indices.
We treat here the quadratic terms in the fermionic action as a bosonic source term
$j_{\alpha \beta}=-j_{\beta \alpha}$ and consider first $Z$ as a functional of arbitrary
fermionic and bosonic sources $\eta$ and $j$. With $W[\eta,j]=\ln Z[\eta, j]$ and using
\begin{equation}
 \langle \tilde{\psi}_{\alpha}\rangle = \psi_{\alpha}
=\frac{\partial W}{\partial \eta_\alpha}
\end{equation}
we can write the fermion propagator as
\begin{equation}\label{eq:prop}
G_{\alpha \beta} = \langle \tilde{\psi}_{\alpha} \tilde{\psi}_{\beta}\rangle
=\frac{\partial W}{\partial j_{\alpha \beta}}=\frac{\partial^2W}
{\partial \eta_{\alpha} \partial \eta_{\beta}}+\frac{\partial W}{\partial \eta_{ \alpha}}
\frac{\partial W}{\partial \eta_{ \beta}}
\end{equation}
Thus $W$ obeys a nonlinear functional differential equation relating the
dependence on $j$ to the one on $\eta$. Similarly, we can write
\begin{eqnarray} \label{eq:sim}
&&\langle \tilde{\psi}_{\alpha} \tilde{\psi}_{\beta}
\tilde{\psi}_{\gamma} \tilde{\psi}_{\delta} \rangle
-\langle  \tilde{\psi}_{\alpha} \tilde{\psi}_{\beta} \rangle
\langle \tilde{\psi}_{\gamma} \tilde{\psi}_{\delta} \rangle\nonumber\\
&&=\frac{\partial^2W}{\partial j_{\alpha \beta}~ \partial j_{\gamma \delta}}\nonumber \\
&&= \frac{\partial^4 W}{\partial \eta_\alpha \partial \eta_\beta
\partial \eta_\gamma \partial \eta_\delta} +
\frac{\partial^2 W}{\partial \eta_\beta \partial \eta_\gamma}
\frac{\partial^2 W}{\partial \eta_\alpha \partial \eta_\delta}\nonumber\\
&&- \frac{\partial^2 W}{\partial \eta_\alpha \partial \eta_\gamma}
\frac{\partial^2 W}{\partial \eta_\beta \partial \eta_\delta}
+\Delta W_B^{(2)}
\end{eqnarray}
with
\begin{eqnarray}
\Delta W_B^{(2)}
 =\frac{\partial^3W}{\partial \eta_{\alpha} \partial \eta_{\beta} \partial \eta_{\gamma}}
\frac{\partial W}{\partial \eta _{\delta}}
&-&\frac{\partial^3W}{\partial \eta_{\alpha} \partial \eta_{\beta} \partial \eta_{\delta}}
\frac{\partial W}{\partial \eta _{\gamma}}\nonumber \\
- \frac{\partial^3W}{\partial \eta_{\beta} \partial \eta_{\gamma} \partial \eta_{\delta}}
\frac{\partial W}{\partial \eta _{\alpha}}
&+&\frac{\partial^3W}{\partial \eta_{\alpha} \partial \eta_{\gamma} \partial \eta_{\delta}}
\frac{\partial W}{\partial \eta _{\beta}}\nonumber \\
-\frac{\partial^2W}{\partial \eta_{\alpha} \partial \eta_{\gamma}}
\frac{\partial W}{\partial \eta _{\beta}}
\frac{\partial W}{\partial \eta _{\delta}}
&+&\frac{\partial^2W}{\partial \eta_{\alpha} \partial \eta_{\delta}}
\frac{\partial W}{\partial \eta _{\beta}}
\frac{\partial W}{\partial \eta _{\gamma}}\nonumber\\
+\frac{\partial^2W}{\partial \eta_{\beta} \partial \eta_{\gamma}}
\frac{\partial W}{\partial \eta _{\alpha}}
\frac{\partial W}{\partial \eta _{\delta}}
&-&\frac{\partial^2W}{\partial \eta_{\beta} \partial \eta_{\delta}}
\frac{\partial W}{\partial \eta _{\alpha}}
\frac{\partial W}{\partial \eta _{\gamma}}
\end{eqnarray}
vanishing for $\eta=0$.

The fermionic effective action in presence of bosonic sources
$j$ obtains by a Legendre transform with respect to the fermionic variables
\begin{equation}
\Gamma_{F}[\psi , j]=-W[\eta,j]+\eta_\alpha \psi_\alpha.
\end{equation}
It obeys (at fixed $j$)
\begin{equation}
\frac{\partial \Gamma_{F}}{\partial \psi_\alpha}=- \eta_\alpha ,
~~~ \Big(\Gamma_{F}^{(2)}\Big)_{\alpha \beta}=
-\frac{\partial^2 \Gamma_{F}}{\partial \psi_\alpha \partial \psi_\beta}
=\Big( W_F^{(2)}\Big) ^{-1}_{\alpha \beta}
\end{equation}
with
\begin{equation}
\Big(W_F^{(2)}\Big)_{\alpha \beta} =
\frac{\partial^2W}{\partial \eta_\alpha \partial \eta_\beta}.
\end{equation}
We may express $ \Gamma_{F}$ as a functional integral in a fermionic background
field $\psi$,
\begin{equation}\label{eq:seven}
\Gamma_{F}[\psi , j]= - \ln \int D \tilde{\psi} \exp(-S_j[\psi +
\tilde{\psi}]+\eta_\alpha \tilde{\psi}_\alpha)
\end{equation}
where we have defined
\begin{equation}\label{eq:defined}
S_j[\psi]=-\frac{1}{2}j_{\alpha \beta} \psi_\alpha \psi_\beta +\frac{1}{24}
\lambda_{\alpha\beta\gamma\delta}
\psi_\alpha \psi_\beta \psi_\gamma \psi_\delta
\end{equation}
\medskip

The bosonic effective action for the composite field $G_{\alpha \beta}$ is constructed
similarly. For the purely bosonic formulation motivated in the introduction we restrict the
formulation to vanishing fermionic sources $\eta_\alpha=0$. The BEA is then defined as
\begin{equation}\label{12}
\Gamma_B [G]=-W[0, j]+\frac{1}{2}j_{\alpha \beta} G_{\alpha \beta}
\end{equation}
and obeys \footnote{Note that due to the antisymmetric index notation
the bosonic matrix multiplication reads $(F\cdot G)_{\alpha  \beta, \alpha^\prime \beta^\prime}
=\frac{1}{2}F_{\alpha \beta, \gamma \delta}G_{\gamma \delta, \alpha^\prime \beta^\prime}$
with unit operator
$I_{\alpha \beta, \alpha^\prime \beta^\prime}
=\delta_{\alpha \alpha^\prime} \delta_{\beta \beta^\prime}
-\delta_{\alpha \beta^\prime} \delta_{\beta \alpha^\prime}.$ This defines
$(W_B^{(2)})^{-1}$. For clarity we often denote contraction of bosonic indices with
a dot in order to distinguish it from fermionic index contraction, i.e.
$j\cdot G=\frac{1}{2}j_{\alpha\beta}G_{\alpha\beta}=-\frac{1}{2}\Tr(jG)$.}
\begin{eqnarray}
\frac{\partial \Gamma_B}{\partial G_{\alpha \beta}}&=& j_{\alpha \beta}~~,\nonumber\\
\left(\Gamma_B^{(2)}\right)_{\alpha \beta, \gamma \delta}
&=&\frac{\partial^2 \Gamma_B}{\partial G_{\alpha \beta} \partial G_{\gamma \delta}}
=\left(W_B^{(2)}\right)^{-1}_{\alpha \beta, \gamma \delta}
\end{eqnarray}
where
\begin{equation}
\left(W_B^{(2)}\right)_{\alpha \beta,\gamma \delta}
=\frac{\partial^2W[0,j]}{\partial j_{\alpha \beta} \partial j_{\gamma \delta}}
\end{equation}
We note that $\Gamma_F$ is an even functional of $\psi$ and contains precisely the same
information as $\Gamma_B$. The bosonic and fermionic effective action are connected by
relations of the type (\ref{eq:prop}),(\ref{eq:sim}). Evaluating them at $\psi=0$ relates
$G_{\alpha \beta}$ to the fermion propagator for vanishing $\psi$,
\begin{equation}\label{eq:vanishing}
G_{\alpha \beta}=(\Gamma_F^{(2)})^{-1}_{\alpha \beta}
\end{equation}
and connects the full bosonic propagator to appropriate (functional)
derivatives of $\Gamma_F$
\begin{eqnarray} \label{eq:functional}
&&\left(\Gamma_B^{(2)}\right)^{-1}_{\alpha \beta, \gamma \delta}=\nonumber\\
&&-\frac{\partial^4\Gamma_F}
{\partial \psi_{\alpha^\prime} \partial \psi_{\beta^\prime}\partial
\psi_{\gamma^\prime}\partial \psi_{\delta^\prime}}
(\Gamma_F^{(2)})^{-1}_{\alpha \alpha^\prime}(\Gamma_F^{(2)})^{-1}_{\beta \beta^\prime}
(\Gamma_F^{(2)})^{-1}_{\gamma \gamma^\prime}(\Gamma_F^{(2)})^{-1}_{\delta \delta^\prime}
\nonumber \\
&&+(\Gamma_F^{(2)})^{-1}_{\beta \gamma}(\Gamma_F^{(2)})^{-1}_{\alpha \delta}
-(\Gamma_F^{(2)})^{-1}_{\alpha \gamma} (\Gamma_F^{(2)})^{-1}_ {\beta \delta}
\end{eqnarray}
The translation rules (\ref{eq:vanishing})(\ref{eq:functional}) -
and similar rules for higher
vertices - permit a mapping between the fermionic effective action
$\Gamma_F$ and the bosonic effective action $\Gamma_B$.
We emphasize that $G_{\alpha\beta}$ carries two fermionic indices and is therefore,
in general, a bilocal bosonic object.

The BEA is simply related to the Helmholtz free energy $F$
\begin{equation}
\Gamma_B=\frac{F}{T}
\end{equation}
The grand canonical partition function $Z(T,\mu,V)$ depends on temperature $T$, volume $V$
and the chemical potential $\mu$ which is a part of the source $j_{\alpha\beta}$.
Similarly, the particle number density is a particular combination of $G_{\alpha\beta}$.
It is sometimes convenient to introduce an auxiliary thermodynamic potential
\begin{equation}\label{eq:equilibrium}
\Gamma_j[G,j]=\Gamma_B-\frac{1}{2}j_{\alpha\beta}G_{\alpha\beta}
=\frac{\hat{F}}{T}\hspace{0.5cm},
\hspace{0.5cm} U_j=\frac{T}{V}\Gamma_j
\end{equation}
such that the equilibrium states correspond to the extrema of $\Gamma_j$. We will call
$\Gamma_j$ a ``free energy'' as well, $\Gamma_j=\hat{F}/T$. The pressure $p$, the energy,
entropy and particle number densities $\epsilon,s$ and $n$ can directly be gained from the
minimum of $\Gamma_j$, i.e. $\Gamma_{j,min}=-W=(F-\mu N)/T, U_{j,min}=\overline{U}$
as
\begin{eqnarray}
p&=&-\overline{U}~,~\epsilon=\overline{U}-T
\frac{\partial\overline{U}}{\partial T}-\mu\frac{\partial\overline{U}}{\partial\mu}~,\nonumber\\
s&=&-\frac{\partial\overline{U}}{\partial T}~,~n=-
\frac{\partial\overline{U}}{\partial\mu}
\end{eqnarray}
Of course, $\Gamma_B$ and $\Gamma_j$ do not only depend on particle number. They are
functionals of all possible fermion propagators $G_{\alpha\beta}$, whereby the physical
propagator is given by the minimum of $\Gamma_j$.

\section{Schwinger-Dyson equation and loop expansion}
\label{schwingerdyson}
In this section we translate the Schwinger-Dyson equation from the fermionic
formulation into a corresponding exact identity for $\Gamma_B$. The lowest order approximation
to the Schwinger-Dyson equation results in a closed equation relating $G$ to $j$. This allows
us to construct explicitely $\Gamma_B$ and the free energy. An iterative solution of the
SD-equation corresponds to a loop expansion for $\Gamma_B$.

Differentiating eq. (\ref{eq:seven}) with respect to $\psi$ yields the exact
Schwinger-Dyson equation for the fermionic effective action
\begin{eqnarray}
\frac{\partial\Gamma_F}{\partial \psi_\beta}
&=&j_{\alpha \beta} \psi_{\alpha}
-\frac{1}{6}\lambda_{\alpha \beta \gamma \delta}
\langle \tilde{\psi}_\alpha \tilde{\psi}_\gamma \tilde{\psi}_\delta \rangle
\nonumber \\
&=&j_{\alpha \beta} \psi_\alpha - \frac{1}{6} \lambda_{\alpha \beta \gamma \delta} \Big\{
\psi_\alpha \psi_\gamma \psi_\delta
+(\Gamma_F^{(2)})^{-1}_{\alpha \gamma} \psi_\delta \nonumber\\
&&-(\Gamma_F^{(2)})^{-1}_{\alpha \delta}
\psi_\gamma+(\Gamma_F^{(2)})^{-1}_{\gamma \delta} \psi_ \alpha
\nonumber \\
&&-(\Gamma_F^{(2)})^{-1}_{\alpha \alpha^\prime}(\Gamma_F^{(2)})^{-1}_{\gamma \gamma^\prime}
(\Gamma_F^{(2)})^{-1}_{\delta \delta^\prime}
\frac{\partial^3 \Gamma_F}
{\partial \psi_{\alpha^\prime} \partial \psi_{\gamma^\prime} \partial \psi_{\delta^\prime}}
\Big\}.
\end{eqnarray}
The fermionic Schwinger-Dyson equation relevant for our purpose
obtains by evaluating a further differentiation with respect to $\psi$ at $\psi=0$ \\
\begin{eqnarray} \label{eq:Schwinger}
\left(\Gamma_F^{(2)}\right)_{\alpha \beta}
&=&-j_{\alpha \beta}
+\frac{1}{2}\lambda_{\alpha \beta \gamma \delta}
\left(\Gamma_F^{(2)}\right)^{-1}_{\gamma \delta}
\nonumber \\
&&+\frac{1}{6}\lambda _{\beta \epsilon^\prime \gamma^\prime \delta^\prime}
\left(\Gamma_F^{(2)}\right)^{-1}_{\epsilon^\prime \epsilon}
\left(\Gamma_F^{(2)}\right)^{-1}_{\gamma^\prime \gamma }
\left(\Gamma_F^{(2)}\right)^{-1}_{\delta^\prime \delta}\nonumber\\
&&\frac{\partial^4 \Gamma_F}{\partial \psi_\alpha \partial \psi_\epsilon
\partial \psi_\gamma \partial \psi_\delta}
\end{eqnarray}

This identity can directly be translated into the bosonic formulation by use
of the identities (\ref{eq:vanishing}) and (\ref{eq:functional}):
\begin{eqnarray} \label{eq:bosonic}
&&G^{-1}_{\alpha \beta}=-j_{\alpha \beta}+\frac{1}{2}\lambda_{\alpha \beta \gamma \delta}
G_{\gamma \delta}+Y_{\alpha \beta}
\nonumber \\
&&Y_{\alpha \beta}=\frac{1}{6}\lambda_{\eta \beta \gamma \delta}G^{-1}_{\alpha \epsilon}
\Big[ (\Gamma_B^{(2)})^{-1}_{\epsilon \eta, \gamma \delta}
+G_{\epsilon \gamma} G_{\eta \delta}-G_{\epsilon \delta} G_{\eta \gamma}  \Big]
\end{eqnarray}
The exact identity (\ref{eq:bosonic}) is a
central equation for this paper. Its iterative solution will lead to a loop expansion
for the BEA. All higher order loop effects are contained in the term $Y_{\alpha\beta}$.
An estimate of its size can therefore be used as a error estimate for the lowest order
approximation.

A first nontrivial approximation neglects the term $Y_{\alpha \beta}$ which
corresponds to the last term in the fermionic Schwinger-Dyson equation
and is of the order $\lambda^2$. For given $j$ and $\lambda$ the lowest order
Schwinger-Dyson equation is a closed equation for $G$
\begin{equation}\label{eq:thesolutionsof}
G^{-1}_{\alpha\beta}=-j_{\alpha\beta}+\frac{1}{2}\lambda_{\alpha\beta\gamma\delta}
G_{\gamma\delta}
\end{equation}
The solutions of eq. (\ref{eq:thesolutionsof}) correspond to extrema of the free energy
$\hat{F}$ in the lowest order approximation.
With $j_{\alpha \beta}=\frac{\partial \Gamma}{\partial G_{\alpha\beta}}$
we can explicitely compute $\Gamma_B$ in the lowest order approximation by integrating
the  differential equation
\begin{equation}\label{eq:differential}
\frac{\partial \Gamma_1}{\partial G_{\alpha \beta}}=-\left(G^{-1}\right)_{\alpha \beta}
+\frac{1}{2}\lambda_{\alpha \beta \gamma \delta} G_{\gamma \delta}
\end{equation}
One obtains (note
$\partial G_{\alpha \beta}/\partial G_{\gamma \delta}
=\delta_{\alpha \gamma} \delta_{\beta \delta}
-\delta_{\alpha \delta} \delta_{\beta \gamma})$
the two loop effective action

\begin{equation}\label{eq:loop}
\Gamma_1=\frac{1}{2}\Tr~\ln~G+\frac{1}{8}
\lambda_{\alpha \beta \gamma \delta}
G_{\alpha \beta} G_{\gamma \delta}+c
\end{equation}

For qualitative estimates of the properties of a fermionic systems the lowest order BEA
(\ref{eq:loop}) will often be sufficient. It is instructive to recall the respective role of the
two terms on the r.h.s. of eq. (\ref{eq:loop}).
The second term corresponds to the classical action (without the source term $\sim j$),
whereas the first one reflects the fermionic fluctuation determinant
$(\frac{1}{2} \Tr~\ln~G=-\frac{1}{2}\ln~\det~G^{-1})$. In a graphical representation the
first term corresponds to a closed fermion loop, whereas the second term involves two loops
connected by the vertex $\lambda$. Here every factor $G_{\alpha\beta}$ corresponds to a
fermion line. A differentiation of $\Gamma_B$ with respect to $G$ corresponds to cutting
a fermion line and lowers the loop order. The two loop BEA is therefore equivalent to the one
loop SD-equation for the fermion propagator.
It may be surprising that the classical contribution appears formally as a two loop
expression. We will see in sect. \ref{freeenergy}
that actually no explicit loop calculation is needed for its evaluation.

It is straightforward to compute the value of the thermodynamic potential $F-\mu T$ or
$\overline{U}$ for given values of the source $j$.
Inserting the solution (\ref{eq:thesolutionsof}) the effective action can be expressed as
\begin{equation}
\Gamma_1=\frac{1}{2}\Tr~\ln~G+\frac{1}{4}j_{\alpha\beta}G_{\alpha\beta}+c^\prime
\end{equation}
where $G$ should be interpreted as a functional of $j$. This yields the lowest order
formula for the free energy
\begin{eqnarray} \label{eq:fieldequation}
\frac{F-\mu N}{T}&=&\Gamma_1-\frac{1}{2}G_{\alpha\beta}j_{\alpha\beta}\nonumber\\
&=&\frac{1}{2}\Tr~\ln~G
-\frac{1}{4}G_{\alpha\beta}j_{\alpha\beta}+c^\prime \nonumber\\
&=&\frac{1}{2}\Tr~\ln~G-\frac{1}{8}\lambda_{\alpha\beta\gamma\delta}
G_{\alpha\beta}G_{\gamma\delta}+\tilde{c}
\end{eqnarray}
If the field equation (\ref{eq:thesolutionsof}) admits more than one solution the expression
(\ref{eq:fieldequation}) can be used in order to determine the one which corresponds  to the
lowest value of the thermodynamic potential $F-\mu T$.
Our formalism therefore is well suited to describe competing order parameters and
first order phase transitions.

A systematic loop expansion for the BEA can be constructed by an iterative solution
for $Y_{\alpha\beta}$. In the lowest order approximation one finds $\Gamma_B^{(2)}$ from
differentiation of eq. (\ref{eq:loop}). This yields
\begin{equation}\label{eq:onehas}
\left(\Gamma^{(2)}_1\right)_{\alpha \beta, \gamma \delta}=
-G^{-1}_{\alpha \gamma} G^{-1}_{\beta \delta}
+G^{-1}_{\alpha \delta} G^{-1}_{\beta \gamma}
+\lambda_{\alpha \beta \gamma \delta}
\end{equation}
and therefore
\begin{eqnarray}\label{eq:andtherefore}
\left(\Gamma_1^{(2)}\right)^{-1}_{\alpha \beta, \gamma \delta}&=&
-G_{\alpha \gamma} G_{\beta \delta}
+G_{\alpha \delta} G_{\beta  \gamma}\nonumber\\
&&-G_{\alpha\rho}G_{\beta\sigma}
\lambda_{\rho\sigma\tau\omega}G_{\tau\gamma}G_{\omega\delta}
\end{eqnarray}
Inserting eq. (\ref{eq:andtherefore}) into the formula for
$Y_{\alpha \beta}$ in eq. (\ref{eq:bosonic}) one obtains
the two loop expression which corresponds to the ``setting'' sun diagram
\begin{equation}\label{eq:settingsun}
Y_{\alpha \beta}=-\frac{1}{6}\lambda_{\alpha \tau \rho \sigma}
\lambda_{\beta \eta \gamma \delta}
G_{\tau \eta} G_{\rho \gamma} G_{\sigma \delta}.
\end{equation}
The three loop effective action therefore receives additional quartic bosonic interactions
\begin{equation}\label{eq:approximation}
\Gamma_2=\Gamma_1-\frac{1}{48}\lambda_{\alpha \tau \rho \sigma}
\lambda_{\beta \eta \gamma \delta}
G_{\alpha \beta} G_{\tau \eta}
G_{\rho \gamma} G_{\sigma \delta}
\end{equation}
The approximation (\ref{eq:approximation}) is valid as long as
$|\lambda G^2|\ll1$ (in a parametric sense). In the opposite extreme
$|\lambda G^2|\gg1$ we may use a ``strong coupling expansion'' which will be discussed
elsewhere.

Let us finally define the two particle irreducible (2PI) part
\begin{equation}\label{31AX}
\widehat{\Gamma}=\Gamma_B-\frac{1}{2}\Tr~\ln~G
\end{equation}
It consists precisely of the sum of 2PI diagrams on the fermionic
level. This is linked in a very direct way to the observation that the self energy in
the SD equation (\ref{eq:Schwinger}) for fermions is one particle irreducible.
Differentiation with respect to $G$ cuts a line and therefore reduces the degree of
reducibility precisely by one. The one particle irreducibility of the self
energy $\partial \widehat{\Gamma}/\partial G=\Gamma_F^{(2)}+j$ is well known.
This would be a contradiction unless $\widehat{\Gamma}$ is 2PI.
The proof of two particle irreducibility of $\widehat{\Gamma}$ becomes therefore extremely
simple in our formulation.

\section{Symmetries}
\label{symmetries}
Many properties of $\Gamma[G;\lambda]$ can be understood in terms of symmetry
transformations. Let us consider a general linear transformation (with
complex $t_{\alpha\beta}$)
\begin{equation}\label{eq:lineartrans}
\tilde{\psi}_\alpha =t_{\alpha \beta} \tilde{\psi}^\prime_\beta.
\end{equation}
We will assume that $t_{\alpha \beta}$ is regular such that the inverse
$t^{-1}_{\alpha \beta}$ exists. If we also transform the sources $j$ and the quartic
coupling $\lambda$ according to
\begin{eqnarray} \label{eq:accordingto}
j_{\alpha \beta}&=&(t^{-1})_{\alpha^\prime \alpha}
(t^{-1})_{\beta^\prime \beta} j^\prime_{\alpha^\prime \beta^\prime}\nonumber\\
\lambda_{\alpha \beta \gamma \delta}
&=&(t^{-1})_{\alpha^\prime \alpha}(t^{-1})_{\beta^\prime \beta}
(t^{-1})_{\gamma^\prime \gamma} (t^{-1})_{\delta^\prime \delta}
\lambda^\prime_ {\alpha^\prime \beta^\prime \gamma^\prime \delta^\prime}
\end{eqnarray}
the action $S_j$ ((eq.\ref{eq:defined})) remains form-invariant,
\begin{equation}
S_j[\psi^\prime, j^\prime, \lambda^\prime]=S_j[\psi, j, \lambda]
\end{equation}
For transformations with unit Jacobian this also holds for the partition function,
$Z[j^\prime,\lambda^\prime]=Z[j,\lambda]$.

\medskip

More generally, for arbitrary regular $t$ the Jacobian of
the functional measure is responsible for an anomaly
\begin{equation}\label{eq:ananomaly}
Z[j^\prime, \lambda^\prime]=\det(t)Z[j,\lambda],~~~
W[j^\prime, \lambda^\prime]=W[j, \lambda]+\ln~\det(t)
\end{equation}
With $G^\prime_{\alpha \beta}=\partial W[j^\prime, \lambda^\prime]/\partial j^\prime_{\alpha \beta}$
related to $G_{\alpha \beta}$ by
\begin{equation}\label{eq:by}
G_{\alpha \beta}=t_{\alpha \alpha^\prime} t_{\beta \beta^\prime} G^\prime_{\alpha^\prime \beta^\prime}
\end{equation}
we find
\begin{equation}
\Gamma_B[G^\prime, \lambda^\prime]=\Gamma_B[G, \lambda]-\ln~\det(t).
\end{equation}
On the other hand, we note the transformation property
\begin{equation}
\det~G=\det(t^2)\det~G^\prime
\end{equation}
which implies
\begin{equation}
\Gamma_B[G^\prime,\lambda^\prime]-\frac{1}{2}\ln~\det~G^\prime
=\Gamma_B[G,\lambda]-\frac{1}{2}\ln~\det~G
\end{equation}
We may therefore write
\begin{equation}\label{eq:thereforewrite}
\Gamma_B[G,\lambda]=\frac{1}{2}\ln~\det~G+\hat{\Gamma}[G,\lambda]
\end{equation}
where the ``reduced effective action'' $\hat{\Gamma}$ is form-invariant
under a simultaneous transformation of $G$ and $\lambda$ according to
eqs. (\ref{eq:by}) (\ref{eq:accordingto}), i.e.
$\hat{\Gamma}[G^\prime, \lambda^\prime]=\hat{\Gamma}[G,\lambda]$. We conclude that
the BEA accounts for possible anomalies in a very simple and explicit form.

The structural information on the possible form of $\Gamma_B$ will be very
useful for possible truncations. For example, let us exploit the invariance
of $\hat{\Gamma}[G, \lambda]$ under simultaneous transformations of $G$ and $\lambda$.
This greatly restricts the possible independent invariants on which $\hat{\Gamma}$
can depend, i. e.
\begin{eqnarray}
I_1&=&\frac{1}{8} \lambda_{\alpha \beta \gamma \delta}G_{\alpha \beta} G_{\gamma \delta},
\nonumber \\
I_2&=&-\frac{1}{48} \lambda_{\alpha \beta \gamma \delta}
\lambda_{\alpha^\prime \beta^\prime \gamma^\prime \delta^\prime}
G_{\alpha \alpha^\prime} G_{\beta \beta^\prime}G_{\gamma \gamma^\prime} G_{\delta \delta^\prime}
\nonumber\\
I_3&=&\frac{1}{8}\lambda_{\alpha \beta \gamma \delta}
\lambda_{\alpha^\prime \beta^\prime \gamma^\prime \delta^\prime}
G_{\alpha \beta} G_{\alpha^\prime \beta^\prime}G_{\gamma \gamma^\prime} G_{\delta \delta^\prime}
\end{eqnarray}
These and other invariants involving higher powers of $G$
are constructed by contracting each index of $\lambda$ with an index of $G$.
We observe that $I_3$ and $I_1^2$ are not 2PI. The three loop BEA must therefore be
proportional to $I_2$.

Let us now come to the symmetries of $\Gamma_B$. These are all transformations acting
on $\tilde{\psi}$ which leave $\Gamma_B$ invariant, with $\lambda$ kept fixed.
Indeed, the full bosonic effective action
$\Gamma_B[G]$ is invariant under all linear transformations (\ref{eq:lineartrans}) which obey $\det(t)=1$
and leave $\lambda_{\alpha \beta \gamma \delta}$ invariant, i. e.
$\lambda_{\alpha^\prime \beta^\prime \gamma^\prime \delta^\prime}~t_{\alpha^\prime\alpha}~t_{\beta^\prime\beta}~
t_{\gamma^\prime \gamma}~t_{\delta^\prime\delta} =\lambda_{\alpha \beta \gamma \delta}$.
This may be a very large symmetry group, much larger than the symmetry leaving
$S_j[\psi]$ invariant for fixed $\lambda$ and $j$. For example, a pointlike
interaction $\overline{\lambda} \int d^dx \big(\overline{\psi}(x)\psi(x) \big)^2$
is invariant under {\em local} $U(1)$ gauge rotations (with opposite phases
for $\psi$ and $\overline{\psi}$) whereas the quadratic fermion kinetic term may
not preserve the local symmetry. Nevertheless, the bosonic effective action
$\Gamma_B$ exhibits the full local symmetry. Actually, $G(y,x)=
\langle \overline{\psi}(y) \psi (x)\rangle$ transforms with different phases at $x$ and $y$,
i. e. $G^\prime(x,y)=\exp~i\big(\alpha(x)-\alpha(y)\big)~G(x,y)$, similar to a
link variable or a string between $x$ and $y$. We will give an example for the possible
practical use of this large symmetry in appendix A. For the pointlike interactions
discussed in the next section the symmetry of $\Gamma_B$ consists of local linear
transformations with unit determinant, combined with appropriate space-transformations,
e.g. the lattice symmetries.

Finally we mention the existence of useful transformations which change the interaction.
For example, models with attractive and repulsive interactions are mapped onto each other
by a transformation (\ref{eq:lineartrans}), (\ref{eq:accordingto}) with the property
$\lambda'_{\alpha\beta\gamma\delta}=-\lambda_{\alpha\beta\gamma\delta}$. We present in
sect. \ref{mappingbetween} a map between the repulsive and attractive Hubbard
model. This establishes, for example, a direct link between antiferromagnetism in the
repulsive Hubbard model and s-wave superconductivity and charge density waves in the
attractive Hubbard model.

\section{Local interactions}
\label{localinteractions}
Several very interesting fermionic models assume a local four-fermion interaction.
We concentrate here on a single spinor field $(\psi_1(x),\psi_2(x))
=(\psi_\uparrow (x),\psi_\downarrow (x))$ with an ``antiparticle''
$(\psi_3(x),\psi_4(x))=(\overline{\psi}_\uparrow(x),
\overline{\psi}_\downarrow(x))$ (or electrons and holes). In a notation with
\begin{eqnarray}
\psi=
\left( \begin{array}{rr}
\psi_1\\
\psi_2
\end{array}\right),
~\overline{\psi}=
\left( \begin{array}{rr}
\overline{\psi}_1\\
\overline{\psi}_2
\end{array}\right)
\end{eqnarray}
the action reads

\begin{equation}\label{eq:interaction}
S_j=-\frac{1}{2}j_{\alpha \beta} \psi_\alpha \psi_\beta + \int_x {\cal L}(x)
\end{equation}
with
\begin{eqnarray}\label{eq:thelimit}
{\cal L}(x)&=&\frac{1}{2} \overline{\lambda} \Big(\overline{\psi}(x) \psi (x)\Big)^2\nonumber\\
&=&\frac{1}{2} \overline{\lambda}\Big(\overline{\psi}_1(x) \psi_1(x)
+\overline{\psi}_2(x) \psi_2(x)\Big)^2
\end{eqnarray}
In this section the integral $\int_x$ corresponds to a sum over points
$\vec{x}$ in a $D$-dimensional lattice and includes a
sum over discrete Euclidean time points
$\tau = \epsilon m_\tau,~m_\tau \in \mathbb{Z}$,
as appropriate for quantum statistics. The spinors are antiperiodic in the
$\tau$-direction with periodicity given by the inverse temperature $T^{-1}$

\begin{equation}
\psi_b(\vec{x}, \tau +T^{-1})
=-\psi_b(\vec{x}, \tau)
\end{equation}
We employ $a,b=1\dots4$ for the internal indices whereas $\alpha, \beta$
count all Grassmann variables, e. g.
$\beta=(b,x)=(b,\tau, \vec{x})$.
The limit $\epsilon \rightarrow 0$
has to be taken at the end \footnote{In a continuum notation one has
$\int_x=\epsilon^{-1}a^{-D} \int d\tau d^D \vec{x}, \delta_{x,y}
=\epsilon a^D \delta(\tau-\tau^\prime) \delta^D
(\vec{x}-\vec{y})$
with $a$ the lattice distance.}. More explicitely, one has
\begin{equation}\label{45AA}
\psi_\alpha\equiv \psi_a(x)=
\left(\begin{array}{c}
\psi(x)\\
\bar{\psi}(x)\end{array}
\right)_a
\end{equation}
and the local interaction corresponds to
\begin{eqnarray} \label{eq:inserting}
\lambda_{\alpha\beta\gamma\delta}&\equiv& \lambda_{abcd}(x,y,z,w)\nonumber\\
&=&\lambda_{abcd}\delta_{y,x}\delta_{z,x}\delta_{w,x}~,\nonumber\\
\lambda_{abcd}&=&-\overline{\lambda}
\epsilon_{abcd}
\end{eqnarray}

The interaction (\ref{eq:thelimit}) has a high degree of
symmetry which can be combined from independent local
$SL(4, {\mathbb C})$-transformations among the
four components $\psi_a(x)$ at every point $x$ and the symmetries of the lattice
(i. e. appropriate translations, rotations and reflections). The group
$SL(4,{\mathbb C})$ consists of all complex $4\times4$ matrices with unit determinant,
i.e. $t_{\alpha\beta}=t_{ab}(x)\delta_{xy}~,~\det t_{ab}=1$.
There is no anomaly
for this symmetry and the bosonic effective action is invariant under the corresponding
transformations of $G_{\alpha \beta}$. Different models like the Hubbard model \cite{HM}
or the Gross-Neveu \cite{GN} model can be obtained by choosing different sources
$j_{\alpha \beta}$
at the end. These sources will reduce the symmetry. Nevertheless, all these models
will be described by the {\em same} bosonic effective action $\Gamma_B$!

Hermiticity of the Hamiltonion is reflected by Osterwalder-Schrader positivity \cite{OS}
of the functional integral. This means that the action $S_j$ should be invariant under the
transformation $(\vartheta^2=1)$
\begin{equation}\label{eq:hermiticity}
\vartheta\Big(\psi(\tau,\vec{x})\Big)=\overline{\psi}(-\tau,\vec{x})
\end{equation}
if accompanied by complex conjugation of all coefficients and total reordering of the
Grassmann variables, e.g.
\begin{eqnarray}
\vartheta\Big(\tilde{\psi}_a(\tau,\vec{x})\tilde{\psi}_b(\tau^\prime,\vec{y})\Big)&=&
\theta_{bb^\prime}\tilde{\psi}_{b^\prime}(-\tau^\prime,\vec{y})\theta_{aa^\prime}
\tilde{\psi}_{a^\prime}(-\tau,\vec{x})\nonumber\\
&=&-\theta_{aa^\prime}\tilde{\psi}_{a^\prime}(x^\prime)
\tilde{\psi}_{b^\prime}(y^\prime)
\theta^T_{b^\prime b}
\end{eqnarray}
In a convenient basis with spin notation (\ref{45AA}) one may have \footnote{
In some other basis the transformation $\theta$ can be appropriately modified.}
\begin{equation}\label{eq:coefficient}
\theta= \left(
\begin{array}{cc}
0&1\\
1&0
\end{array}\right)
\end{equation}
For ``even sources'' obeying
\begin{equation}
j_{ab}(x,y)=-\theta^T_{aa^\prime}j_{a^\prime b^\prime}^{*}(x^\prime,y^\prime)\theta_{b^\prime b}
\end{equation}
the action $S_j$ is indeed invariant and $W[j]$ therefore real. Discarding spontaneous
breaking of the $\vartheta$-symmetry this implies
\begin{eqnarray}\label{eq:vartheta}
\vartheta\Big(G_{ab}(x,y)\Big)&=&-\theta_{aa^\prime}G_{a^\prime b^\prime}(x^\prime,y^\prime)
\theta^T_{b^\prime b}(x,y)\nonumber\\
&=&G^*_{ab}(x,y)
\end{eqnarray}
and we may restrict the bosonic fields $G_{\alpha\beta}$ to those obeying eq.
(\ref{eq:vartheta}) without changing our previous constructions. For the choice
(\ref{eq:coefficient}), (\ref{eq:vartheta}) one finds
\begin{eqnarray}
\langle\overline{\tilde{\psi}}(x)\tilde{\psi}(x)\rangle&=&
\langle\overline{\tilde{\psi}}(x^\prime)\tilde{\psi}(x^\prime)\rangle^* \nonumber\\
\langle\overline{\tilde{\psi}}(x)\vec{\tau}\tilde{\psi}(x)\rangle&=&
\langle\overline{\tilde{\psi}}(x^\prime)\vec{\tau}\tilde{\psi}(x^\prime)\rangle^* \nonumber\\
\langle\overline{\tilde{\psi}}_1(x)\overline{\tilde{\psi}}_2(x)\rangle&=&
-\langle\tilde{\psi}_1(x^\prime)\tilde{\psi}_2(x^\prime)\rangle^*
\end{eqnarray}

\bigskip
\begin{center}
{\bf Hubbard model}
\end{center}

In the following we concentrate on the Hubbard model
while the Gross-Neveu model and its
relation to the Hubbard model are discussed in appendix A.
The Hubbard model has a repulsive local interaction $U>0$ with action $S$ given by
\begin{eqnarray}\label{eq:Hubbard}
S= \int\limits^\beta_0
d\tau\Big[&&{\sum_{i}} \left\{\overline{\psi}_i\partial_\tau \psi_i
-\mu \overline{\psi}_i \psi_i
+\frac{1}{2}U\left(\overline{\psi}_i \psi_i\right)^2\right\}\nonumber\\
+&&\sum_{ij}\overline{\psi}_i \widehat{T}_{ij}\psi_j\Big]
\end{eqnarray}
Here $i,j$ denote the lattice sites in a $D$-dimensional cubic lattice. We concentrate
on the simplest version where $\widehat{T}_{ij}=-t$ for next neighbours and zero otherwise. The
identification with our previous discussion holds for $\overline{\lambda}=\epsilon U$.
A local source term is given by the chemical potential $\mu$ which vanishes only for
half filling
\begin{equation}\label{halffilling}
j_{13}(x,x)=j_{24}(x,x)=-\mu \epsilon
\end{equation}
Another source term arises from the next neighbour interactions (or hopping term)
\begin{eqnarray} \label{hopping}
j_{13}(\vec{x},\tau ;\vec{x}\pm \vec{a},\tau)&=&j_{24}(\vec{x}, \tau ;\vec{x}\pm\vec{a},\tau )
\nonumber \\
=-j_{31}(\vec{x},\tau ;\vec{x} \pm \vec{a},\tau)
&=&-j_{42}(\vec{x},\tau ;\vec{x} \pm \vec{a}, \tau)
=-t\epsilon
\end{eqnarray}
Here $\vec{a}$ is the unit lattice vector, $a=|\vec{a}|$ the lattice distance and
$x=(\tau, \vec{x}),\vec{x}=a\vec{m}$ with $\vec{m}$ a set of integers associated to $i$.
Finally, the $\tau$-derivative in $\int d \tau \overline{\psi} \partial_\tau \psi$ is
expressed by
\begin{eqnarray}\label{setof}
&&j_{13}(\vec{x},\tau;\vec{x},\tau +\epsilon)
=-j_{13}(\vec{x},
\tau;\vec{x},\tau-\epsilon)= \nonumber \\
&&j_{24}(\vec{x}, \tau;\vec{x}, \tau+\epsilon)
=-j_{24}(\vec{x},\tau;\vec{x},\tau-\epsilon)= \nonumber \\
&&j_{31}(\vec{x},\tau;\vec{x},\tau+\epsilon)
=-j_{31}(\vec{x},\tau;\vec{x},\tau-\epsilon)= \nonumber \\
&&j_{42}(\vec{x},\tau;\vec{x},\tau+\epsilon)
=-j_{42}(\vec{x},\tau;\vec{x},\tau-\epsilon)= -\frac{1}{2}
\end{eqnarray}
One should take $\epsilon \rightarrow 0$ at the end.

\section{Gap equation}
\label{gapequation}
For local interactions the last term in eq. (\ref{eq:thesolutionsof}) is local. In the
lowest order approximation the inverse fermionic propagator can therefore be written as
the sum of the inverse classical propagator, $-j$, and a local ``gap'' $\Delta$,
\begin{equation}\label{eq:equal}
G^{-1}_{ab}(x,y)=-j_{ab}(x,y)+\Delta_{ab}(x)\delta_{xy}
\end{equation}
where $\delta_{xy}$ is an appropriate generalization of
the Kronecker symbol. Indeed, a nonzero $\Delta$ in certain channels often
induces an effective mass gap. In terms of the fermion propagator at equal arguments
\begin{equation}\label{eq:auxiliary}
g_{ab}(x)=G_{ab}(x,x)
\end{equation}
the generalized gap equation obtains by inversion of eq. (\ref{eq:equal})
\begin{equation}\label{eq:gap}
g_{ab}(x)=(-j+\Delta)^{-1}_{ab}(x,x)
\end{equation}
We further note the leading order relation (\ref{eq:thesolutionsof})
between $\Delta$ and $g$
\begin{equation}
\Delta_{ab}(x)=-\frac{\overline{\lambda}}{2} \epsilon_{abcd} g_{cd}(x)
\end{equation}
This turns eq. (\ref{eq:gap}) into a closed ``gap equation'' for $\Delta$
\begin{equation}\label{eq:thelocalgap}
\Delta_{ab}(x)=-\frac{\overline{\lambda}}{2}\epsilon_{abcd}(-j+\Delta)^{-1}_{cd}(x,x)
\end{equation}

It is instructive to write this gap equation in Fourier space. We define
\begin{eqnarray}
G_{ab}(x,y)&=&\int_p\int_q e^{-ipx} e^{iqy}G_{ab}(p,q)~,\nonumber\\
G_{ab}(p,q)&=&-G_{ba}(-q,-p)
\end{eqnarray}
where the momentum integration reads
\begin{eqnarray}
\int_p&\equiv&\epsilon a^D\int\frac{d^dp}{(2\pi)^d}~,\nonumber\\
\delta_{pq}&=&\left(\epsilon a^D\right)^{-1}\left(2\pi\right)^d
\delta^d(p-q)
\end{eqnarray}
On a cubic lattice the integration interval is restricted by
$|p_k|\leq \pi/a$ and $\delta^D(p-q)$ has to be taken modulo $2 \pi/a$.
Finally, for $\tau$ on a torus with circumference $T^{-1}$, as appropriate for the
Matsubara formalism with $p_0=2 \pi nT$, we take
\begin{eqnarray}
\int_p&\equiv&\epsilon a^DT\sum_n\int\frac{d^D\vec{p}}{(2\pi)^D}~,\nonumber\\
\delta_{pq}&=&\left(\epsilon a^DT\right)^{-1} \delta_{mn} \left(2\pi\right)^D
\delta^D(\vec{p}-\vec{q})
\end{eqnarray}
The $p_0$-integration is bound by $|p_0|<\pi/ \epsilon$ or,
equivalently, the Matsubara sum extends over a range $|n|\leq 1/(2\epsilon T)$ with
$\delta_{mn}$ taken modulo $(\epsilon T)^{-1}$. (The largest possible value for
$\epsilon$ is $T^{-1}$.) The gap equation involves the momentum integration
characteristic for a one loop expression
\begin{eqnarray}\label{eq:inversion}
g_{ab}(Q)&=&\int \limits_x e^{-iQx} g_{ab}(x)\nonumber\\
&=&\int \limits_q(-j+\Delta)^{-1}_{ab}(q-Q,q)=-g_{ba}(Q)
\end{eqnarray}

We will concentrate on translation invariant sources
\begin{equation}
j_{ab}(p,q)=J_{ab}(q)\delta_{pq}~,~J_{ab}(q)=-J_{ba}(-q)
\end{equation}
A particularly simple situation arises for a translation invariant gap
$\Delta_{ab}(x)=\overline{\Delta}_{ab}$,
$\Delta_{ab}(p,q)=\overline{\Delta}_{ab}\delta_{pq}$ where $(-j+\Delta)$ is diagonal
in momentum space and can therefore easily be inverted. One finds that also $g_{ab}$
is translation invariant, $g_{ab}(x)=\overline{g}_{ab}$, and the homogeneous gap
equation
\begin{equation}
\overline{\Delta}_{ab}=-\frac{\overline{\lambda}}{2}\epsilon_{abcd}
\int\limits_q\left(-J(q)+\overline{\Delta}\right)^{-1}_{cd}
\end{equation}
only needs the inversion of a $4x4$ matrix for every $q$ separately.
More generally, the gap can reflect spontaneous breaking of translation
symmetry if $\Delta(Q)\neq 0$ for $Q\neq 0$, where $Q=q-p$ and
\begin{eqnarray}
\Delta(p,q)=\Delta(q-p)&=&\Delta(Q)\nonumber\\
=\int_x e^{-iQx} \Delta(x)&=&-\Delta^T(Q)
\end{eqnarray}
We will see in sect. \ref{antiferromagnetism} how to solve the gap equation
(\ref{eq:inversion}) in the case of an
inhomogeneous gap.

\section{Free energy for local gaps}
\label{freeenergy}
In general, the free energy functional (\ref{eq:equilibrium}) is a rather complicated object. We will show here that
in the local gap approximation (corresponding to the leading order SD-equation) it can be
reduced to a comparatively simple functional of the local gap $\Delta(x)$, namely
\begin{equation}\label{eq:secondterm}
\frac{\hat{F}_1}{T}=\frac{\hat{F}_0}{T}-\frac{1}{2}\Tr~\ln(-j+\Delta)+\tilde{c}
\end{equation}
where
\begin{equation}\label{eq:exact}
\frac{\hat{F}_0}{T}=\frac{1}{8\overline{\lambda}}\epsilon_{abcd}\int_x\Delta_{ab}(x)\Delta_{cd}(x)
\end{equation}
The first term $\hat{F}_0$ is a simple mass like term for the local field $\Delta$, whereas the
second term in eq. (\ref{eq:secondterm}) reflects the fermionic fluctuation determinant in
presence of the gap. The explicit expression (\ref{eq:exact}) assumes the minimal local
interaction discussed in sect. \ref{localinteractions}.
The construction given below can easily be generalized to more complex interactions.

We recall at this point that a reconstruction of the free energy by integration of the gap
equations (cf. \cite{BR}) is problematic. In general, one may have the knowledge about the
gap equations in several directions $\bar{\Delta}_i$ in the space of all possible
$\Delta$. This amounts to a set of partial differential equations which are equivalent to
$\partial F/\partial\bar{\Delta}_i=0$. However, the integration of a given gap equation
yields a function $T(\bar{\Delta}_i)$ about which one only knows that it has a partial
extremum in the $\bar{\Delta}_i$ direction at a value $\bar{\Delta}^{(0)}_i$ which
coincides with an extremum of $F$. This is clearly insufficient for a comparison of
different local minima. The unsatisfactory situation may be somewhat improved by
combining gap equations for more than one direction and by the inclusion of sources.
Nevertheless, the available information is often even insufficient for a check of
the integrability condition
$\partial^2 F/\partial\bar{\Delta}_i\partial\bar{\Delta}_j=\partial^2 F/\partial
\bar{\Delta}_j\partial\bar{\Delta}_i$. Our direct construction of the free energy functional
avoids all these problems.

Let us next derive the relations (\ref{eq:secondterm}), (\ref{eq:exact}) and discuss their
applicability. For this purpose we
treat the local gap approximation (\ref{eq:equal}) as an ansatz for $G^{-1}$.
In addition to the local field $\Delta$ we also use the local field $g$ as defined by eq.
(\ref{eq:auxiliary}). Since the lowest order relation (\ref{eq:gap}) is not exact we treat
$\Delta$ and $g$ as independent variables. We will see below that the gap equation
(\ref{eq:gap}) results as an extremum condition for $\hat{F}$ in leading order. Our
construction can be generalized beyond leading order where the relation between
$\Delta$ and $g$ becomes more complex. The free energy
\begin{equation}
\frac{\hat{F}_1}{T}=\Gamma_1-\frac{1}{2}G_{\alpha\beta}j_{\alpha\beta}
\end{equation}
obtains from eq. (\ref{eq:loop}) and can be written in the form (\ref{eq:secondterm}) with
\begin{equation}
\frac{\hat{F}_0}{T}=\frac{1}{2}Tr\{Gj\}+\frac{1}{8}\lambda_{abcd}
\int_xg_{ab}(x)g_{cd}(x)+c
\end{equation}
We next use $j=-G^{-1}+\Delta$ and the locality of $\Delta$. Up to a shift in the irrelevant
additive constant this yields
\begin{equation}\label{eq:reinsert}
\frac{\hat{F}_0}{T}=-\frac{1}{2}\int_x\Delta_{ab}(x)g_{ab}(x)+\frac{1}{8}
\lambda_{abcd}\int_xg_{ab}(x)g_{cd}(x)+\tilde{c}
\end{equation}

The functional $\hat{F}_1$ depends on two independent local variables $\Delta$ and $g$ as
well as on $j$. We observe that the variation with respect to $\Delta$ at
fixed $g$ and $j$ precisely yields the gap equation (\ref{eq:gap}). (A similar procedure will
fix $\Delta(g)$ also beyond the leading order.) Inserting $\Delta(g)$ the free energy
$\hat{F}$ becomes a functional of $g$ and one may then look for its minimum. For practical
computations of the extrema it is more convenient to solve first the field
equation for $g$ as a functional of $\Delta$ (from the variation of $\hat{F}_1$ with respect
to $g$ at fixed $\Delta$) and reinsert the solution $g_s[\Delta]$ into
eq. (\ref{eq:reinsert}). From
$\partial \hat{F}_1/\partial g=0$ one obtains
\begin{equation}\label{82}
-\Delta_{ab}(x)+\frac{1}{2}\lambda_{abcd}g_{cd}(x)=0
\end{equation}
or, inserting (\ref{eq:inserting})
\begin{equation}\label{eq:reinserting}
g_{ab}(x)=-\frac{1}{2\overline{\lambda}}\epsilon_{abcd}\Delta_{cd}(x)
\end{equation}
Reinserting (\ref{eq:reinserting}) into the formula (\ref{eq:reinsert}) for $\hat{F}_0$ yields the
promised simple quadratic form (\ref{eq:exact}).

\medskip

It can easily be checked that the variation of $\hat{F}_1$ (\ref{eq:secondterm}) with respect
to $\Delta(x)$ yields the local gap equation (\ref{eq:thelocalgap}). The value of $\hat{F}_1$
at the minimum also coincides with the second formula in eq. (\ref{eq:fieldequation}). It is
important, however, that the simple form (\ref{eq:secondterm}) holds not only for the value
of $\hat{F}_1$ at the minimum, but can also be used to search the extrema of $\hat{F}_1$! This turns the
free energy formula (\ref{eq:secondterm}) into a powerful tool for a comparison of different
extrema of the free energy - as needed, for example, for the description of a first order
phase transition. Nevertheless, even in lowest order of the loop expansion for local gaps
we have only shown the validity of the free energy (\ref{eq:secondterm}),(\ref{eq:exact})
for the location of the extrema and the values of $\hat{F}$ at the extrema. This is
sufficient for a computation of the ground state in lowest order, but not for the bosonic
masses and suceptibilities which involve second derivatives at the minimum. We believe,
however, that our simple formula (\ref{eq:secondterm}),(\ref{eq:exact}) can serve as a
good approximation for the bosonic masses and interactions in many practical situations.

The free energy (\ref{eq:secondterm}),(\ref{eq:exact})
also establishes a close analogy between the BEA and partial
bosonization where the term quadratic in the bosonic field results from the Hubbard-
Stratonovich transformation. In sect. \ref{partialbosonization} we will discuss
a definition of a free energy suitable for partial bosonization and make the
comparison more explicit. In fact, the SD-approach and MFT lead to the same gap equation
and the same free energy provided the partial bosonization is chosen precisely
such that the quadratic term $\hat{F}_0$ coincides in both formulations. This provides
for a very simple criterion for the possible choices of partial bosonization which are
consistent with the Hartree-Fock approximation.

We finally observe that the ansatz
\begin{eqnarray}
&&G_{ab}(x,x)=g_{ab}(x)~,\nonumber\\
&&G_{ab}(x,y\neq x)=(-j+\Delta)^{-1}_{ab}(x,y)
\end{eqnarray}
can be used for a simple estimate of the size of the higher order corrections in eq.
(\ref{eq:settingsun}) (\ref{eq:approximation}). The three loop contribution in eq.
(\ref{eq:approximation}) has a local term $\sim \lambda^2g^4$ and a nonlocal term
\footnote{The part $\sim(-j+\Delta)^{-4}(x,x)$ has to be substracted in order to avoid
double counting.} $\sim\lambda^2(-j+\Delta)^{-4}$. In particular, the size of the local
correction can easily be compared with the leading order term $\sim\lambda g^2$ once $g$
has been computed in leading order.

\section{Antiferromagnetism in the Hubbard model}
\label{antiferromagnetism}
For the Hubbard model with repulsive coupling one does not expect a homogenous gap.
For example, antiferromagnetic behavior corresponds to a flip of sign for
neighboring lattice sites
\begin{eqnarray}\label{eq:condensate}
\langle\overline{\psi} (x)\vec{\tau} \psi(x)\rangle
&=&-\langle\overline{\psi}(x+a)\vec\tau \psi(x+a) \rangle\nonumber\\
&=&\frac{\pi^2}{h_a}\vec{a}~ \exp(-iQ_ax)
\end{eqnarray}
with $Q_a=(0,\pi /a,\pi /a, \dots)$. (We consider here a $D$-dimensional cubic
lattice $(d=D+1)$ with lattice distance $a$
and have introduced the ``Yukawa coupling'' $h_a$ for later convenience.)
For the condensate (\ref{eq:condensate}) the translations by two lattice distances
2$a$ do not change the state. We therefore investigate gaps with the reduced
symmetry of translations by 2$a$:
\begin{eqnarray}\label{eq:inhomogenousgap}
\Delta(Q)&=&\Delta_h~\delta_{Q,0}+\Delta_a~\delta_{Q,Q_a}~,\nonumber\\
\Delta(p,q)&=&\Delta_h~\delta_{p,q}+\Delta_a~\delta_{p+Q_a,q}.
\end{eqnarray}
Below the homogenous gap $\Delta_h$ will be connected to charge density and the
inhomogeneous gap $\Delta_a$ to antiferromagnetism.

With respect to the fundamental translations $t_x,t_y$ by one lattice distance
the gaps $\Delta_h=-\Delta_h^T$ and $\Delta_a=-\Delta_a^T$ have even and odd parity
whereas $J$ is even
\begin{equation}
t_x(J)=J~,~t_x(\Delta_h)=\Delta_h~,~t_x(\Delta_a)=-\Delta_a.
\end{equation}
This tells us immediately that for a similar decomposition
$(g_h=-g_h^T,~g_a=-g^T_a)$
\begin{equation}
g(Q)=g_h~\delta_{Q,0}+g_a~\delta_{Q,Q_a}
\end{equation}
the ``antiferromagnetic part'' $g_a$ must involve an odd power of $\Delta_a$.
For the Hubbard model with next neighbour interactions the nonvanishing sources read
\begin{eqnarray}
J_{13}(-q)&=&J_{24}(-q)=-J_{31}(q)=-J_{42}(q)\nonumber\\
&=&[2\pi inT-\mu -2t\sum^D_{k=1}~\cos(aq_k)]\epsilon
\end{eqnarray}
Our task is the inversion of the matrix $(-j+\Delta)$ in the gap equation (\ref{eq:inversion})
for the case of the inhomogeneous gap (\ref{eq:inhomogenousgap}).

Let us restrict the discussion to a gap
\begin{eqnarray}\label{eq:localgap}
&&(\Delta_h)_{13}=(\Delta_h)_{24}=\tilde{\rho}\epsilon , \\
&&(\Delta_a)_{13}=-(\Delta_a)_{24}=-\tilde{a}_3\epsilon,~
(\Delta_a)_{23}=-\tilde{a}_1\epsilon+i\tilde{a}_2\epsilon,~ \nonumber\\
&&(\Delta_a)_{14}=-\tilde{a}_1\epsilon-i\tilde{a}_2\epsilon\label{eq:localgap1}
\end{eqnarray}
where $\tilde{\rho}$ transforms as a charge density and
$\vec{a}=(\tilde{a}_1,\tilde{a}_2,\tilde{a}_3)$ is a real antiferromagnetic spin vector.
We find that the pieces $\sim\tilde{a}_1,\tilde{a}_3$ commute with $-J+\Delta_h$
whereas the one $\sim \tilde{a}_2$ anticommutes. On the other hand, the contributions
$\sim\tilde{a}_1,\tilde{a}_3$ are real and the one $\sim\tilde{a}_2$ is purely
imaginary. This implies for real $\tilde{a}_i,\tilde{\rho}$
\begin{equation}
\Delta_a^*(-J(q)+\Delta_h)^*=(-J(q)+\Delta_h)\Delta_a
\end{equation}
The inversion of the matrix $(-j+\Delta$) in eq. (\ref{eq:inversion}) can now be achieved
by using the identity
\begin{eqnarray}
&&(-j+\Delta)^{-1}(p,q)=\nonumber\\
&&-{\cal N}^{-1}(p)
\Big\{\Big(J(p+Q_a)-\Delta_h\Big)^*\delta_{pq}+\Delta^*_a\delta_{p+Q_a,q}\Big\}
\end{eqnarray}
with
\begin{eqnarray}
{\cal N}(p)&=&\Big(J(p+Q_a)-\Delta_h\Big)^*\Big(J(p)-\Delta_h\Big)-\Delta^*_a\Delta_a \nonumber \\
&=&\epsilon^2\{{\cal A}(p)+4\pi inT\mu_{\textup{eff}} ~\textnormal{diag}(1,1,-1,-1)\} \nonumber \\
{\cal A}(p)&=&(2\pi nT)^2+\alpha-\mu^2_{\textup{eff}} +4t^2\Big(\sum_k~\cos~ap_k\Big)^2
\end{eqnarray}
depending on
\begin{equation}
\mu_{\textup{eff}} =\mu+\tilde{\rho}~,~\alpha=\tilde{a}_j \tilde{a}_j
\end{equation}
Here ${\cal N}^{-1}(p)$ plays the role of an effective squared fermion propagator

\begin{eqnarray}
{\cal N}^{-1}(p)&=&{\cal M}^{-1}(p){\cal N}^*(p)\epsilon^{-4}~,~ \nonumber \\
\nonumber \\
{\cal M}(p)&=&{\cal N}^*(p){\cal N}(p)\epsilon^{-4}\nonumber \\
&=&\Big[(2\pi nT)^2+\alpha +4t^2(\sum_k~\cos~ap_k)^2-\mu^2_{\textup{eff}} \Big]^2\nonumber\\
&&+4\mu^2_{\textup{eff}}(2\pi nT)^2
\end{eqnarray}

\begin{figure}
\centering
\psfrag{AF}{AF}
\psfrag{T/U[K/eV]}{$T/U[K/eV]$}
\psfrag{hrhorho}{$\mu_{\textup{eff}}$}
\includegraphics[scale=0.37]{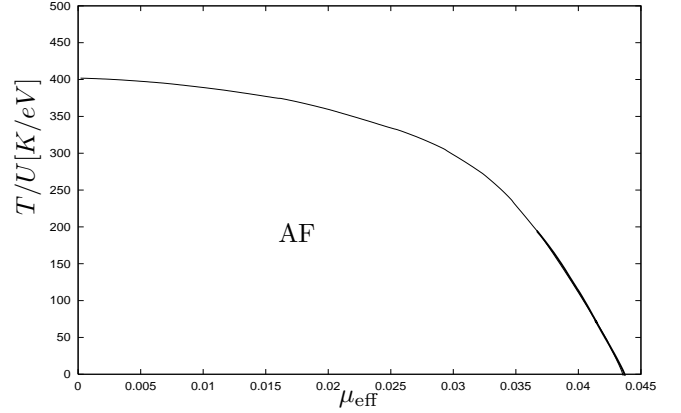}
\caption[]{Phase boundary for the antiferromagnetic (AF) phase in the Hubbard
model. We consider next neighbour interactions with $U=t$. The modifications of this leading
order diagram due to bosonic fluctuations are important \cite{BBWN}, see sect.
\ref{functionalrenormalization}.}
\label{fig:rho_T}
\end{figure}

For $\alpha >0$ we observe that ${\cal M}(p)$ is strictly positive for all $p$, even for
$T=0$. An antiferromagnetic condensate therefore acts as mass gap for the fermions.
Inserting our results in (\ref{eq:inversion}) one finds that $g_{ab}(Q)$ vanishes except
for
\begin{eqnarray}
g(0)&=&-\int_q {\cal N}^{-1}(q)\Big(J(q+Q_a)-\Delta_h\Big)^* \nonumber \\
g(Q_a)&=&-\int_q {\cal N}^{-1}(q)\Delta^*_a\nonumber\\
&=&\frac{1}{\epsilon}\int_q {\cal M}^{-1}(q)
{\cal A}(q)
\left(\begin{array}{cc}
0&\tilde{a}_k\tau_k\\
-\tilde{a}_k\tau_k^*&0
\end{array}\right)
\end{eqnarray}
In leading order one has ($\overline{\lambda}=U\epsilon)$
\begin{eqnarray}
g(Q_a)=U^{-1}
\left(\begin{array}{cc}
0&\tilde{a}_k\tau_k\\
-\tilde{a}_k\tau_k^*&0
\end{array}\right)
\end{eqnarray}
This yields the antiferromagnetic gap equation
\begin{equation}
\frac{U}{\epsilon} \int_q {\cal M}^{-1}(q){\cal A}(q)=1
\end{equation}
where we recall $\int_q=\epsilon a^DT \sum\limits_n(2\pi)^{-D}\int d^Dq,~q_k^2\leq (\pi /a)^2$
and $n$ is half integer. As it should be, this gap equation
becomes independent of $\epsilon$
\begin{equation}\label{eq:centralresult}
a^D UT \sum_n \int \dfrac{d^D \vec{q}}{(2\pi)^D}
\dfrac{{\cal A}(q)}{{\cal A}^2(q)+(4\pi nT\mu_{\textup{eff}})^2}=1
\end{equation}
The solution of the gap equation (\ref{eq:centralresult})
determines the antiferromagnetic order parameter $\alpha$ as a function of $T$ and
$\mu_{\textup{eff}}$. In particular, it allows us to determine the region in $T$ and $\mu$ for
which the antiferromagnetic state $(\alpha > 0)$ occurs. We show this region in
fig. \ref{fig:rho_T}. By construction the gap equation (\ref{eq:centralresult}) is one loop
exact and therefore consistent with Hartree-Fock \cite{HFH}. Nevertheless, we have
implemented the spin rotation symmetry in a simple and direct way and can therefore proceed
to a straightforward computation of the free energy for the antiferromagnetic state (below)
and the spin wave excitations above the antiferromagnetic ground state \cite{BBW2,BBWN}.

Eq. (\ref{eq:centralresult}) coincides with the gap equation in the mean field approximation
derived in the ``colored Hubbard model'' \cite{BBW} if we identify
(cf. eq. (\ref{eq:condensate}))
$\tilde{\alpha}=h^2_a\alpha=h_a^2\vec{a}\vec{a},~\mu_{\textup{eff}} =h_\rho\rho,~
U=2h_a^2/\pi^2$. In contrast to the mean field approximation where the relation between
$h_a^2$ and $U$ depends on free parameters $\lambda_i$ which reflect the details of the
bosonization procedure $(h_a^2=(2\lambda_1+\lambda_2-3\lambda_3+1)\pi^2U/3)$,
the present Schwinger-Dyson equation gives a unique relation between the four
fermion coupling $U$ and the effective Yukawa coupling $h_a$. (The latter characterizes
the interaction between the composite boson fields $\vec{a}(x)$ responsible for
antiferromagnetism and the fermions.)
In particular, the part of the phase diagram shown in fig. \ref{fig:rho_T} corresponds
to one of the phase diagrams shown in \cite{BBW}, now for a particular value
$h_a^2=(\pi^2/2)U$. For half filling $(\mu=0),t=U$ and $U\approx 1\textup{eV}$ the transition
temperature is well compatible with the observed range of transition temperatures
for the antiferromagnetic phase in materials which exhibit high temperature superconductivity
in the case of doping (e.g. $\mu \neq 0)$. The inclusion of the bosonic fluctuations will
lower the critical temperature (sect. \ref{functionalrenormalization}).

For the free energy of the antiferromagnetic state we can directly use the formula
(\ref{eq:exact}) and insert eqs.(\ref{eq:localgap})(\ref{eq:localgap1}), resulting in
\begin{equation}\label{94}
\frac{\hat{F}_0}{T}=\frac{1}{U}\int d\tau\sum_i(\alpha-\tilde{\rho}^2)
\end{equation}
The term $\sim \alpha$ coincides with the result of partial bosonization \cite{BBW}
precisely for the choice $h^2_a=(\pi^2/2)U$. For given $\mu_{\textup{eff}}$ the computation of
the free energy can therefore be directly inferred from \cite{BBW}. On the other hand, the
negative sign of the term $\sim \tilde{\rho}^2$ indicates that there exists no consistent
choice of partial bosonization which reproduces the gap equation of BEA for real
$\tilde{\rho}$. (A consistent choice must have a positive quadratic term.) This demonstrates
in a simple fashion that it is not always possible to define a well defined
Hubbard-Stratonovich transformation such that the mean field approximation
is consistent with the Hartree-Fock result!

A similar investigation can be done for the possibility of superconductivity.
A local $d_{x^2-y^2}$-wave superconducting order parameter reads
\begin{eqnarray}
i\langle \psi (x)\tau_2 \psi(x)\rangle&=&
\frac{\pi^2}{2h_d}d \sum^2_{k=1}~\exp(-iQ^k_dx_k) \nonumber \\
i\langle\overline{\psi} (x)\tau_2 \overline{\psi}(x)\rangle&=&
-\frac{\pi^2}{2h_d}\overline{d} \sum^2_{k=1}~\exp(-iQ^k_dx_k)
\end{eqnarray}
with $Q^1_d=(0,\pi/a,0,0,\dots),~Q_d^2=(0,0,\pi/a,0,\dots)$. The discrete symmetry
$\vartheta$ (cf. eqs (\ref{eq:hermiticity}),(\ref{eq:coefficient}) implies
\footnote{This could be different
for other realizations of $\vartheta$. Replacing the unit in eq. (\ref{eq:coefficient}) by $\tau_3$
would lead to $\overline{d}=-d^*$.} $\overline{d}=d^*$.
Similarly, a $d_{xy}$-wave or $s$-wave superconducting order parameter takes the form
\begin{equation}
i\langle \psi(x)\tau_2 \psi(x)\rangle=
\frac{\pi^2}{h_c}c~\exp\Big(-i(Q_ax)\Big)
\end{equation}
or
\begin{eqnarray}
i\langle \psi(x)\tau_2 \psi(x)\rangle&=&
\frac{\pi^2}{h_s}s
\end{eqnarray}
Here $i\tau_2$ assures the antisymmetry and all above order parameters are
invariant under spin rotations.

In particular, for a pure $d_{x^2-y^2}$-superconductor only $d$ differs from zero and we
may consider a local gap
\begin{equation}
\Delta(p,q)=\Delta_h\delta_{p,q}+\frac{1}{2}\Delta_d\sum_{k=1}^2\delta_{p+Q^k_d,q}
\end{equation}
with $\Delta_h$ given by eq.(\ref{eq:localgap}) and $(\Delta_d)_{12}=d,~
(\Delta_d)_{34}=-d^*$ or
\begin{equation}\label{eq:localgap2}
\Delta_d=i\epsilon
\left(\begin{array}{lr}
\tilde{d}\tau_2&0\\
0&-\tilde{d}^*\tau_2
\end{array}\right)
\end{equation}
The computation can now be performed in analogy to the antiferromagnetic gap.
For the local gap (\ref{eq:localgap2}) the ``classical free energy'' reads
\begin{equation}\label{107}
\frac{\hat{F}_0}{T}=-\frac{1}{2U}\int d\tau\sum_i \tilde{d}^*\tilde{d}
\end{equation}
Due to the negative sign in eq. (\ref{107}) one finds no solution for the
lowest order gap equation with $d^* d\neq 0$.

Nonlocal superconducting gaps are well motivated by functional renormalization group
studies \cite{MEZ,16A,16B}. They have been investigated in the framework of partial
bosonization in \cite{BBW}. Unfortunately, the Fierz ambiguity influences strongly
the MFT phase diagram. A computation of the BEA would therefore be quite useful.
However, in lowest order all nonlocal superconducting gaps vanish, i.e. $G_{12}(x,y\neq x)=0$, in
the case of local interactions. A nontrivial gap equation for a nonlocal superconducting
order parameter must therefore proceed to the next to leading order equation based on eqs.
(\ref{eq:bosonic}), (\ref{eq:settingsun}). As an alternative, we notice that the
renormalization flow of the fermionic interactions induces nonlocal interactions once
the fluctuations with $q^2>k^2$ are included. This is precisely the aim of the renormalized
gap equation derived in sect. \ref{renormalizedgapequation}.
Within the BEA the integration of the high momentum fluctuations with $q^2>k^2$ can
be performed by an approximate solution of the exact functional renormalization group
equations which will be derived in sect. \ref{exactfermionic}.
A proposal for the stabilization of
d-wave super conductivity in an approach similar to the BEA can be found in \cite{NM}.

\section{Symmetries of the Hubbard model}
\label{symmetrieshubbard}
In this section we consider more closely the symmetries of the Hubbard model. They will
give important constraints for the general form of the free energy. We will restrict
the discussion here to those transformations that also leave the nonlocal source terms
(\ref{hopping})(\ref{setof}) invariant. On the other hand, the chemical potential
(local source (\ref{halffilling})) will only appear through the field equation
as a possible symmetry breaking term. Besides the discrete lattice symmetries the next-
neighbour Hubbard model exhibits a global $SU(2)\times SU(2)$ symmetry.

Consider first local infinitesimal transformations
\begin{equation}\label{SH1}
\delta\psi_a(x)=iL_{ab}(x)\psi_b(x)
\end{equation}
The interaction term is invariant provided $\Tr L=0$ and invariance of the term in eq.
(\ref{eq:Hubbard}) involving $\partial_\tau$,
\begin{equation}\label{SH2}
{\cal L}_\tau=\bar{\psi}\partial_\tau \psi=\frac{1}{2}\psi_a K_{ab}\partial_\tau\psi_b~,~
K=\left({0~1}\atop{1~0}\right)
\end{equation}
requires
\begin{equation}\label{SH3}
L^TK+KL=0
\end{equation}
Both conditions are obeyed of $L$ generates the symmetry
$SU(2)\times SU(2)$, with
\begin{equation}\label{SH4}
L(x)=\vec{\alpha}(x)\vec{S}+\vec{\beta}(x)\vec{U}
\end{equation}
where $\vec{S}$ are the spin generators
\begin{equation}\label{SH5}
\vec{S}_1=\frac{1}{2}\left(
\begin{array}{cc}
\vec{\tau}&0\\
0&-\vec{\tau}^T
\end{array}\right)
\end{equation}
and
\begin{eqnarray}\label{SH6}
U_1&=&\frac{1}{2}\left(
\begin{array}{ll}
0&\tau_2\\
\tau_2&0
\end{array}\right)~,~
U_2=\frac{1}{2}\left(
\begin{array}{cc}
0&-i\tau_2\\
i\tau_2&0
\end{array}\right)~,~ \nonumber\\
U_3&=&\frac{1}{2}
\left(
\begin{array}{cc}
1&0\\
0&-1
\end{array}\right)
\end{eqnarray}
obey the commutation relation $[U_i,U_j]=i\epsilon_{ijk}U_k,$ ~ $[S_i,U_j]=0$. The electric
charge generator is given by $Q=-2U_3$ whereas $U_1$ and $U_2$ change electrons into holes.
We note that the chemical potential respects the spin and charge symmetry
$SU(2)\times U(1)$ while
it is not invariant under the transformations generated by $U_1$ and $U_2$.

For the Hubbard model on a quadratic (cubic) lattice with next neighbour
interactions we may divide the lattice
into two sublattices $A,B$, such that the hopping term $\sim t$ connects spinors on different
sublattices $\sim\bar{\psi}_A\psi_B$.
The next neighbour interaction remains invariant under
a global $SU(2)\times SU(2)$ symmetry with
\begin{equation}\label{SH7}
\vec{\alpha}_A=\vec{\alpha}_B,~
\beta_{3A}=\beta_{3B},\beta_{1A}=-\beta_{1B},~
\beta_{2A}=-\beta_{2B}
\end{equation}
The Hubbard model with next neighbour interaction has no further continuous symmetry
beyond $SU(2)\times SU(2)$. Adding diagonal interactions reduces this symmetry to
$SU(2)\times U(1)$.

The spinors $\psi_a$ transform as the $(2,2)$ representation of
$SU(2)\times SU(2)$, which takes
its canonical form in the basis
\begin{equation}\label{SH8}
\phi=\left(
\begin{array}{rrr}
\psi_4&,&-i\psi_1\\
-\psi_3&,&-i\psi_2
\end{array}\right)  ~,~
\delta\phi=\frac{1}{2}(\vec{\alpha}\vec{\tau}\phi-\vec{\beta}\phi\vec{\tau})
\end{equation}
Equivalently, we can also use
\begin{equation}\label{SH9}
\tilde{\phi}=\left(
\begin{array}{rrr}
\psi_2&,&-\psi_1\\
i\psi_3&,&i\psi_4
\end{array}\right)
=i\tau_2\phi^T\tau_2~,~
\delta\tilde{\phi}=\frac{1}{2}
(\vec{\beta}\vec{\tau}\tilde{\phi}-\vec{\alpha}\tilde{\phi}\vec{\tau})
\end{equation}

In consequence, the spinor bilinears $g_{ab}$ belong to two irreducible representations
$(3,1)+(1,3)$. The magnetization
\begin{equation}\label{SH10}
\vec{m}(x)=\langle\bar{\psi}(x)\vec{\tau}\psi(x)\rangle=
\langle\psi_a(x)(K\vec{S})_{ab}
\psi_b(x)\rangle
\end{equation}
is a vector with respect to the spin rotations and a singlet with respect to the
generators $\vec{U}$. On the other hand,
\begin{equation}\label{SH11}
\vec{u}(x)=\langle\psi_a(x)(K\vec{U})_{ab}\psi_b(x)\rangle
\end{equation}
is a vector with respect to the $\vec{U}$-rotations and a spin singlet. In particular,
an antiferromagnetic state corresponds to $\vec{m}(x)\sim\vec{a}\exp (-iQ_ax)$ and
the charge density $\rho(x)$ is given by $\rho(x)\equiv u_3(x)=
\langle\bar{\psi}(x)\psi(x)\rangle$. The six local fermion bilinears $g_{ab}(x)$
can be written as linear combinations of $m_i(x)$ and $u_i(x)$. In particular, the
electron-pairs (or hole pairs)
\begin{eqnarray}\label{SH12}
\pi(x)&=&i\langle\psi(x)\tau_2\psi(x)\rangle
=2\psi_1(x)\psi_2(x)~,~\nonumber\\
\pi^*(x)&=&-i\langle\bar{\psi}(x)\tau_2\bar{\psi}(x)\rangle
=-2\psi_3(x)\psi_4(x)
\end{eqnarray}
can be written as
\begin{equation}\label{SH13}
\pi=u_2+iu_1~,~\pi^*=u_2-iu_1
\end{equation}
Consistently with eq. (\ref{eq:vartheta}) the fields $\vec{m}$ and $\vec{u}$ are real.
As a consequence of the $SU(2)\times SU(2)$ symmetry the effective action can only depend
on invariants as $u_k(x)u_k(x)$ etc. In particular, the free energy
(\ref{94}) can  be extended to
\begin{equation}\label{SH14}
\frac{\hat{F}_0}{T}=\frac{1}{U}\int d\tau\sum_i(\alpha-\gamma)~,~
\gamma=\tilde{\rho}^2+\pi^*\pi=u_ku_k
\end{equation}
(for homogeneous $\pi\sim s$). For $\mu\neq0$, the degeneracy between the three components
of $\vec{u}$ will be lifted by the solution to the field equation, since the source
$\sim\mu$ points in the three-direction and therefore breaks the invariance  with respect
to $\vec{U}$-transformations.

\section{Mapping between repulsive and attractive Hubbard model}
\label{mappingbetween}
In this section we discuss a mapping from the repulsive
$(U>0)$ to the attractive $(U<0)$
Hubbard model. The automorphism
\begin{equation}\label{SH15}
T_A:\phi\rightarrow i\tilde{\phi}~,~\tilde{\phi}\rightarrow -i\phi
\end{equation}
maps the generators $\vec{S}$ and $\vec{U}$ onto each other. (On the level of
individual components one has \\
$T_A:\psi_4\rightarrow i\psi_2~,~\psi_2\rightarrow-i\psi_4~,~
\psi_1\rightarrow\psi_1~,~\psi_3\rightarrow\psi_3.)$ \\
It changes the sign of the interaction term $(U\rightarrow -U)$ and leaves the term
${\cal L}_\tau$ (\ref{SH2}) invariant. We may consider a second automorphism
\begin{equation}\label{SH16}
T_B:\phi\rightarrow-i\tau_3\tilde{\phi}\tau_3~,~
\tilde{\phi}\rightarrow i\tau_3\phi\tau_3
\end{equation}
that also changes the sign of the interaction and leaves ${\cal L}_\tau$ invariant.
(It reads in components $T_B:\psi_4\rightarrow-i\psi_2,~\psi_2\rightarrow i\psi_4,~
\psi_1\rightarrow\psi_1,~\psi_3\rightarrow\psi_3.$) This transformation differs
from $T_A$ by an additional flip of sign of the ``lower spin components''
$\psi_2,\psi_4$. The next neighbour coupling of the Hubbard model remains invariant if
$T_A$ acts on the sublattice $\psi_A$ and $T_B$ acts on $\psi_B$. For $\mu=0$ we have
therefore established a transformation that maps the attractive on the repulsive
Hubbard model and vice versa.

In order to investigate the interesting consequences of this map we should have a
look at the action on the local fermion bilinears $\vec{m}$ and $\vec{u}$. For the sites
of the sublattices A and B one has different transformation properties
\begin{eqnarray}\label{SH17}
&&\vec{m}\rightarrow\vec{u}~,~\vec{u}\rightarrow\vec{m}\hspace{0.3cm}\text{for}~A\nonumber\\
&&\vec{m}\rightarrow\tilde{\vec{u}}~,~
\vec{u}\rightarrow\tilde{\vec{m}}\hspace{0.3cm}\text{for}~B
\end{eqnarray}
with $\tilde{u}_{1,2}=-u_{1,2},\tilde{u}_3=u_3$ and similar for $\tilde{m}$.
In consequence, the antiferromagnetic state of the repulsive Hubbard model is mapped into
a charge density wave with
$\langle\bar{\psi}(x)\psi(x)\rangle\sim
(-1)^P|\vec{a}|$ in the attractive Hubbard model. (Here $P=1$ on $A$ and
$P=-1$ on $B$.) It is convenient to introduce the field $\vec{v}(x)$ as
\begin{equation}\label{MH1}
v_3(x)=u_3(x)~,~v_{1,2}(x)=
\exp (iQ_ax)u_{1,2}(x),
\end{equation}
which obeys
\begin{equation}\label{MH2}
\vec{v}(x)=\left\{
\begin{array}{lll}
\vec{u}(x)&\text{for}&A\\
\tilde{\vec{u}}(x)&\text{for}&B
\end{array}\right.
\end{equation}
In terms of this field the combined automorphism $\hat{T}$ (i.e.
$\hat{T}=T_A$ on $A,~\hat{T}=T_B$ on $B$) simply maps
\begin{equation}\label{MH3}
\hat{T}~:~\vec{m}(x)\leftrightarrow \vec{v}(x)
\end{equation}
With respect to the global $\vec{U}$-transformations (\ref{SH7}) the field
$\vec{v}(x)$ indeed transforms as a global vector (i.e. the same $SU(2)$-transformation
for every $x$), in complete analogy to $\vec{m}(x)$ with respect to the
$\vec{S}$-transformations. (For the transformations $U_1,U_2$ the minus sign in eq.
(\ref{SH7}) for $B$ is absorbed by the minus in the definition of $\tilde{u}_{1,2}$.)

The mapping $\hat{T}$ (\ref{MH3}) allows a simple interpretation of several properties
of the Hubbard model. For half filling $(\mu=0)$ the low temperature state of the
repulsive Hubbard model is antiferromagnetic. Without loss of generality we may choose the
$3$-direction $\langle\vec{m}(x)\rangle=m_3(x)$
\begin{equation}\label{MH4}
m_3(x)=\bar{a}_3\exp(iQ_ax).
\end{equation}
This implies, for the same low temperature $T$ and the same $|U|$, that the attractive
Hubbard model is characterized by a charge density wave
\begin{equation}\label{MH5}
\rho(x)=\bar{\rho}\exp(iQ_a x)~,~\bar{\rho}=\bar{a}_3
\end{equation}
The $SU(2)$-spin-symmetry of the repulsive Hubbard model exhibits a degeneracy of the
order parameter in the $(a_1,a_2,a_3)$-space. Correspondingly, the $\vec{U}$-symmetry
of the attractive Hubbard model predicts a degeneracy in the space
$(v_1,v_2,v_3)$. The $SU(2)$-partners of the charge density wave
(\ref{MH5}) are the homogeneous $s$-wave Cooper-pairs
\begin{equation}\label{MH6}
\pi(x)=\bar{\pi}~,~\bar{\pi}^*\bar{\pi}=a^2_3
\end{equation}
Thus the low temperature state can also be $s$-wave super-conductivity. This degeneracy
between the charge-density wave and the $s$-wave superconducting state for $\mu=0$ seems
compatible with existing numerical results \cite{Sca}, \cite{Lacaze}. Furthermore, we predict
for $\mu=0$ the existence of two (real) massless Goldstone bosons for the ordered
state, both for $U>0$ {\em and} $U<0$.

A recent renormalization group study \cite{BBWN} has computed the temperature
dependence of the antiferromagnetic order parameter $\vec{a}$ for the repulsive
Hubbard model in two dimensions. These results can be taken over one to one for the
temperature dependence of $\bar{\rho}$ or $|\bar{\pi}|$ for the attractive Hubbard model.
The mapping $\hat{T}$ can also be applied for the discussion of the correlation length
$\xi$ for $T>T_c$. Close to the critical temperature $T_c$ one finds \cite{JuW}
\begin{equation}\label{MH7}
\xi\approx c\exp \left(20.7\frac{T_c}{T}\right)
\end{equation}
Here $T_c$ depends on the size of the experimental probe and eq. (\ref{MH7}) holds
for a size of 1 cm.
The computation \cite{BBWN} yields
for $|U|/t=3$ a value $T_c\approx 0.115 t$. However, the value of $T_c$ is not universal
and depends on details of the model (and the computation).
Consistent with the Mermin-Wagner
theorem $T_c$ vanishes logarithmically in the limit where the size of the probe becomes
infinite. The same behaviour of the correlation length governs the
(anti)ferromagnetic correlation function
$\langle m(x)m(y)\rangle\exp i\left(Q_a(y-x)\right)$ for $U>0$ and the charge density
wave and superconducting correlation functions $\langle\rho(x)\rho(y)\rangle$
$\exp i\left(Q_a(y-x)\right)$ or $\langle\pi^*(x)\pi(y)\rangle$ for $U<0$.

For the attractive Hubbard model away from half filling $(\mu\neq 0)$ one expects that the
degeneracy is lifted and the low temperature state is $s$-wave superconducting, with
$\langle\pi(x)\rangle=\bar{\pi}\neq 0$. The phase transition for $d=2$ is expected to
be of the Kosterlitz-Thouless \cite{KT} type, such that in the limit of an infinite
sample only a suitable renormalized order parameter differs from zero for
$T<T_c$ \cite{RGKT}. Indications for this transition have been seen in simulations
\cite{Sca}, \cite{Lacaze}. These properties can be mapped to the repulsive Hubbard
model in a homogeneous magnetic field. In fact, the transformation $\hat{T}$ maps
a source term with a chemical potential into one with a constant magnetic field:
\begin{equation}\label{MH8}
\hat{T}:~ \mu\rho(x)=\mu v_3(x)\leftrightarrow B_3 m_3(x)~,~B_3=\mu
\end{equation}
We conclude that the low temperature state of the repulsive Hubbard model in an
homogeneous magnetic field $\vec{B}$ (for $\mu=0)$ is antiferromagnetic with
direction orthogonal to $\vec{B}$, i.e. for $\vec{B}=(0,0,B_3)$ one has
$a_{1,2}\neq0,a_3=0$. The spin symmetry is now reduced to the $U(1)$-spin-rotations
around the three axis (which is mapped by $\hat{T}$ to the charge-symmetry for
$U<0$). In complete analogy to the superfluid for $U<0$ one predicts for
$U>0$ and $B_3\neq0$ the characteristic properties of a spontaneously broken
$U(1)$-symmetry, namely the existence of one Goldstone boson and of stable vortex
solutions.

The $\hat{T}$-transformation offers many interesting perspectives. Results from
numerical simulations of the attractive Hubbard model can be directly taken over to
appropriate quantities in the repulsive Hubbard model. In the other direction, the
presumed $d$-wave superconductivity for the repulsive Hubbard model sufficiently
away from half filling can be mapped into a corresponding spin wave in the attractive
Hubbard model in an homogeneous magnetic field.

\section{Partial Bosonization}
\label{partialbosonization}
The BEA is economical in the sense that redundancy is avoided. In principle,
$\Gamma_B$ contains the complete information about all Greens functions. Many
physical situations, however, are well approximated by the coexistence of effective
local bosonic and fermionic degrees of freedom. These aspects of locality are not
very apparent in the BEA, the latter being based on the bilocal fields $G_{\alpha\beta}$.
For situations dominated by local bosonic degrees of freedom - as often for the critical
behavior near a second order phase transition - an effective locality can be recovered by
restricting the BEA to local fields $g_{ab}(x)$. In contrast, if local fermions are relevant,
it may be advantageous to keep explicitely the local fermion fields $\psi(x)$ as well.

\bigskip
\begin{center}
{\bf Mixed effective action}
\end{center}

We may construct a ``mixed effective action'' $\Gamma_M$ which depends both
on $\psi_\alpha$ and $G_{\alpha\beta}$, thereby accepting a certain redundancy of the
description. This task is easily achieved by a Legendre transformation of
$W[\eta ,j]$ (cf. sect. \ref{bosoniceffective}) with respect to both the fermionic
and bosonic sources
\begin{equation}\label{PB1}
\Gamma_M[\psi,G]=-W[\eta,j]
+\eta_\alpha\psi_\alpha+
\frac{1}{2}j_{\alpha\beta}G_{\alpha\beta}
\end{equation}
with
\begin{equation}\label{PB2}
\psi_\alpha[\eta,j]=
\frac{\partial W[\eta,j]}{\partial\eta_\alpha}~,
G_{\alpha\beta}[\eta,j]=
\frac{\partial W[\eta,j]}{\partial j_{\alpha\beta}}
\end{equation}
The field equations read now
\begin{equation}\label{PB3}
\frac{\partial\Gamma_M}{\partial\psi_\alpha}=-\eta_\alpha~,~
\frac{\partial\Gamma_M}{\partial G_{\alpha\beta}}=j_{\alpha\beta}
\end{equation}
and the relation $W^{(2)}\Gamma^{(2)}_M=1$ is now a matrix relation in the combined
space of fermionic and bosonic fields.

The BEA is simply related to $\Gamma_M$ by restriction to $\psi=0$,
\begin{equation}\label{PB4}
\Gamma_B[G]=\Gamma_M[\psi=0,G]
\end{equation}
This follows from the fermionic character of $\psi$ which implies that $\Gamma_M$ can
only involve even powers of $\psi$. In turn,
$\eta_\alpha=-\partial\Gamma_M/\partial\psi_\alpha$ vanishes for $\psi_\alpha=0$
and the definition (\ref{PB1}) coincides for with eq. (\ref{12}). Furthermore, the
mixed derivatives $\partial^2\Gamma_M/(\partial\psi\partial G)$ vanish for
$\psi=0$. The matrix of second functional derivatives $\Gamma^{(2)}_M$ therefore
becomes block diagonal in fermion and boson space if $\psi=0$, and similar for
$W^{(2)}$ if $\eta=0$. We emphasize that the simple relation (\ref{PB4}) is
closely connected to the fermionic character of $\psi$ and bosonic character of $G$.
It would not hold automatically for the two-particle irreducible effective action
for a fundamental
bosonic theory. We also point out that the fermionic effective action $\Gamma_F[\psi]$
does not obtain for $G=0$, but rather obeys
\begin{equation}\label{PB5}
\Gamma_F[\psi;j]=
\Gamma_M\Big[\psi,G^0[\psi;j]\Big]-j\cdot
G^0[\psi;j]
\end{equation}
Thereby $G^0[\psi;j]$ is a functional of $\psi$ which is determined by the solution of the
field equation $\partial\Gamma_M/\partial G=j$ in presence of arbitrary $\psi$. With
\begin{equation}\label{PB6}
\frac{\partial\Gamma_F}{\partial\psi_\alpha}_{|j}=
\frac{\partial\Gamma_M}{\partial\psi_\alpha}_{|G}
\end{equation}
one infers
\begin{equation}\label{PB7}
\frac{\partial^2\Gamma_F}{\partial\psi_\alpha\partial\psi_\beta}=
\frac{\partial^2\Gamma_M}{\partial\psi_\alpha\partial\psi_\beta}-
\frac{\partial^2\Gamma_M}{\partial\psi_\beta\partial G_{\gamma\delta}}
\frac{\partial G^0_{\gamma\delta}[\psi]}{\partial\psi_\alpha}
\end{equation}
(For $\psi=0$ the second term on the r.h.s of eq. (\ref{PB7}) vanishes.)
Taking two more derivatives and evaluating at $\psi=0$ relates the terms quartic in
$\psi $ in $\Gamma_F$ and $\Gamma_M$
\begin{eqnarray}\label{PB8}
&&\frac{\partial^4\Gamma_F}
{\partial\psi_\epsilon\partial\psi_\eta\partial\psi_\alpha\partial\psi_\beta}=
\frac{\partial^4\Gamma_M}
{\partial\psi_\epsilon\partial\psi_\eta\partial\psi_\alpha\partial\psi_\beta}\nonumber\\
&&+\frac{\partial^3\Gamma_M}
{\partial\psi_\alpha\partial\psi_\beta\partial G_{\gamma\delta}}
\frac{\partial^2 G^0_{\gamma\delta}[\psi]}
{\partial\psi_\epsilon\partial\psi_\eta}
-\frac{\partial^3\Gamma_M}
{\partial\psi_\eta\partial\psi_\beta\partial G_{\gamma\delta}}
\frac{\partial^2 G^0_{\gamma\delta}[\psi]}
{\partial\psi_\epsilon\partial\psi_\alpha}\nonumber\\
&&+\frac{\partial^3\Gamma_M}
{\partial\psi_\epsilon\partial\psi_\beta\partial G_{\gamma\delta}}
\frac{\partial^2 G^0_{\gamma\delta}[\psi]}
{\partial\psi_\eta\partial\psi_\alpha}
\end{eqnarray}
The last three terms on the r.h.s. of eq. (\ref{PB8}) correspond to ``trees'' from
the exchange of $G$-bosons, as can be verified by taking suitable derivatives of eq.
(\ref{PB2}).

The redundancy of the mixed description is reflected by exact identities for the
derivatives of $\Gamma_M$. The relation (\ref{eq:prop}) holds for arbitrary $j$ and $\eta$.
It results in a functional identity for $\Gamma_M$ which holds for arbitrary
$\psi$ and $G$
\begin{equation}\label{PB9}
\left((\Gamma^{(2)}_M)^{-1}_{FF}\right)_{\alpha\beta}=
G_{\alpha\beta}-\psi_\alpha\psi_\beta
\end{equation}
Here $\left(\Gamma^{(2)}_M\right)^{-1}_{FF}$ denotes the fermion-fermion component
of the inverse of the matrix of second derivatives $\Gamma^{(2)}_M$. For $\psi=0$ one has
(cf. eq. (\ref{PB7}))
\begin{equation}\label{PB10}
\left(\Gamma^{(2)}_M\right)^{-1}_{FF}
[\psi=0,G]=\left(\Gamma^{(2)}_F\right)^{-1}
[\psi=0;j]
\end{equation}
and we recover from eq. (\ref{PB6}) the relation (\ref{eq:vanishing}). Further identities
can be obtained by taking partial derivatives of eq. (\ref{PB9}) with respect to $\psi$ or
$G$. The identity
\begin{equation}\label{PB11}
\frac{\partial^2\Gamma_M}{\partial\psi_\alpha\partial\psi_\beta}_{|\psi=0}
=-\left(G^{-1}\right)_{\alpha\beta}
\end{equation}
can be used for a direct evaluation of $G^{-1}$ without the need to solve a field equation.

The redundancy of $\Gamma_M$ can be greatly reduced by resticting the most general sources
$j_{\alpha\beta}$ to a small subset. For example, we may only consider local sources
$j_{ab}(x)$ and the associated local bosonic fields
$g_{ab}(x)=\delta W/\delta j_{ab}(x)$. Performing the Legendre transform similar to eq.
(\ref{PB1}) one defines an effective action $\hat{\Gamma}_{FB}$ which only depends on local
fermionic and bosonic fields
\begin{equation}\label{PBX1}
\hat{\Gamma}_{FB}=-W+\int_x
\left(\eta_a\psi_a+\frac{1}{2}j_{ab}g_{ab}\right)
\end{equation}
with field equation
\begin{equation}\label{PB12}
\frac{\delta\hat{\Gamma}_{FB}}{\delta\psi_a(x) }=-\eta_a(x)~,~
\frac{\delta\hat{\Gamma}_{FB}}{\delta g_{ab}(x)}=j_{ab}(x)
\end{equation}

The relation between $\hat{\Gamma}_{FB}$ and $\Gamma_M$ proceeds by splitting
\begin{equation}\label{PB13}
G_{\alpha\beta}=g_{ab}(x)\delta_{xy}+\hat{G}_{ab}(x,y)
\end{equation}
Next one solves the field equations for $\hat{G}_{ab}(x,y)$ as a functional of arbitrary
local fields $g_{ab}(x),\psi_a(x)$. Inserting this solution into $\Gamma_M$ yields
$\hat{\Gamma}_{FB}$. Using the redundancy of $\Gamma_M$ we may also compute
$\hat{G}_{ab}(x,y)$ from the fermionic part of $\Gamma_M$
\begin{equation}\label{PB14}
\hat{G}_{ab}(x,y)=
\left(\left(\Gamma^{(2)}_M\right)^{-1}_{FF}\right)_{ab}
(x,y)-g_{ab}\delta_{xy}
\end{equation}
As an example, we can compute the two loop effective action $\hat{\Gamma}_{FB}$ from eq.
(\ref{eq:approximation}). Using a local interaction one has
\begin{eqnarray}\label{PB15}
\hat{\Gamma}_{FB,2}&=&\frac{1}{8}\int_x\lambda_{abcd}g_{ab}(x)g_{cd}(x)
-\frac{1}{2}\Tr \ln\left(\Gamma^{(2)}_M\right)_{FF}\nonumber\\
&&-\frac{1}{48}\int_x\int_y\lambda_{abcd}
\lambda_{efgh}\nonumber\\
&&\left(\left(\Gamma^{(2)}_M\right)^{-1}_{FF}\right)_{ae}(x,y)
\left(\left(\Gamma^{(2)}_M\right)^{-1}_{FF}\right)_{bf}(x,y)\nonumber\\
&&\left(\left(\Gamma^{(2)}_M\right)^{-1}_{FF}\right)_ {cg}(x,y)
\left(\left(\Gamma^{(2)}_M\right)^{-1}_{FF}\right)_{dh}(x,y)
\end{eqnarray}
with $\Gamma^{(2)}_M$ depending on $g$ and $\psi$. We note that the full information
about all n-point functions is contained in $\hat{\Gamma}_{FB}$ due to its dependence
on the fermionic fields. The restriction to the bosonic part by setting $\psi=0$ is
now a functional of the local bosonic fields $g_{ab}(x)$ only.

\bigskip
\begin{center}
{\bf Local fermion-boson-model}
\end{center}

The partially bosonized effective action $\hat{\Gamma}_{FB}$ can be equivalently
computed from a functional integral involving fermionic and bosonic degrees of freedom.
This becomes apparent by a generalized Hubbard-Stratonovich \cite{HS} transformation.
We restrict the discussion here to a local four fermion interaction (\ref{45AA})
and rewrite the partition function in an equivalent form as
\begin{eqnarray}\label{PB16}
Z[\eta,j]&=&\int{\cal D}\tilde{\psi}{\cal D}\tilde{g}\exp
\Bigg\{-\int_x\Bigg[\left(\frac{\bar{\lambda}}{24}\epsilon_{abcd}-
\frac{1}{24}S_{abcd}\right)\nonumber\\
&&\left(\tilde{g}_{ab}(x)-\tilde{\psi}_a(x)\tilde{\psi}_b(x)
-\tilde{j}_{ab}(x)\right)\nonumber\\
&&\left(\tilde{g}_{cd}(x)-\tilde{\psi}_c(x)\tilde{\psi}_d(x)
-\tilde{j}_{cd}(x)\right)\nonumber\\
&&-\eta_a(x)\tilde{\psi}_a(x)\Bigg]
-S_j[\psi]\Bigg\}
\end{eqnarray}
Here the totally antisymmetric part of $S_{abcd}$ vanishes - otherwise
$S_{abcd}=S_{cdab}=-S_{bacd}=-S_{abdc}$ is arbitrary, reflecting the ``Fierz
ambiguity'' \cite{IW,JW}.
The integral over $\tilde{g}$ is a well defined Gaussian integral. It
only yields an irrelevant multiplicative constant provided that
\begin{eqnarray}\label{PB17}
{\cal L}_m&=&-\frac{1}{24}\tilde{\lambda}_{abcd}
\tilde{g}_{ab}(x)\tilde{g}_{cd}(x),\nonumber\\
\tilde{\lambda}_{abcd}&=&-\bar{\lambda}\epsilon_{abcd}+S_{abcd}
\end{eqnarray}
is a positive definite quadratic form. This can be established by the use of appropriate
involutions acting on the $\psi$ as complex conjugation. This involution relates the real
and imaginary parts of $g$ and correspondingly of $\tilde{g}$.

The ``insertion of a unit'' has been chosen such that the terms quartic
in $\tilde{\psi}$ cancel the quartic interaction in $S_j[\psi]$.
It is obvious that this cancellation does not depend on the choice of
$S_{abcd}$ - different choices of $S_{abcd}$ ``distribute'' the original four fermion
interaction into different channels of exchange of local bosons. The
remaining fermionic action is quadratic in $\tilde{\psi}$
\begin{eqnarray}\label{PB18}
Z&=&\int {\cal D}\tilde{\psi}{\cal D}\tilde{g}\exp
\left\{-(S_m+S_y+S_{F,kin}+S_J+S_\eta)\right\}\nonumber\\
S_y&=&\frac{1}{12}\tilde{\lambda}_{abcd}\int_x\tilde{\psi}_a(x)
\tilde{\psi}_b(x)\tilde{g}_{cd}(x),\nonumber\\
S_{F,kin}&=&\frac{1}{2}\int_x\tilde{\psi}_a(x)\tilde{\psi}_b(x)J_{ab}(x)\nonumber\\
&&-\frac{1}{2}\int_x\int_y
j_{ab}(x,y)\tilde{\psi}_a(x)\tilde{\psi}_b(y)\nonumber\\
&=&\frac{1}{2}\int_x\int_y\tilde{\psi}_a(x)K_{ab}(x,y)\tilde{\psi}_b(y),\nonumber\\
S_\eta&=&-\int_x\eta_a(x)\tilde{\psi}_a(x)
\end{eqnarray}
Here we have introduced the sources
\begin{equation}\label{PB19}
J_{ab}(x)=-\frac{1}{6}\tilde{\lambda}_{abcd}\tilde{j}_{cd}(x)~,~
\end{equation}
which appear in appropriate linear combinations in eq. (\ref{PB16})
\begin{equation}
\tilde{j}_{ab}(x)=\rho_{abcd}J_{cd}(x)
\end{equation}
The ``bosonic'' pieces of the action are then given by
\begin{eqnarray}\label{PB20}
S_m&=&\int_x{\cal L}_m~,\nonumber\\
S_J&=&-\int_x
\left(\frac{1}{2}J_{ab}(x)\tilde{g}_{ab}(x)-\frac{1}{8}\rho_{abcd}
J_{ab}(x)J_{cd}(x)\right)
\end{eqnarray}
According to our condition on $\tilde{\lambda}$ the matrix
$\rho_{ab,cd}$ is positive definite such that after diagonalization all
``mass eigenvalues'' in ${\cal L}_m$ are positive.

For a suitable choice of the local sources $J_{ab}(x)$ one may cancel the local part
of the bilinear source $j_{ab}(x,y)$, e.g. mass terms for the fermions or a chemical
potential. These pieces are now reflected by terms linear in $\tilde{g}$. The nonlocal part
of $j_{ab}(x,y)$, i.e. the kinetic operator, remains, however, and we end with a kinetic
operator
\begin{equation}\label{146A}
K_{ab}(x,y)=-j_{ab}(x,y)+J_{ab}(x)\delta_{xy}.
\end{equation}
We recognize in eq. (\ref{PB18}) the standard formulation of a partition function with
bosonic and fermionic fields, coupled by a Yukawa interaction $S_y$. We emphasize that
both the Yukawa couplings in $S_y$ and the bosonic quadratic term (``boson mass term'') in
$S_m$ depend on the choice of $S_{abcd}$. We know from our construction that all exact
results for correlation functions must be independent of $S_{abcd}$. On the other hand,
simple approximation schemes like mean field theory (which integrates out the fermion
fluctuations for a given local ``background field'' $g$) typically show a strong
dependence on $S_{abcd}$. This often reflects a severe shortcoming
dubbed ``Fierz ambiguity''
\cite{IW}. The validity of an approximation can be (partially) tested by
establishing the independence of results on $S_{abcd}$.

In complete analogy with the standard treatment we may define a
free energy $W_{FB}$ as a functional of the local
sources $J_{ab}(x)$ and $\eta_a(x)$, while keeping $K_{ab}(x,y)$ as a fixed part of
the action
\begin{equation}\label{PB19a}
W_{FB}[\eta,J]=\ln Z[\eta,J]+\frac{1}{8}\rho_{abcd}\int_x J_{ab}(x)J_{cd}(x)
\end{equation}
Defining the effective action $\Gamma_{FB}$ by the usual Legendre transformation
\begin{eqnarray}\label{PB20a}
\Gamma_{FB}[\psi,\bar{g}]&=&-W_{FB}+\int_x
(\eta_a\psi_a+\frac{1}{2}J_{ab}\bar{g}_{ab}),\nonumber\\
\bar{g}_{ab}(x)&=&\frac{\delta W_{FB}}{\delta J_{ab}(x)}~,~
\psi_a(x)=\frac{\delta W_{FB}}{\delta\eta_a(x)}
\end{eqnarray}
we recognize that the only difference between $\Gamma_{FB}$ and $\hat{\Gamma}_{FB}$
(cf. eq. (\ref{PBX1})) results from the term quadratic in $J$ in eq. (\ref{PB19}) and the
difference between $J_{ab}\bar{g}_{ab}$ and $j_{ab}g_{ab}$.

The precise definition of $j_{ab}$ in eq. (\ref{PBX1}) and $J_{ab}$ in eqs. (\ref{PB18}),
(\ref{PB19}) depends on the splitting of $j_{ab}(x,y)$ into a part kept fixed
$(K_{ab}(x,y))$ and a part with respect to which the Legendre transform is performed.
Indeed, the redundancy in the construction of the effective action
$\hat{\Gamma}_{FB}$ allows us to put a local fermion-bilinear either into the
source for $g_{ab}$ or into a part of the action. Let us incorporate the kinetic
term $K_{ab}(x,y)$ (\ref{146A}) into the action such that the source term in the
second equation (\ref{PB12}) becomes $j_{ab}(x)\rightarrow J_{ab}(x)$. Comparing
$\partial\ln Z/\partial j_{ab}=g_{ab}$ and eqs. (\ref{PB19}), (\ref{PB20}) yields then the
simple relation
\begin{eqnarray}\label{PB21}
\bar{g}_{ab}(x)&=&\langle\tilde{g}_{ab}(x)
\rangle=g_{ab}(x)+\frac{1}{2}\rho_{abcd} J_{cd}(x)\nonumber\\
&=&\langle\tilde{\psi}_a(x)\tilde{\psi}_b(x)\rangle+
\frac{1}{2}\rho_{abcd}J_{cd}(x)
\end{eqnarray}
This relates the fermion-boson model introduced via the generalized Hubbard-Stratonovich
transformation to the partial bosonization from the BEA
\begin{eqnarray}\label{PB22}
\Gamma_{FB}&=&\hat{\Gamma}_{FB}+\frac{1}{8}\rho_{abcd}\int_x
J_{ab}(x)J_{cd}(x)\nonumber\\
&=&\hat{\Gamma}_{FB}+\frac{1}{8}\rho_{abcd}\int_x
\frac{\delta\hat{\Gamma}_{FB}}{\delta g_{ab}(x)}
\frac{\delta\hat{\Gamma}_{FB}}{\delta g_{cd}(x)}
\end{eqnarray}
Both $\hat{\Gamma}_{FB}$ and $\Gamma_{FB}$ obey the same field equation
\begin{equation}\label{PB23}
\frac{\delta\hat{\Gamma}_{FB}}{\delta g_{ab}(x)}=
\frac{\delta\Gamma_{FB}}{\delta \bar{g}_{ab}(x)}=
J_{ab}(x)
\end{equation}
If we interprete $\bar{g}_{ab}(x)$ as an extremum of
$\Gamma_{J,FB}=\Gamma_{FB}-\frac{1}{2}\int_xJ_{ab}\bar{g}_{ab}$ (at fixed $J$) then
$g_{ab}$ is a corresponding extremum of
$\hat{\Gamma}_{J,FB}=\hat{\Gamma}_{FB}-\frac{1}{2}\int_x J_{ab}g_{ab}$.

Away from the extrema, however, the functionals $\Gamma_{J,FB}$ and $\hat{\Gamma}_{J,FB}$
differ. For example, the second functional derivatives at the extrema do not coincide.
This has important consequences for the mass matrices of the composite
bosons $\bar{g}$ or $g$.
Indeed, one obtains
\begin{eqnarray}\label{154RA}
&&\frac{\delta^2\Gamma_{FB}}{\delta\bar{g}_{ab}(x)\delta\bar{g}_{cd}(y)}=
\frac{\delta^2\hat{\Gamma}_{FB}}{\delta g_{ab}(x)\delta g_{cd}(y)}\nonumber\\
&&-\frac{1}{2}\rho_{efgh}
\int_z\frac{\partial^2\Gamma_{FB}}{\partial \bar{g}_{ab}(x)\partial\bar{g}_{ef}(z)}
\frac{\partial^2\hat{\Gamma}_{FB}}{\partial g_{gh}(z)\partial g_{cd}(y)}
\end{eqnarray}
by differentiating $(\partial\Gamma_{FB}/\partial\bar{g})(g+\rho J)=
(\partial\hat{\Gamma}_{FB}/\partial g)(g)=J(g)$, eq. (\ref{PB23}), with respect to $g$.
The difference between the two point functions for the bosons $\bar{g}$ and $g$ simply
reflects the fact that the choice of bosonic fields is not unique. The bosonic field
$\bar{g}$ is a nonlinear function of the fermion bilinear $g$ according to eq.
(\ref{PB21}), i.e.
\begin{equation}\label{154RB}
\bar{g}_{ab}(x)=g_{ab}(x)+\frac{1}{2}\rho_{abcd}
\frac{\delta\hat{\Gamma}_{FB}}{\delta g_{cd}(x)}[g]
\end{equation}
Of course, the computation of the correlation functions for $g$ and $\bar{g}$ yields
the same results if we either start from the mixed effective action or from partial
bosonization via a Hubbard-Stratonovich transformation.

We conclude this section with a simple but important observation. In general, the functional
$\hat{\Gamma}_{FB}[g]$ is not bounded from below. On the other hand, the functional
$\Gamma_{FB}[\bar{g}]$ contains a positive definite ``classical part'' inherited from the
Hubbard-Stratonovich transformation (cf. eqs. (\ref{PB17}), (\ref{PB18}))
\begin{equation}\label{154RC}
\Gamma^{(cl)}_{FB}=-\frac{1}{24}
\tilde{\lambda}_{abcd}\int_x\bar{g}_{ab}(x)
\bar{g}_{cd}(x)
\end{equation}
In many cases the fluctuation effects become unimportant for large values of
$|\bar{g}|$. Then $\Gamma_{FB}$ grows quadratically for large $|\bar{g}|$. This
is reflected by the positive definite second term in eq. (\ref{PB22}). We also note
that for any logarithmic divergence $\hat{\Gamma}_{FB}\sim \ln g$, where
$\hat{\Gamma}_{FB}$ may get arbitrarily large negative values, the
positive second term in eq.(\ref{PB22}) will dominate such that $\Gamma_{FB}$ gets
large positive values. We conclude that in contrast to $\hat{\Gamma}_{FB}$ the
functionals $\Gamma_{FB}$ and $\Gamma_{J,FB}$ are typically bounded from below.
The boundedness of $\Gamma_{J,FB}$ may be helpful for the classification of extrema.

In particular, we may identify $\Gamma_{J,FB}$ as a suitable definition of a free energy.
The solutions of the field equation correspond to its extrema. They are stable if the
matrix of second derivatives is positive semidefinite. As we have argued
$\Gamma_{J,FB}$ is typically bounded from below. On the other hand
its precise form depends on the particular choice of
$S_{abcd}$ (\ref{PB17}). This corresponds to the dependence of the variable
$\bar{g}$ on $S_{abcd}$ according to eq. (\ref{154RB}). As mentioned already before,
the independence of the physical quantities on the choice of $S_{abcd}$ can be
used as a test of the validity of approximations. In sect. \ref{freeenergy} we
have formulated a simple criteria for the choice of $S_{abcd}$ which ensures that the
solution of the lowest order gap equation agrees with the Hatree-Fock-approximation.
As shown explicitely in sect. \ref{antiferromagnetism} this can be achieved at best
for some of the directions in field space.

\section{Exact fermionic renormalization group equation}
\label{exactfermionic}
So far we have used the Schwinger-Dyson equation as a main tool for a computation of the
bosonic effective action. This equation does not reduce to a closed set of equations
when we go beyond the leading order gap equation. The term $\sim Y_{\alpha \beta}$ in
eq. (\ref{eq:bosonic}) involves the bosonic propagator $(\Gamma_B^{(2)})^{-1}$
and we need, in turn,
an equation for this propagator. It is obvious that the correction term
$\sim Y_{\alpha\beta}$ becomes important whenever a bosonic bound state becomes
(almost) massless. In this case $\Gamma_B^{(2)}$ has a zero eigenvalue and the bosonic
propagator $(\Gamma_B^{(2)})^{-1}$ becomes very large in some range of momenta. In
particular, the lowest order Schwinger-Dyson equation is not expected to give a reliable
description of critical phenomena associated to a second order phase transition.

Furthermore, the expansion of $\Gamma_B$ in powers of the coupling involves the
microscopic or ``bare'' coupling $\lambda$. This is not very satisfying from a
renormalization group viewpoint. In fact, it should be possible to integrate out a
certain range of short distance fluctuations $k^2<q^2<\Lambda^2$ such that a new action
$S_j$ takes into account the effects of these fluctuations.
As a result, an effective
coupling $\lambda_k$ will replace $\lambda$. One therefore would like to derive some
type of ``renormalized gap equation'' which involves $\lambda_k$ rather than $\lambda$.
In this and the next sections
we will show how both problems can be attacked by the formulation of an exact
renormalization group equation for the BEA.

We will first base
our investigation on the successful exact flow equation for the fermionic
effective average action
and translate it from the original formulation in terms of fermion fields to the
bosonic language used in this work. For this purpose we introduce a fermionic cutoff
$R^{(F)}_k$ such that only modes with momenta $q^2>k^2$ are effectively included in the
functional integral. (The momentum $q$ may measure the distance from the Fermi surface.)
Our formalism will result in an exact renormalized gap
equation which takes precisely the form of the lowest order equation (\ref{eq:thesolutionsof}).
Only the couplings and sources are replaced by renormalized couplings and sources
$\lambda_k$ and $j_k$
\begin{equation}\label{107AA}
G^{-1}_{\alpha\beta}=-(j_k)_{\alpha\beta}+\frac{1}{2}
(\lambda_k)_{\alpha\beta\gamma\delta}G_{\gamma\delta}+\left(R^{(F)}_k\right)_{\alpha\beta}
\end{equation}
For $k\rightarrow0$ the cutoff term $R^{(F)}_k$ vanishes.

We will also derive from the exact renormalization group equation for the BEA an
approximate formulae for the $k$-dependence of the renormalized coupling
$\lambda_k$
\begin{eqnarray}\label{107BB}
\partial_k(\lambda_k)_{\alpha\beta\gamma\delta}&=&
\{(\lambda_k)_{\alpha\gamma\epsilon\mu}(\lambda_k)_{\beta\delta\eta\nu}
-(\lambda_k)_{\alpha\delta\epsilon\mu}(\lambda_k)_{\beta\gamma\eta\nu}\}\nonumber\\
&&\bar{G}_{\mu\nu}\bar{G}_{\epsilon\rho}
\left(\partial_kR^{(F)}_k\right)_{\rho\sigma}\bar{G}_{\sigma\eta}
\end{eqnarray}
as well as for the renormalized source term
\begin{eqnarray}\label{155AX}
\partial_k(j_k)&=&\frac{1}{2}\big\{(\lambda_k)_{\alpha\gamma\epsilon\mu}
(\lambda_k)_{\beta\delta\eta\nu}-(\lambda_k)_{\alpha\delta\epsilon\mu}
(\lambda_k)_{\beta\gamma\eta\nu}\big\}\nonumber\\
&&\bar{G}_{\gamma\delta}\bar{G}_{\mu\nu}\bar{G}_{\epsilon\rho}
\partial_k\left(R^{(F)}_k\right)_{\rho\sigma}\bar{G}_{\sigma\eta}
\end{eqnarray}
On the r.h.s. the vertices $\lambda_k$ are contracted with fermion propagators $\bar{G}$
evaluated at the minimum of $\Gamma_j$, i.e. solutions of eq. (\ref{107AA}). The first
equation therefore takes a one loop and the second a two loop (setting sun diagram) form.
The $k$-dependence arises through the cutoff $R^{(F)}_k$.
The system of equations (\ref{107AA}-\ref{155AX}) is closed and can be used as a practical
tool. A more explicite contact to critical phenomena
will be established in sect. \ref{bosonicrenormalizationgroup}
by varying an infrared cutoff for the composite bosonic fluctuations.

In this section we make use of the functional renormalization group for the
effective average action for fermionic systems
\cite{ERGE}. We recapitulate the main features here briefly. In order to derive an
exact flow equation for the fermionic effective action $\Gamma_F$ we add
an infrared cutoff $R_k^{(F)}$ quadratic in the fermion fields
\begin{equation}\label{108}
j_{\alpha\beta}=\overline{j}_{\alpha\beta}-\left(R_k^{(F)}\right)_{\alpha\beta}
\end{equation}
Typically, the cutoff $R^{(F)}_k$ modifies the fermion propagator for a momenta
$q^2<k^2$ such that only the fluctuations with $q^2>k^2$ contribute effectively in the
$k$-dependent partition function. After the Legendre transformation we
substract the cutoff piece in the definition of the effective action ($W=\ln Z$)
\begin{equation}\label{159}
\Gamma_{F,k}=-W+\eta_\alpha\psi_\alpha-\frac{1}{2}\psi_\alpha
\left(R_k^{(F)}\right)_{\alpha\beta}\psi_\beta
\end{equation}
In eq.  (\ref{108}) $\overline{j}_{\alpha\beta}$ are the ``physical sources''
and the cutoff scale
$k$ is varied, with $R_k^{(F)}$ vanishing for $k\rightarrow 0$. The ``physical
effective action'' is then recovered as $\Gamma_F=\Gamma_{F,k=0}$.
For large $k$ the fermionic fluctuations are suppressed and $\Gamma_{F,k}$
can be evaluated perturbatively.

An exact flow or renormalization group
equation describes the dependence of $\Gamma_{F,k}$ on the scale $k$ in
terms of the exact fermionic propagator in presence of the cutoff \cite{ERGE}
\begin{equation}\label{eq:cutoff}
\partial_k\Gamma_{F,k}[\psi,\overline{j}]=-\frac{1}{2}\partial_k
\left(R_k^{(F)}\right)_{\alpha\beta}
\left(\Gamma_{F,k}^{(2)}+R_k^{(F)}\right)^{-1}_{\beta\alpha}
\end{equation}
This can be used to extrapolate from ``microphysics'' at large $k$ to the full
(quantum-)effective action for $k\rightarrow 0$. One possibility uses $R^{(F)}_k$
in order to implement an infrared cutoff \cite{ERGE}. Instead of an infrared
cutoff, $R_k^{(F)}$ may alternatively cut off the momenta which are close to
the Fermi surface. For example, the temperature in the effective fermionic
propagator may be replaced by $(T^n+k^n)^{1/n}$, cf. \cite{BBW2,BBWN}.
Approximative solutions of eq. (\ref{eq:cutoff}) yield interesting results for chiral
symmetry breaking in models with local four quark interactions \cite{EV,AO,JuW}. Also the
onset of instabilities in the Hubbard model  has been discussed in this framework \cite{16B}
and in a closely related approach \cite{16A}.

The exact flow equation has proven to be a powerful tool for renormalization group studies.
One therefore may ask what is its correspondence in the formulation of the BEA.
In this context we emphasize that the bosonic effective action $\Gamma_B$ is
valid for arbitrary sources. It therefore makes no distinction between a ``physical''
source $\overline{j}$ and an ``unphysical'' cutoff term $R_k^{(F)}$. In consequence,
the addition of the cutoff term manifests itself only in the $k$-dependence of the
{\em solution} of the field equation
\begin{equation}
\frac{\partial\Gamma_B}{\partial G_{\alpha\beta}}=\overline{j}_{\alpha\beta}
-\left(R_k^{(F)}\right)_{\alpha\beta}
\end{equation}
while $\Gamma_B$ itself is independent of $k$.
The renormalization group flow corresponds now to a trajectory in the space of
solutions for $k$-dependent sources!

This may be visualized by defining
\begin{equation}
\Gamma_{\overline{j},k}=\Gamma_B-\frac{1}{2}G_{\alpha\beta}\overline{j}_{\alpha\beta}
+\frac{1}{2}G_{\alpha\beta}\left(R_k^{(F)}\right)_{\alpha\beta}
\end{equation}
such that the solution $\overline{G}_k$ of the field equation corresponds to an
extremum of $\Gamma_{\overline{j},k}$ for all $k$
\begin{equation}\label{eq:identity}
\frac{\partial\Gamma_{\overline{j},k}}{\partial G_{\alpha\beta}}{~\atop |}_{\overline{G}_k}=0
\end{equation}
The $k$-dependence of $\Gamma_{\overline{j},k}$ at fixed $G$ is simply given by
\begin{equation}\label{eq:sec1}
\partial_k\Gamma_{\overline{j},k}=-\frac{1}{2}\partial_k\left(R_k^{(F)}\right)_{\alpha\beta}
G_{\beta\alpha}
\end{equation}
corresponding to eq. (\ref{eq:cutoff}). This is, however, not the main quantity of
interest.

A central quantity is the $k$-dependent location of the minimum of
$\Gamma_{\overline{j},k}$ which defines the ``average fermion propagator''
$\bar{G}_k$ as a solution of eq. (\ref{eq:identity}) for a fixed source $j$.
The relevant exact flow equation can be obtained by taking a
total $k$-derivative of the identity (\ref{eq:identity}),
\begin{equation}
\partial_k\frac{\partial\overline{\Gamma}_j}{\partial G_{\alpha\beta}}{~\atop |}_{\overline{G}}
+\frac{1}{2}\frac{\partial^2\overline{\Gamma}_j}
{\partial G_{\alpha\beta}\partial G_{\epsilon \eta}}
\partial_k\overline{G}_{\epsilon \eta}=0
\end{equation}
and reads
\begin{equation}\label{eq:sec2}
\partial_k\overline{G}_{\epsilon \eta}=-\frac{1}{2}
\left(\Gamma_B^{(2)}\right)^{-1}_{\epsilon \eta,\alpha\beta}
\partial_k\left(R^{(F)}_k\right)_{\alpha\beta}
\end{equation}
The equivalence of eq.(\ref{eq:sec2}) with the flow of the inverse fermion
propagator can be established by using $\overline{G}^{-1}=\Gamma_F^{(2)}+R_k^{(F)}$
and the identity (\ref{eq:functional}), i. e.
\begin{eqnarray}
\partial_k\left(\Gamma_F^{(2)}\right)_{\alpha\beta}&=&
-\left(\overline{G}^{-1}\partial_k\overline{G}~\overline{G}^{-1}\right)_{\alpha\beta}
-\partial_k\left(R_k^{(F)}\right)_{\alpha\beta}\nonumber \\
&=&\frac{1}{2}\left(\Gamma_F^{(4)}\right)_{\alpha\beta\gamma\delta}
\overline{G}_{\gamma\epsilon}
\overline{G}_{\delta\eta}
\partial_k\left(R^{(F)}_k\right)_{\epsilon \eta}
\end{eqnarray}
This is precisely what follows by taking two derivatives of eq.(\ref{eq:cutoff})
with respect to $\psi$, evaluated at $\psi=0$.

The r.h.s. of eq. (\ref{eq:sec2}) involves the bosonic propagator $(\Gamma_B^{(2)})^{-1}$.
We therefore also want to know the $k$-dependence of $\Gamma_B^{(2)}$ and, more
generally, of the $n$-point functions corresponding to a given source
$\overline{j}$. This can be found by decomposing $G=\overline{G}_k+\Delta G$
and differentiating $\Gamma_k[\Delta G]=\Gamma_{\overline{j}}(\overline{G}+\Delta G)$
with respect to $\Delta G$. As an example, we may investigate the $k$-dependence of
the inverse bosonic propagator
\begin{eqnarray} \label{168}
&&\partial_k \left(\frac{\partial^2\Gamma_B}{\partial\Delta G_{\alpha\beta}
\partial\Delta G_{\gamma\delta}} \right){~\atop|}_{\Delta G=0}\nonumber=\\
&&\frac{1}{2}\frac{\partial^3\Gamma_B}{\partial G_{\alpha\beta}\partial G_{\gamma\delta}
\partial G_{\epsilon \eta}}{~\atop|}_{G=\overline{G}} \partial_k\overline{G}_{\epsilon \eta}
\end{eqnarray}
In a bosonic matrix notation this reads\footnote{Recall that the bosonic indices correspond
the fermionic double indices $(\alpha\beta)$ such
that $G$ and $R_k^{(F)}$ are vectors and $\lambda$ or
$\Gamma_B^{(2)}$ are matrices,
with $A\cdot B=\frac{1}{2}A_{\alpha\beta}B_{\alpha\beta}$ and similar for matrices.}
\begin{equation}\label{eq:sec3}
\partial_k\Gamma_B^{(2)}=-\Gamma_B^{(3)}\cdot(\Gamma_B^{(2)})^{-1}
\cdot\partial_kR_k^{(F)}
\end{equation}
and similar flow equations hold for higher couplings $\Gamma_B^{(n)}$.
(At first sight, eq.(\ref{eq:sec3})
bears little resemblance with the fermionic flow equation. In order to establish
the equivalence one needs to establish an identity similar to eq.(\ref{eq:functional})
relating $\Gamma_B^{(3)}$ to derivatives of $\Gamma_F$.)
In summary, we have obtained a sequence of exact flow equations
(\ref{eq:sec1}), (\ref{eq:sec2}), (\ref{eq:sec3}) etc.

One could evaluate eq.(\ref{eq:sec3}) in the approximation (\ref{eq:onehas}),
(\ref{eq:andtherefore}) where
$\Gamma_B^{(3)}=\Gamma_{\det}^{(3)}$ is given by
\begin{eqnarray} \label{eq:isgivenby}
(\Gamma^{(3)}_B)_{\alpha\beta\gamma\delta\epsilon\eta}
=&-&G^{-1}_{\alpha\epsilon} G^{-1}_{\gamma\eta} G^{-1}_{\beta\delta}
+ G^{-1}_{\alpha\eta} G^{-1}_{\gamma\epsilon} G^{-1}_{\beta\delta}\nonumber \\
&-&G^{-1}_{\alpha\gamma} G^{-1}_{\beta\epsilon} G^{-1}_{\delta\eta}
+ G^{-1}_{\alpha\gamma} G^{-1}_{\beta\eta} G^{-1}_{\delta\epsilon}\nonumber \\
&+&G^{-1}_{\alpha\epsilon} G^{-1}_{\delta\eta} G^{-1}_{\beta\gamma}
- G^{-1}_{\alpha\eta} G^{-1}_{\delta\epsilon} G^{-1}_{\beta\gamma}\nonumber \\
&+&G^{-1}_{\alpha\delta} G^{-1}_{\beta\epsilon} G^{-1}_{\gamma\eta}
- G^{-1}_{\alpha\delta} G^{-1}_{\beta\eta} G^{-1}_{\gamma\epsilon}
\end{eqnarray}
and obtain
\begin{eqnarray}\label{eq:169}
\partial_k\Big(&\Gamma_B^{(2)}&
(\overline{G})\Big)_{\alpha\beta\gamma\delta}=\nonumber\\
&-&\overline{G}^{-1}_{\alpha\gamma}\partial_k\left(R_k^{(F)}\right)_{\beta\delta}
-\overline{G}^{-1}_{\beta\delta}\partial_k\left(R_k^{(F)}\right)_{\alpha\gamma}\nonumber \\
&+&\overline{G}^{-1}_{\alpha\delta}\partial_k\left(R_k^{(F)}\right)_{\beta\gamma}
+\overline{G}^{-1}_{\beta\gamma}\partial_k\left(R_k^{(F)}\right)_{\alpha\delta}\nonumber \\
&+&\frac{1}{2}\overline{G}_{\rho\epsilon}\partial_t\left(R_k^{(F)}\right)_{\epsilon\eta}
\overline{G}_{\eta\sigma}
\left\{\lambda_{\rho\sigma\beta\delta}\overline{G}^{-1}_{\alpha\gamma}\right.\nonumber\\
&+&\left.\lambda_{\rho\sigma\alpha\gamma}\overline{G}^{-1}_{\beta\delta}
- \lambda_{\rho\sigma\beta\gamma}\overline{G}^{-1}_{\alpha\delta}
- \lambda_{\rho\sigma\alpha\delta}\overline{G}^{-1}_{\beta\gamma}\right\}
\end{eqnarray}
This reflects, however, only the change of the part in $\Gamma_B^{(2)}$
which results from the term $\Gamma_{\det}=\frac{1}{2}\ln~\det~G$ in
eq. (\ref{eq:thereforewrite}). More interesting is the change in the 2PI-piece
$\widehat{\Gamma}^{(2)}$ as we will show below.

\section{Renormalized gap equation}
\label{renormalizedgapequation}
In this section we derive the renormalized gap
equation (\ref{107AA}) which has the structural form of the lowest
order Schwinger-Dyson equation. However, using appropriately renormalized sources
and couplings $j_k$ and $\lambda_k$ this equation becomes exact.

Let us first argue that the 2PI-part (\ref{31AX})
$\widehat{\Gamma}^{(2)}[\overline{G}]$ corresponds to an effective
$k$-dependent renormalized coupling $\lambda_k$.
Indeed, the expansion of $\Gamma_{\overline{j},k}$ around the $k$-dependent minimum
$\overline{G}_k$
\begin{equation}
\Gamma_{\overline{j},k}=\frac{1}{8}(\Gamma_B^{(2)}
[\overline{G}_k])_{\alpha\beta\gamma\delta}
\Delta G_{\alpha\beta}\Delta G_{\gamma\delta}+\cdots
\end{equation}
\begin{equation}\label{123}
(\Gamma_B^{(2)}[\overline{G}])_{\alpha\beta\gamma\delta}=
-\overline{G}^{-1}_{\alpha\gamma} \overline{G}^{-1}_{\beta\delta}
+ \overline{G}^{-1}_{\alpha\delta} \overline{G}^{-1}_{\beta\gamma}
+(\widehat{\Gamma}^{(2)}[\overline{G}])_{\alpha\beta\gamma\delta}
\end{equation}
can be compared with eqs. (\ref{eq:loop}), (\ref{eq:onehas}).
This demonstrates that $\widehat{\Gamma}^{(2)}[\overline{G}_k]$ plays
the role of a running coupling and we define
\begin{equation}\label{eq:ZR99}
(\lambda_k)_{\alpha\beta\gamma\delta}=\widehat{\Gamma}^{(2)}_{\alpha\beta,\gamma\delta}
[\overline{G}]
\end{equation}
We emphasize that the renormalized coupling $\lambda_k$ will typically not
describe a purely local interaction, but rather exhibit a nontrivial momentum dependence
(form factors). This holds even if we start from a strictly local microscopic
interaction.

For large values of $k$ the renormalized coupling $\lambda_k$ equals
the microscopic coupling $\lambda$.
The lowest order Schwinger-Dyson equation (\ref{eq:differential})
\begin{equation}
\overline{G}^{-1}_{\alpha\beta}=-\overline{j}_{\alpha\beta}
+\left(R^{(F)}_k\right)_{\alpha\beta}
+\frac{1}{2}\lambda_{\alpha\beta\gamma\delta}
\overline{G}_{\gamma\delta}
\end{equation}
becomes always a good approximation for large enough $k$ where the contribution from
fluctuations is suppressed due to the effective cutoff $R_k^{(F)}$.
Indeed, for $k\rightarrow\infty~,~R_k^{(F)}\rightarrow\infty$ we may write
$\overline{G}=(-\overline{j}+R_k^{(F)}+\Delta)^{-1}$ and solve the lowest order
gap equation (\ref{eq:differential})
\begin{equation}
\Delta_{\alpha\beta}=\frac{1}{2}\lambda_{\alpha\beta\gamma\delta}
\left(-\overline{j}+R_k^{(F)}+\Delta\right)^{-1 }_{\gamma\delta}~,~
\Delta=\lambda\cdot\overline{G}
\end{equation}
From the asymptotic behavior
\footnote{We choose $R_k^{(F)}$ such that all eigenvalues diverge for $k\rightarrow\infty$.}
$\overline{G}\sim(R_k^{(F)})^{-1}~,~
\overline{G}\cdot\lambda\cdot\overline{G}\sim(R_k^{(F)})^{-2}$ it is
apparent that the lowest order Schwinger-Dyson equation becomes exact
\footnote{This implies for
$k\rightarrow\infty$ the lowest order expression
$\Gamma_{\overline{j},k}=\frac{1}{2}\Tr~\ln\Big\{G(R_k^{(F)}-j)\Big\}
+(R_k^{(F)}-j)\cdot G+\frac{1}{2}G\cdot\lambda\cdot G+c(k)$ with
$c$ independent of $G$.} in this limit.
For large enough $k_0$ we may therefore
start with the ``initial condition'' $\lambda_{k_0}=\lambda$.

We next use our definition of $\lambda_k$ (\ref{eq:ZR99}) in order to
write down an exact gap
equation which is valid for arbitrary scales $k$. Indeed, defining also
the ``renormalized sources''
\begin{equation}\label{eq:renormalization}
(j_k)_{\alpha\beta}=\overline{j}_{\alpha\beta}
-\frac{\partial\widehat{\Gamma}}{\partial G_{\alpha\beta}}[\overline{G}]
+\frac{1}{2}(\lambda_k)_{\alpha\beta\gamma\delta}
\overline{G}_{\gamma\delta}
\end{equation}
the minimum $\overline{G}$  obeys the {\em exact}
renormalized Schwinger-Dyson equation for all $k$
\begin{equation}\label{eq:Schwinger-Dyson}
\overline{G}^{-1}_{\alpha\beta}=
-(j_k)_{\alpha\beta}+(R_k^{(F)})_{\alpha\beta}
+\frac{1}{2}(\lambda_k)_{\alpha\beta\gamma\delta}
\overline{G}_{\gamma\delta}
\end{equation}
This has the generic form of the lowest order equation but the complexity
of the problem is now encoded in the need to compute the renormalized source and
coupling $j_k$ and $\lambda_k$. For $k\rightarrow 0$ the cutoff $R^{(F)}_k$ vanishes.
If one succeeds to compute $\lambda_{k\rightarrow 0}$ and
$j_{k\rightarrow 0}$ in dependence on the
initial values $\lambda$ and $\bar{j}$ one ends with an exact gap equation
relating the ``physical sources''
$\overline{j}$ to the full fermion propagator $\overline{G}$. One can then use the
lowest order Schwinger equation (\ref{eq:Schwinger}) with renormalized parameters
$j_{k\rightarrow 0},\lambda_{k\rightarrow 0}$
instead of the microscopic parameters $j,\lambda$. Instead of
the completely neglected higher order terms in a truncation of eq. (\ref{eq:vanishing}) with
$Y_{\alpha\beta}=0$ the approximations are now linked to the use of approximative
values for $j_k$ and $\lambda_k$. Even if only approximative flow equations for $\lambda_k$
and $j_k$ can be solved in practice,
a large part of the contribution $\sim Y_{\alpha\beta}$ can be absorbed into renormalized
parameters in this way.

So far, the exact renormalized Schwinger-Dyson equation (\ref{eq:Schwinger-Dyson}) is only
a formal organization of the field equation (\ref{eq:identity}). Its practical use
depends on the ability to compute $\lambda_k$ and $j_k$. In order to get some
intuition for the renormalized coupling $\lambda_k$ we first compare it to the
$1PI$-four fermion vertex $\tilde{\lambda}_k$. Indeed, the
renormalized four fermion coupling $\tilde{\lambda}_k$ can be related to
$\lambda_k$ by use of eq. (\ref{eq:functional}), i.e.
\begin{eqnarray}\label{177YA}
(\tilde{\lambda}_k)_{\alpha\beta\gamma\delta}&=&
\frac{\partial^4\Gamma_{F,k}}
{\partial\psi_\alpha\partial\psi_\beta\partial\psi_\gamma\partial\psi_\delta}_{|\psi=0}\nonumber\\
&=&-\bar{G}^{-1}_{\alpha\gamma}\bar{G}^{-1}_{\beta\delta}+
\bar{G}^{-1}_{\alpha\delta}\bar{G}^{-1}_{\beta\gamma}\nonumber\\
&&-\left(\Gamma^{(2)}_B\right)^{-1}_{\alpha'\beta'\gamma'\delta'}
\bar{G}^{-1}_{\alpha'\alpha}
\bar{G}^{-1}_{\beta'\beta}
\bar{G}^{-1}_{\gamma'\gamma}
\bar{G}^{-1}_{\delta'\delta}
\end{eqnarray}
(The presence of the cutoff $R^{(F)}_k$ does not change the identity
(\ref{eq:functional}).) In a matrix notation we write eq. (\ref{123}) as
\begin{equation}\label{177YB}
\Gamma^{(2)}_B={\cal G}+\lambda_k={\cal G}({\mathbbm 1}+{\cal G}^{-1}\lambda_k)
\end{equation}
with
\begin{eqnarray}\label{177YC}
{\cal G}_{\alpha\beta,\gamma\delta}
&=&-\bar{G}^{-1}_{\alpha\gamma}
\bar{G}^{-1}_{\beta\delta}+
\bar{G}^{-1}_{\alpha\delta}
\bar{G}^{-1}_{\beta\gamma}\nonumber\\
{\cal G}^{-1}_{\alpha\beta,\gamma\delta}&=&-
\bar{G}_{\alpha\gamma}
\bar{G}_{\beta\delta}+
\bar{G}_{\alpha\delta}
\bar{G}_{\beta\gamma}
\end{eqnarray}
This yields the exact relation
\begin{eqnarray}\label{177YD}
(\tilde{\lambda}_k)_{\alpha\beta\gamma\delta}&=&{\cal G}_{\alpha\beta,\gamma\delta}
-\left[({\mathbbm 1}+{\cal G}^{-1}\lambda_k)^{-1}
{\cal G}^{-1}\right]_{\alpha'\beta',\gamma'\delta'}\nonumber\\
&&\bar{G}^{-1}_{\alpha'\alpha}
\bar{G}^{-1}_{\beta'\beta}
\bar{G}^{-1}_{\gamma'\gamma}
\bar{G}^{-1}_{\delta'\delta}\nonumber\\
&=&{\cal G}_{\alpha\beta,\gamma\delta}-
\left[{\cal G}({\mathbbm 1}+{\cal G}^{-1}\lambda_k)^{-1}\right]_{\alpha\beta,\gamma\delta}
\nonumber\\
&=&\left[\lambda_k({\mathbbm{1}}+
{\cal G}^{-1}\lambda_k)^{-1}\right]_{\alpha\beta,\gamma\delta}
\end{eqnarray}
An expansion in $\lambda_k$ shows that $\tilde{\lambda}_k$ agrees with
$\lambda_k$ only in lower order
\begin{equation}\label{177YE}
\tilde{\lambda}_k=\lambda_k-\lambda_k
{\cal G}^{-1}\lambda_k+\lambda_k
{\cal G}^{-1}\lambda_k
{\cal G}^{-1}\lambda_k-\dots
\end{equation}

The difference between $\lambda_k$ and $\tilde{\lambda}_k$ is crucial in the region
where one of the composite bosons becomes massless. Then $\Gamma^{(2)}_B$ has a zero
eigenvalue and we learn from eq. (\ref{177YA}) that $\tilde{\lambda}_k$ diverges
- as expected from the diagrams with exchange of a massless boson in the partially
bosonized language. This divergence of $\tilde{\lambda}_k$ has been observed in the
study of flow equations for the Hubbard model \cite{MEZ,16A,16B} or quark models models
for strong interactions \cite{EV,JuW,AO}. We emphasize that the divergence of
$\tilde{\lambda}_k$ is inevitable if massless bosons are exchanged. However, the
coupling $\lambda_k$ may remain finite even for for
$\tilde{\lambda}_k\rightarrow\infty$
\begin{equation}\label{177YF}
\lambda_k=(1-\tilde{\lambda}_k{\cal G}^{-1})^{-1}\tilde{\lambda}_k
\end{equation}
Indeed, a zero eigenvalue of $\Gamma^{(2)}_B$ suggests a
finite $\lambda_k=\hat{\Gamma}^{(2)}$ in eq. (\ref{123}).
The important consequences for the running of $\lambda_k$ will be discussed in the
next section.

Finally, we also should acquire
some intuition for the definition (\ref{eq:renormalization}) of the
renormalized sources $j_k$. For all $k$ we may approximate
\begin{eqnarray}\label{eq:weapproximate}
\Gamma_{\overline{j},k}&=&\frac{1}{2}\ln~\det~G+\frac{1}{8}(\lambda_k)_{\alpha\beta\gamma\delta}
G_{\alpha\beta}G_{\gamma\delta}\nonumber\\
&&-\frac{1}{2}(j_k)_{\alpha\beta}G_{\alpha\beta}
+\frac{1}{2}\left(R^{(F)}_k\right)_{\alpha\beta}G_{\alpha\beta}
\end{eqnarray}
Then the renormalized sources are defined such that the
minimum $\overline{G}$ of the approximated form
(\ref{eq:weapproximate}) coincides with the true minimum of $\Gamma_{\overline{j},k}$. The
renormalization of the source term has its direct correspondence in the renormalization
of the kinetic and mass terms in the fermionic language. Furthermore, using
the definition $\lambda_k=\widehat{\Gamma}^{(2)}[\overline{G}_k]$ the flow equation
for $j_k$ is directly related to the one for $\lambda_k$
(cf. eq.(\ref{eq:renormalization}))
\begin{equation}\label{130}
\partial_k(j_k)_{\alpha\beta}=\frac{1}{2}\partial_k(\lambda_k)_{\alpha\beta\gamma\delta}
\overline{G}_{\gamma\delta}
\end{equation}
We will see that $j_k$ differs from $\bar{j}$ only by the effects of 2PI-two-loop
diagrams.

\section{Gap improved running of couplings}
\label{runningcoupling}
Based on the results of the preceeding section we propose in this section
approximate formulae for the
flow of $\lambda_k$ and $j_k$ and an approximate renormalized gap equation that can be
used in practice. In this context we recover a nonperturbative flow equation for the
renormalized four fermion interaction $\tilde{\lambda}_k$ that has been extensively studied
in the past. We compare the flow equations for $\lambda_k$ and
$\tilde{\lambda}_k$ and argue that the diseases of the flow of
$\tilde{\lambda}_k$ are cured for the flow of $\lambda_k$.

Let us next first turn to the flow of the renormalized coupling $\lambda_k$. The
change of $\lambda_k$ reflects the higher order terms in
$\widehat{\Gamma}$.
We can write eq. (\ref{168}) in the form
\begin{eqnarray}\label{178AX}
\partial_k\Gamma^{(2)}_B&=&\partial_k\Gamma^{(2)}_{\det}+\partial_k\hat{\Gamma}^{(2)}\nonumber\\
&=&\Gamma^{(3)}\partial_t\bar{G}=\Gamma^{(3)}_{\det}\partial_t\bar{G}+
\hat{\Gamma}^{(3)}\partial_t\bar{G}
\end{eqnarray}
The part $\partial_k\Gamma^{(2)}_{\det}=\Gamma^{(3)}_{\det}\partial_t\bar{G}$ is
described in eqs. (\ref{eq:isgivenby}), (\ref{eq:169}) and we remain with the exact flow
equation $\partial_k\lambda_k=\hat{\Gamma}^{(3)}\partial_t\bar{G}$ or
\begin{equation}\label{131}
\partial_k\lambda_k=-\widehat{\Gamma}^{(3)}\cdot
(\Gamma^{(2)}_{\det}+\lambda_k)^{-1}\cdot
\partial_kR_k^{(F)}
\end{equation}
Here $(\lambda_k)_{\alpha\beta,\gamma\delta}
=(\widehat{\Gamma}^{(2)}[\overline{G}])_{\alpha\beta,\gamma\delta}$
is interpreted as a matrix and
\begin{eqnarray}\label{eq:matrix}
\Gamma_{\det}&=&\frac{1}{2}\ln~\det~\overline{G}~,\nonumber\\
\left(\Gamma^{(2)}_{\det}\right)_{\alpha\beta,\gamma\delta}&=&
-\overline{G}^{-1}_{\alpha\gamma}
\overline{G}^{-1}_{\beta\delta}
+\overline{G}^{-1}_{\alpha\delta}
\overline{G}^{-1}_{\beta\gamma}\nonumber\\
&=&{\cal G}_{\alpha\beta,\gamma\delta}
\end{eqnarray}
At this level the flow equations (\ref{130}), (\ref{131}) remain exact. The system
of differential equations is not closed, however.
In order to proceed we need an estimate of $\widehat{\Gamma}^{(3)}[\overline{G}]$.

We next will derive an approximation for $\hat{\Gamma}^{(3)}$ that leads
to a version of the nonperturbative flow equation for the quartic coupling
$\tilde{\lambda}_k$ which has been extensively studied in the past
\cite{EV,AO,16A,16B}. In the same approximation the 2PI-character of the flow
of $\lambda_k$ will lead to important improvements if the renormalized gap equation
is exploited. As a starting point we use the
definition (\ref{eq:bosonic}) of $Y_{\alpha\beta}=\partial\widehat{\Gamma}/
\partial G_{\alpha\beta}-\frac{1}{2}\lambda_{\alpha\beta\gamma\delta}G_{\gamma\delta}$
and obtain the exact identity
\begin{eqnarray}
&&\frac{\partial^2\widehat{\Gamma}}{\partial G_{\alpha\beta}\partial G_{\gamma\delta}}
=\lambda_{\alpha\beta\gamma\delta}
+\frac{\partial Y_{\alpha\beta}}{\partial G_{\gamma\delta}}\\
&&=\frac{1}{3}\lambda_{\alpha\beta\gamma\delta}
+\frac{1}{6}\lambda_{\rho\beta\mu\nu}
\frac{\partial}{\partial G_{\gamma\delta}}
\Big\{G^{-1}_{\alpha\sigma}(\Gamma_B^{(2)})^{-1}_{\sigma\rho\mu\nu}\Big\}\nonumber\\
&&=\frac{1}{3}\lambda_{\alpha\beta\gamma\delta}
+\frac{1}{6}\lambda_{\rho\beta\mu\nu}
\Big\{(G^{-1}_{\alpha\gamma}
G^{-1}_{\sigma\delta}
-G^{-1}_{\alpha\delta}
G^{-1}_{\sigma\gamma})
\left(\Gamma_B^{(2)}\right)^{-1}_{\sigma\rho\mu\nu}\nonumber\\
&&\hspace{0.4cm}-\frac{1}{4}G^{-1}_{\alpha\sigma}
\left(\Gamma_B^{(2)}\right)^{-1}_{\sigma\rho\sigma^\prime\rho^\prime}
\frac{\partial^3\Gamma}{\partial G_{\sigma^\prime\rho^\prime}
\partial G_{\mu^\prime\nu^\prime}\partial G_{\gamma\delta}}
\left(\Gamma_B^{(2)}\right)^{-1}_{\mu^\prime\nu^\prime\mu\nu}\Big\} \nonumber
\end{eqnarray}

\medskip
\noindent
With
\begin{equation}
\left(\Gamma_B^{(2)}\right)^{-1}_{\alpha\beta\gamma\delta}=
-G_{\alpha\gamma}G_{\beta\delta}
+G_{\alpha\delta}G_{\beta\gamma}
+\tilde{\Delta}_{\alpha\beta\gamma\delta}
\end{equation}
and eq. (\ref{eq:isgivenby}) this yields the exact expression
\begin{equation}\begin{split}\label{eq:furtherdiff}
&\frac{\partial^2\widehat{\Gamma}}{\partial G_{\alpha\beta}\partial G_{\gamma\delta}}=
\lambda_{\alpha\beta\gamma\delta}\\
&+\frac{1}{6}\lambda_{\rho\beta\mu\nu}
\Big\{(G^{-1}_{\alpha\gamma}
G^{-1}_{\sigma\delta}
-G^{-1}_{\alpha\delta}
G^{-1}_{\sigma\gamma})
\tilde{\Delta}_{\sigma\rho\mu\nu} \\
&+\dfrac{1}{4}G^{-1}_{\alpha\sigma}\Big[\tilde{\Delta}_{\sigma\rho\sigma^\prime\rho^\prime}
\dfrac{\partial^3\Gamma_{\det}}
{\partial G_{\sigma^\prime\rho^\prime}
\partial G_{\mu^\prime\nu^\prime}
\partial G_{\gamma\delta}}\\
&\hspace{0.4cm}(G_{\mu^\prime\mu}G_{\nu^\prime\nu}-G_{\mu^\prime\nu}G_{\nu^\prime\mu})\\
&+(G_{\sigma\sigma^\prime}G_{\rho\rho^\prime}
-G_{\sigma\rho^\prime}G_{\rho\sigma^\prime})
\dfrac{\partial^3\Gamma_{\det}}
{\partial G_{\sigma^\prime\rho^\prime}
\partial G_{\mu^\prime\nu^\prime}
\partial G_{\gamma\delta}}
\tilde{\Delta}_{\mu^\prime\nu^\prime\mu\nu} \\
&-\tilde{\Delta}_{\sigma\rho\sigma^\prime\rho^\prime}
\dfrac{\partial^3\Gamma_{\det}}
{\partial G_{\sigma^\prime\rho^\prime}
\partial G_{\mu^\prime\nu^\prime}
\partial G_{\gamma\delta}}
\tilde{\Delta}_{\mu^\prime\nu^\prime\mu\nu} \\
&-\left(\Gamma_B^{(2)}\right)^{-1}_{\sigma\rho\sigma^\prime\rho^\prime }
\dfrac{\partial^3\widehat{\Gamma}}
{\partial G_{\sigma^\prime\rho^\prime}
\partial G_{\mu^\prime\nu^\prime}
\partial G_{\gamma\delta}}
\left(\Gamma_B^{(2)}\right)^{-1}_{\mu^\prime\nu^\prime\mu\nu}\Big]\Big\}
\end{split}\end{equation}
Therefore the difference between $\lambda_k$ and $\lambda$
vanishes for $\tilde{\Delta}=0$ and $\widehat{\Gamma}^{(3)}=0$.

One possible approach would be a systematic perturbative expansion for small $\lambda$.
Comparison with eq. (\ref{eq:andtherefore}) shows $\tilde{\Delta}\sim\lambda G^4$ as
long as $\lambda G^2$ remains small. In a perturbative expansion the difference
$\lambda_k-\lambda$ is of the order $\lambda^2$. Indeed, the exact
Schwinger-Dyson equation for $\widehat{\Gamma}^{(3)}$ is obtained by a further
differentiation of eq. (\ref{eq:furtherdiff}). We see that it is not closed
since it involves $\widehat{\Gamma}^{(4)}$. An approximate solution can
be obtained by neglecting terms $\sim\widehat{\Gamma}^{(4)}$ and
$\widehat{\Gamma}^{(3)}$ on the r. h. s., thus expressing $\widehat{\Gamma}^{(3)}$
in terms $\widehat{\Gamma}^{(2)},\overline{G}$ and $\lambda$. For small $\lambda G^2$
one finds $\hat{\Gamma}^{(3)}\sim\lambda^2 G$.

We propose here an even simpler approximation which is not restricted to weak coupling.
For this purpose we take the lowest order
expression (\ref{eq:settingsun}) for $Y_{\alpha\beta}$ such that
\begin{eqnarray}\label{eq:184}
&&\frac{\partial^2\widehat{\Gamma}}{\partial G_{\alpha\beta}\partial G_{\gamma\delta}}
=\lambda_{\alpha\beta\gamma\delta}\nonumber\\
&&-\frac{1}{2}(\lambda_{\alpha\gamma\rho\sigma}
\lambda_{\beta\delta\mu\nu}
-\lambda_{\alpha\delta\rho\sigma}
\lambda_{\beta\gamma\mu\nu})
G_{\rho\mu}G_{\sigma\nu}
\end{eqnarray}
Simple differentiation leads to an expression for $\hat{\Gamma}^{(3)}$
\begin{eqnarray}\label{eq:185}
&&\frac{\partial^3\widehat{\Gamma}}
{\partial G_{\alpha\beta}\partial G_{\gamma\delta}\partial G_{\epsilon\eta}}
=-(\lambda_{\alpha\gamma\epsilon\sigma}
\lambda_{\beta\delta\eta\nu}
-\lambda_{\alpha\gamma\eta\sigma}
\lambda_{\beta\delta\epsilon\nu}\nonumber\\
&&-\lambda_{\alpha\delta\epsilon\sigma}
\lambda_{\beta\gamma\eta\nu}
+\lambda_{\alpha\delta\eta\sigma}
\lambda_{\beta\gamma\epsilon\nu})
G_{\sigma\nu}
\end{eqnarray}
After a renormalization group improvement $\lambda\rightarrow\lambda_k$
on the r.h.s. this yields the flow equation for $\lambda_k$
\begin{eqnarray} \label{eq:flowequation}
&&\partial_k(\lambda_k)_{\alpha\beta\gamma\delta}
=\frac{1}{2}\Big\{(\lambda_k)_{\alpha\gamma\epsilon\mu}
(\lambda_k)_{\beta\delta\eta\nu}\nonumber\\
&&-(\lambda_k)_{\alpha\delta\epsilon\mu}
(\lambda_k)_{\beta\gamma\eta\nu}\Big\}
\overline{G}_{\mu\nu}(\overline{\Gamma}_B^{(2)})^{-1}_{\epsilon\eta\sigma\tau}
\partial_k\left(R_k^{(F)}\right)_{\sigma\tau}
\end{eqnarray}
with (\ref{123})
\begin{equation}\label{139}
\left(\overline{\Gamma}_B^{(2)}\right)_{\epsilon\eta\sigma\tau}
=-\overline{G}_{\epsilon\sigma}
\overline{G}^{-1}_{\eta\tau}
+\overline{G}^{-1}_{\epsilon\tau}
\overline{G}^{-1}_{\eta\sigma}
+(\lambda_k)_{\epsilon\eta\sigma\tau}
\end{equation}

The flow equations (\ref{eq:sec2}) and (\ref{eq:flowequation}) form now a closed system
of differential equations
for the running of $\overline{G}_k$ and $\lambda_k$.  The one loop character of the
flow equation for $\lambda_k$ becomes obvious in the form
\begin{eqnarray}\label{140}
&&\partial_k(\lambda_k)_{\alpha\beta\gamma\delta}=\\
&&-\Big\{(\lambda_k)_{\alpha\gamma\epsilon\mu}
(\lambda_k)_{\beta\delta\eta\nu}
-(\lambda_k)_{\alpha\delta\epsilon\mu}
(\lambda_k)_{\beta\gamma\eta\nu}\Big\}
\overline{G}_{\mu\nu}\partial_k\overline{G}_{\epsilon\eta}\nonumber
\end{eqnarray}
Correspondingly, the flow (\ref{130}) of $j_k$ is given
by the ``setting sun'' two loop diagram.

The exact $SD$-equation (\ref{eq:Schwinger-Dyson}) for the $k$-dependent value of $\bar{G}$
and the exact flow equation (\ref{eq:sec2}) for $\partial_k\bar{G}$ are closely related -
the latter follows by differentiation of the former using eqs. (\ref{123}), (\ref{130}).
Instead of solving eq. (\ref{eq:sec2}) one may therefore evaluate at every $k$-step the new
values of $\lambda_k$ and $j_k$ by the differential equations (\ref{140}), (\ref{130}). For
this purpose one can insert on the r.h.s. of eq. (\ref{140}) the lowest order\footnote{
Corrections to eq. (\ref{140AA}) are $\sim\lambda_k\bar{G}^2\partial_k\bar{G}$,
cf. eq. (\ref{130}) and could be treated iteratively.} formula
(cf. eq. (\ref{eq:Schwinger-Dyson}))
\begin{equation}\label{140AA}
\partial_k\bar{G}_{\epsilon\eta}=-\bar{G}_{\epsilon\alpha}\partial_k
\left(R^{(F)}_k\right)_{\alpha\beta}
\bar{G}_{\beta\eta}
\end{equation}
This leads to the flow equations (\ref{107BB}) and (\ref{155AX}).
In this form the close analogy to previously investigated flow equations
for the four-fermion-coupling \cite{MEZ,16A,16B,EV,AO,21A} is apparent.
Replacing $\bar{G}$ by the ``classical'' average propagator $(-\bar{j}+R^{(F)}_k)^{-1}$
yields precisely a structure similar to the
well-known closed one loop equation for $\tilde{\lambda}_k$.

Actually, despite the structural similarity the one loop flow of $\tilde{\lambda}_k$
differs from the 2PI-flow of $\lambda_k$ by an additional term even in lowest order in
$\lambda_k$. Using eq. (\ref{177YE}) we find in lowest order
\begin{equation}\label{198QA}
\partial_k\tilde{\lambda}_k=\partial_k\lambda_k-\lambda_k\partial_k{\cal G}^{-1}\lambda_k
\end{equation}
In the approximation (\ref{140}) this yields
\begin{eqnarray}\label{198QB}
\partial_k(\tilde{\lambda}_k)_{\alpha\beta\gamma\delta}&=&-\bar{G}_{\mu\nu}\partial_k
\bar{G}_{\epsilon\eta}
\{(\lambda_k)_{\alpha\gamma\epsilon\mu}(\lambda_k)_{\beta\delta\eta\nu}\\
&&-(\lambda_k)_{\alpha\delta\epsilon\mu}(\lambda_k)_{\beta\gamma\eta\nu}
-(\lambda_k)_{\alpha\beta\epsilon\mu}(\lambda_k)_{\gamma\delta\eta\nu}\}\nonumber
\end{eqnarray}
and we observe that the (last) additional term as compared to
$\partial_k\lambda_k$ (\ref{140}) leads to a total antisymmetrization of the r.h.s. in the
indices $(\alpha\beta\gamma\delta)$ as appropriate for a four-fermion vertex. Inserting on
the r.h.s the lowest order relation $\tilde{\lambda}_k=\lambda_k$ and
$\bar{G}=(-\bar{j}+R_k)^{-1}$ reproduces the usual one loop flow for
$\tilde{\lambda}_k$ which typically diverges for some finite $k$ if the temperature
is sufficiently low\footnote{For quantitative investigations of this
or similar flow equations see
\cite{EV,AO} for Nambu-Jona-Lasinio type quark models and \cite{MEZ,16A,16B} for the
Hubbard model.}. In the past, this divergence at finite $k$ has been a major
obstacle for the exploration of the low temperature phase using eq. (\ref{198QB}).

It is possible (but not known) that the flow of $\lambda_k$ is better behaving
if one replaces eq. (\ref{198QB}) by eq. (\ref{140}) while keeping
$\bar{G}=(-\bar{j}+R_k)^{-1}$. Here we will advocate to cure the disease by
a more accurate of $\bar{G}$. Indeed, we want to exploit the 2PI-character of our
formalism and propose to evaluate $\bar{G}$ by an approximation to the exact renormalized
Schwinger-Dyson equation (\ref{eq:Schwinger-Dyson}).

In many circumstances a reasonable approximation neglects
the difference between $j_k$ and $\bar{j}$. Indeed, due to the 2PI setting, the running
of $j_k$ occurs only at two loop order as apparent from the combination of
eqs. (\ref{130}) and (\ref{107BB}) which yields eq. (\ref{155AX}). All one loop effects
for the propagator-renormalization (and much more!) are contained in the ``gap''
$\sim\lambda G$ present in the implicit equation (\ref{eq:Schwinger-Dyson}) for the
approximate average propagator
\begin{equation}\label{189AX}
\bar{G}=(-\bar{j}+R^{(F)}_k+\lambda_k\cdot\bar{G})^{-1}
\end{equation}
Of course, the solution of the ``gap equation'' (\ref{189AX}) has now to be evaluated
for every value of $k$ and typically involves additional approximations. Nevertheless,
this procedure offers an important advantage:
one can check at every $k$-step for the possible appearance
of additional solutions of the gap equation. In fact, in many situations one expects that a
symmetry breaking gap $\Delta$ vanishes for large $k$ where only a small part of the
fluctuations is included. The flow can then produce a characteristic  bifurcation
phenomenon where below a critical scale $k_{bf}$ both solutions with $\Delta=0$ and
$\Delta\neq 0$ coexist. In this case the computation of $\lambda_k$ (and $j_k$) should
follow the solution with lowest free energy.

The dominant features of the term $\sim\lambda\cdot\bar{G}$ in eq. (\ref{189AX}) can
often be accounted for by a {\em local} gap as discussed in sect. \ref{gapequation}, i.e.
\begin{equation}\label{189BX}
\bar{G}=(-\bar{j}+R^{(F)}_k+\Delta)^{-1}
\end{equation}
Here the local gap $\Delta_{\alpha\beta}=\Delta_{ab}(x)\delta_{xy}$ involves
now the renormalized coupling $\lambda_k$. In the approximation of local interactions
the gap equation is analogous to eq. (\ref{eq:thelocalgap}) with $\bar{\lambda}$
replaced by a corresponding renormalized coupling - more generally $\Delta$
corresponds to the local part of the term $\sim\lambda\cdot G$ in the inverse of eq.
(\ref{189AX}), similar to eq. (\ref{eq:thesolutionsof}). Consider now the case of
spontaneous symmetry breaking. Once $\lambda_k$ has grown large enough there
will be a solution with $\Delta\neq 0$. We define $k_{cr}$ such that for $k<k_{cr}$ the
solution with $\Delta\neq 0$ has the lowest free energy. For the running of
$\lambda_k$ one should therefore insert eq. (\ref{189BX}) into eq.
(\ref{107BB}) with $\Delta=0$ for $k>k_{cr}$ and $\Delta\neq 0$ for
$k<k_{cr}$. Typically, $\lambda_k$ will increase towards a finite large value
as $k$ approaches $k_{cr}$ from above.
There is no reason for a further strong increase of $\lambda_k$ in the
SSB-regime for $k<k_{cr}$. One therefore expects in this case that $\lambda_k$
remains finite as $k\rightarrow 0$, perhaps in contrast to an approximation which neglects
$\Delta$.

There may still remain some cases where $\lambda_k$ grows very large at some scale
$k_s$. Then, for large
$\lambda_k\bar{G}^2$, one may doubt the validity of our approximations. Nevertheless, in
many situations one may still extract useful information from the renormalized gap
equation (\ref{eq:Schwinger-Dyson}). For example,
the value of the gap $\Delta$ may already settle to an almost $k$-independent
value for scales substantially above $k_s$. In order to explore such a possibility one may
assume that for $k$ smaller than some value $\bar{k}$ the effects of the change of
$\lambda_k$ and $j_k$ on the value of $\bar{G}$ are small and can be neglected. (This
assumption is used implicitely for {\em all} $k$ for the lowest order SD-equation,
whereas we use it here only for $k<\bar{k}$.) We can then evaluate eq.
(\ref{eq:Schwinger-Dyson}) for $k=0$ and use the approximations
$j_0=j_{\bar{k}}~,~\lambda_0=\lambda_{\bar{k}}$. This
yields the renormalized gap equation (\ref{107AA}) with $R^{(F)}_k$ set to zero.
The combination of the gap equation
with the renormalization flow becomes now a very powerful tool. A computation can proceed
in two steps: First one solves approximately in two steps: First one solves approximately
the flow equations and computes $j_{\bar{k}},\lambda_{\bar{k}}$. The second step uses the
renormalized coupling and source for the gap equation. The validity of the approximation can
be (partly) checked by testing if the results are independent  of the particular choice of
$\bar{k}$. Clearly, this requires that $R^{(F)}_{\bar{k}}$
has only a small influence on the value of the gap $\Delta$.

The renormalization group improved gap equation (\ref{107AA}) should lead to a much better
grasp of the physical situation whenever the running of couplings has an important effect.
This is certainly the case for the Hubbard model \cite{{MEZ,16A,16B}}. In particular,
the four fermion interaction in the Hubbard model becomes nonlocal for $k$ sufficiently
below the ultraviolet cutoff (e.g. inverse lattice distance). As mentioned in the
preceeding section, this allows for the solution of a nontrivial gap equation for a nonlocal
superconducting order parameter. Furthermore, the error estimated from the dependence on
$\bar{k}$ constitutes a lower bound for the error from the neglection of higher order terms
$\sim Y_{\alpha\beta}$. More precisely, this concerns the contributions that cannot be
absorbed in the renormalization group running. Last but not least
the renormalized gap equation (\ref{107AA}) offers a new perspective on
the conceptual use and validity of the lowest order SD-equation: The lowest order gap
equation should be thought of using optimally renormalized couplings and sources, such
that residual errors are minimized. This remark is particularly important if one is merely
interested in the qualitative structure of the results and less in the quantitative values
of $\lambda_k$ and $j_k$. The errors of neglected fluctuations (higher order terms) may be
much smaller than naively estimated from the use of the unrenormalized equation!

We close this section by a cautionary remark on the correct use of the renormalized gap
equation. In this context we note that eq. (\ref{107AA}) (or eq. (\ref{189BX}) or similar)
has the same ultraviolet
cutoff as the leading order gap equation (\ref{eq:thesolutionsof}), i.e. the physical
microscopic scale (lattice distance or similar). In particular, the scale $k$ does not
act as an UV cutoff in the renormalized gap equation, implying that fluctuations with all
momenta (not only $q^2<k^2$) are effectively included. In fact, the gap equation often
receives a dominant contribution from the short distance fluctuations. One therefore has
to treat the momentum dependence of the renormalized coupling $\lambda_k$ carefully.
For high ``external momenta'' $Q$ the flow of $\lambda_k$ is severely suppressed if
$k^2\ll Q^2$. For large $Q^2$ one therefore typically finds $\lambda_k\approx \lambda$.
On the other side, the running of $\lambda_k$ may be important for $Q^2\ll k^2$,
modifying the contributions of fluctuations with small momenta in the gap equation.

\section{Bosonic renormalization group equation}
\label{bosonicrenormalizationgroup}
Critical behavior is associated to an infinite correlation length for
composite bosonic fields. In our formalism this corresponds to vanishing
eigenvalues of the inverse bosonic propagator $\Gamma_B^{(2)}=\Gamma_{\det}^{(2)}+
\widehat{\Gamma}^{(2)}=\Gamma^2_{\det} +\lambda_k$. Obviously, the
lowest order Schwinger-Dyson equation breaks down in the critical region since the
correction
\begin{eqnarray}
Y_{\alpha\beta}&=&\frac{1}{6}\lambda_{\eta\beta\gamma\delta}G^{-1}_{\alpha\epsilon}
H_{\epsilon\eta,\gamma\delta}~, \nonumber\\
H&=&\left(\Gamma_B^{(2)}\right)^{-1}-
\left(\Gamma^{(2)}_{\det}\right)^{-1}\nonumber\\
&=&-\left(\Gamma^{(2)}_{\det}\right)^{-1}
\cdot\lambda_k\cdot\left(\Gamma^{(2)}_{\det}+\lambda_k\right)^{-1}
\end{eqnarray}
grows very large. Renormalization group equations are the
standard tool to approach the critical behavior.
Even though formally exact the renormalized gap equation (\ref{eq:Schwinger-Dyson})
seems not ideal for this purpose since the infrared cutoff $R_k^{(F)}$ acts only on the fermionic
fluctuations. In the context of critical behavior it seems more appropriate to introduce
an infrared cutoff for the composite bosons.

Let us therefore add to the action $S_j$ a ``bosonic infrared cutoff'' in the form
\begin{equation}\label{eq:191}
\Delta^{(B)}_kS=\frac{1}{8}\left(R_k^{(B)}\right)_{\alpha\beta\gamma\delta}
\tilde{\psi}_\alpha\tilde{\psi}_\beta\tilde{\psi}_\gamma\tilde{\psi}_\delta
\end{equation}
This term is quadratic in the fermionic bilinears in analogy to the piece containing the
fermion cutoff $R^{(F)}_k$ which is quadratic in the fermionic fields. We will discuss the
properties of $R^{(B)}_k$ required for an IR cutoff below. Adding the term
(\ref{eq:191}) amounts to a change of the effective quartic coupling $\lambda$. Defining
$W_k=\ln~Z_k$ for the system in presence of the cutoff and subtracting the
cutoff again after the Legendre transform, the effective average action reads in
the bosonic language (with $j=j[G])$
\begin{equation}\label{eq:becomesin}
\Gamma_{B,k}=-W_k[0,j]+\frac{1}{2}j_{\alpha\beta}G_{\alpha\beta}
-\frac{1}{8}\left(R_k^{(B)}\right)_{\alpha\beta\gamma\delta}
G_{\alpha\beta}G_{\gamma\delta}
\end{equation}
The exact flow equation for the $k$-dependence of $\Gamma_{B,k}$
is now easily derived in analogy to \cite{ERGE}
\begin{eqnarray} \label{eq:cutoffsubstracted}
\partial_k\Gamma_{B,k}{~\atop|}_G
&=&-\partial_kW_k{~\atop|}_j
-\frac{1}{8}\partial_k\left(R_k^{(B)}\right)_{\alpha\beta\gamma\delta}
G_{\alpha\beta}G_{\gamma\delta}\nonumber\\
&=&\frac{1}{8}\partial_k(R_k^{(B)})_{\alpha\beta\gamma\delta}
(\langle\tilde{\psi}_\alpha\tilde{\psi}_\beta\tilde{\psi}_\gamma \tilde{\psi}_\delta\rangle
-G_{\alpha\beta}G_{\gamma\delta})\nonumber\\
&=&\frac{1}{8}\partial_k(R_k^{(B)})_{\alpha\beta\gamma\delta}
(W_k^{(2)})_{\alpha\beta\gamma\delta}
\end{eqnarray}
Here we have used eq. (\ref{eq:sim}). In our matrix notation we have
$W_k^{(2)}=(\Gamma_B^{(2)}+R_k^{(B)})^{-1}$ and we recover the well
known exact flow equation for bosonic systems \cite{ERGE}
\begin{equation}\label{eq:flowequation1}
\partial_k\Gamma_{B,k}=\frac{1}{2}\widetilde{\textup{tr}}\Big\{\partial_kR_k^{(B)}\cdot
(\Gamma^{(2)}_{B,k}+R_k^{(B)})^{-1}\Big\}
\end{equation}
We note, however, that the trace $\widetilde{\textup{tr}}$ acts on
bosonic indices $(\alpha\beta)$. Since the bosonic fields are bilocal it involves a double
momentum integration. Despite the similarity in the form this constitutes an important
practical difference as compared to the more familiar exact flow equation for a finite
number of {\em local} bosonic fields.
We will see below how this difficulty can be partly overcome for a suitable
choice of $R_k^{(B)}$.

The bosonic flow equation makes no distinction between a model with ``fundamental''
bosonic degrees of freedom and composite bosonic fields. This is a nice feature since the
conceptual difference between ``fundamental bosons'' and ``composite bosons'' is not clear,
anyhow. The well developed formalism
of the effective average action and nonperturbative flow equations can therefore directly
be used in our formulation. Furthermore, the bosonic cutoff $R_k^{(B)}$ can be combined with
a fermionic cutoff $R_k^{(F)}$. The ``initial value'' of the flow for large $k$ (or
$k\rightarrow\infty$) depends on the choice of the cutoff $R_k^{(B)}$ (and possibly
$R_k^{(F)}$) and will be discussed below.

It should be noticed that the flow actually only concerns $\widehat{\Gamma}$ in eq.
(\ref{eq:thereforewrite}). The part $\Gamma_{\det}$ remains unaffected since the flow
equation (\ref{eq:flowequation1}) is manifestly covariant with respect to the
transformations (\ref{eq:accordingto})(\ref{eq:by}). Nevertheless, the contribution
from the fermionic fluctuation determinant is implicitely present in the flow of
$\widehat{\Gamma}$
\begin{equation}\label{eq:flowequation2}
\partial_k\widehat{\Gamma}_k=\frac{1}{2}\widetilde{\textup{tr}}\Big\{\partial_kR_k^{(B)}
(\widehat{\Gamma}_k^{(2)}+\Gamma^{(2)}_{\det}+R^{(B)}_k)^{-1}\Big\}
\end{equation}
In the very simple truncation
\begin{equation}\label{eq:truncation}
\widehat{\Gamma}_k=\frac{1}{2}G\cdot\lambda_k\cdot G-\Delta j_k \cdot G
\end{equation}
one has
\begin{eqnarray}
&&\partial_k(\lambda_k)_{\alpha\beta\gamma\delta}= \\
&&\frac{1}{2}\widetilde{\textup{tr}}
\Big\{\partial_kR_k^{(B)}
\frac{\partial^2}{\partial G_{\alpha\beta}\partial G_{\gamma\delta}}
[\lambda_k+\Gamma^{(2)}_{\det}+R^{(B)}_k]^{-1}\Big\}_{\stackrel{~}{|}G=\overline{G}}
\nonumber
\end{eqnarray}
and we observe that the only contributions arise from the effective bosonic cubic and
quartic couplings associated to the $G$-derivatives of $\Gamma_{\det}^{(2)}$ (cf. eqs.
(\ref{eq:matrix}), (\ref{eq:isgivenby})). There is, however, no need to restrict the
truncation of $\widehat{\Gamma}$ to the form (\ref{eq:truncation}). Then contributions
to the bosonic cubic and quartic coupling arise also from $G$-derivatives of
$\widehat{\Gamma}_k^{(2)}$. We emphasize that the exact flow equation
(\ref{eq:flowequation1}), possibly combined with a contribution from the $k$-dependence
of a fermionic cutoff to the running of $\overline{G}$ (cf. eq. (\ref{eq:sec2})), has to be
equivalent with the exact flow equation for partially bosonized versions of the fermionic
model that will be briefly discussed in the next section. In particular,
the partially bosonized version offers a simple starting point for the
computation of $\Gamma_{B,k}$ for large values of $k$.

The simple and suggestive form of the
exact flow equation (\ref{eq:flowequation1}) can be used as a convenient starting point for
approximations. For a practical use one has to reduce the complexity to a finite number of
local bosonic fields and to truncate the corresponding bosonic effective action. We also
have to define a suitable infrared cutoff $R^{(B)}_k$ and to determine the initial condition
for large $k$.

A crucial point is the ``initial condition'' for the flow for $k=\Lambda$. (Instead of
$k=\Lambda$ one may also use $k\rightarrow\infty$.) One would like to
choose $R_\Lambda^{(B)}$ such that $\Gamma_{B,\Lambda}$ can be computed reliably and takes
a simple form. A particularly interesting choice is (in a matrix notation)
\begin{equation}
R^{(B)}_\Lambda=-\lambda
\end{equation}
In this case the bosonic cutoff precisely cancels the interaction term such that
$W_\Lambda$ is given by a free theory. Since the cutoff is substracted again in the
definition (\ref{eq:becomesin}) of $\Gamma_{B,\Lambda}$ one has the exact solution
\begin{equation}\label{eq:exactsolution}
\Gamma_{B,\Lambda}=\Gamma_1
\end{equation}
At the scale $\Lambda$ the lowest order SD-equation becomes therefore exact!

The missing fluctuations are next included by ``switching on'' the coupling in the
higher order fluctuation effects as $k$ is lowered from $\Lambda$ to zero. For
$R^{(B)}_{k=0}$ the full BEA - including all fluctuations and all interaction effects - is
recovered from the solution $\Gamma_{B,k\rightarrow 0}$.

We next have to define a suitable cutoff $R_k^{(B)}$ for intermediate $k$. For this
purpose it is useful to interprete $G_{\alpha\beta}$ as an infinite
number of bosonic fields labeled by the relative coordinate $z=y-x$ and to perform a Fourier
transform with respect to the ``center coordinate'' $s=(y+x)/2$. In this basis one has
\begin{equation}
G_{\alpha\beta}=G_{ab}(Q,z)=\int_se^{-iQs}G_{ab}(s,z)
\end{equation}
and the local gap discussed in sect. \ref{gapequation} corresponds to the field labeled by
$z=0,G_{ab}(s,z)=g_{ab}(s)\delta(z)$. For local interactions it is sufficient to have the
cutoff acting only on $g_{ab}(s)$ and we preserve translation invariance by
\begin{eqnarray}\label{152}
&&(R^{(B)}_k)_{ab,cd}(Q,z,Q^\prime,z^\prime)=\nonumber\\
&&\int_z\int_{z^{\prime
}}e^{iQz}e^{-iQ^\prime z^\prime}(R_k^{(B)})_{ab,cd}(s,z,s^\prime,z^\prime)\nonumber\\
&&=(R_k^{(B)})_{abcd}(Q)\delta(Q-Q^{\prime})\delta(z)\delta(z^{\prime})
\end{eqnarray}
with $(R_\Lambda^{(B)})_{abcd}(Q)=-\lambda_{abcd}$. We note that for this cutoff
the trace in eq. (\ref{eq:flowequation1}) reduces to a single momentum integral (plus
summation over internal indices).

For the minimal set of spinors, $\lambda_{abcd}=-\overline{\lambda}\epsilon_{abcd}$, the
initial effective action $\widehat{\Gamma}_\Lambda$ depends \footnote{This property will
not be preserved for $\widehat{\Gamma}_k,k<\Lambda$, due to the contribution from
$\Gamma^{(2)}_{\det}$ in the flow equation (\ref{eq:flowequation2}).} only on six complex
bosonic fields $g_{ab}(s)$. By virtue of the discrete symmetry $\vartheta$
(sect. \ref{localinteractions})
this can be reduced to four real and one complex fields.
With respect to the $SO(4)\hat{=}SU(2)_s\times SU(2)_c$ symmetry
transformations these fields transform as $(3,1)+(1,3)$. The triplet $\vec{m}$ with
respect to the spin transformations \footnote{We omit the tilde in the notations for the
spinors here.}
\begin{eqnarray}
\vec{m}(s)&=&\langle\overline{\psi}(s)\vec{\tau}\psi(s)\rangle~,~ \nonumber\\
m_1&=&g_{32}+g_{41}~,~m_2=-i(g_{32}-g_{41})~,~\nonumber\\
m_3&=&g_{31}-g_{42}
\end{eqnarray}
and the triplet $\vec{c}$ with respect to the ``charge'' transformations $SU(2)_c$
\begin{eqnarray}
c_1&=&-(g_{12}-g_{34})~,~c_2=-i(g_{12}+g_{34})~,~\nonumber\\
c_3&=&\langle\overline{\psi}\psi\rangle
=g_{31}+g_{42}
\end{eqnarray}
span the irreducible representations. The third generator  of $SU(2)_c$
corresponds to electric charge. The charged boson corresponds to the electron pair
$g_{12}=\langle\psi i\tau_2\psi\rangle=-(c_1-ic_2)/2$ with an associated hole pair
$g_{34}=\langle\overline{\psi}
i\tau_2\overline{\psi}\rangle=(c_1+ic_2)/2$. Labeling by $s,t$ the irreducible
bosonic representations we write
\begin{equation}
R_k^{(B)}(Q^2)=\hat{\lambda}_sr_s\left(\frac{Q^2}{k^2},\frac{k^2}{\Lambda^2}\right)
\delta_{st}
\end{equation}
where $r_s(k=\Lambda)=1$ and $\hat{\lambda}_s$ are determined such that $R_\Lambda^{(B)}
=-\lambda$. We also require that $r_s$ vanishes fast for $k^2\ll \tilde{Q}_s^2$ such that
the fluctuations with high $\tilde{Q}^2_s$ are already effectively included in
$\Gamma_{B,k}$. Here $\tilde{Q}_s$ is the deviation of the momentum $Q$ from the
momentum $\overline{Q}_s$ for which the free energy of the boson $s$ is minimal \footnote{
If the free energy at a given scale $k$ is minimal for a homogenous field $g_s$ one has
$\tilde{Q}_s=Q$. For a preferred antiferromagnet one would rather have $\tilde{Q}_s=
Q-Q_a$.}.

As a possible truncation for practical computations we propose to keep two sets of local
bosonic fields. One set is selected from the $g_{ab}(s)$ whereas the other set corresponds
to local gaps $\Delta_{ab}(s)$ according to the ansatz
\begin{equation}\label{156}
G^{-1}_{ab}(s,z)=-(j_k)_{ab}(z)\delta(s)+\Delta_{ab}(s)\delta(z)
\end{equation}
This allows us to express $\Gamma_{\det}$ as a functional of $\Delta$, and similar for
$\Gamma^{(2)}_{\det}$ in the flow equation (\ref{eq:flowequation2}) for $\widehat{\Gamma}_k$.
For $\widehat{\Gamma}_k$ we make the ansatz
\begin{equation}\label{eq:secondterm1}
\widehat{\Gamma}_k=\Gamma^\prime_k[g,\Delta]-\frac{1}{2}[(j_k)_{\alpha\beta}
-j_{\alpha\beta}]G_{\alpha\beta}
\end{equation}
and derive the flow equation for $j_k$ from the first $G$-derivative of
eq.(\ref{eq:flowequation2}), evaluated at $s=0~,~z\neq 0$. The remaining flow equation
\footnote{Note that the second term in eq. (\ref{eq:secondterm1}) does not contribute
to $\widehat{\Gamma}_k^{(2)}$.} for $\Gamma^\prime_k$
\begin{equation}\label{158}
\partial_k\Gamma^\prime_k=\frac{1}{2}\textup{tr}\{\partial_kR_k^{(B)}
(\Gamma^{\prime(2)}_k+\Gamma^{(2)}_{\det}[\Delta]+R_k^{(B)})^{-1}\}
\end{equation}
obtains by restricting eq. (\ref{eq:flowequation2}) to fields depending only on $s$.
Corresponding to the choice of cutoff (\ref{152}) the trace tr involves now only one momentum integration and summation
over the internal bosonic indices $(ab)$. At this stage,
the inversion of the matrix of second
functional derivatives has still to be done in a larger space, however.

We have now arrived at a flow equation for $\Gamma^\prime_k$ which is already rather similar
to the well studied flow equation for a finite number of bosonic fields. The truncations
developed in this context, e.g. a derivative expansion \cite{DE} can be applied. The only
unusual contribution is the part $\sim\Gamma^{(2)}_{\det}[\Delta]$ which accounts for the
fermion loops in presence of the gap. \footnote{Note that in absence of a fermionic cutoff
$R_k^{(F)}$ the quadratic part of the fermion fluctuations is already included in
$\Gamma_{B,\Lambda}$ (\ref{eq:exactsolution}). On the other hand, for
$R_\Lambda^{(F)}\rightarrow \infty$ the minimum $\overline{G}$ is in a region where fermion
fluctuations are effectively absent for $k=\Lambda$.}
In order to relate $g$ and $\Delta$ one
minimizes the free energy $\Gamma_j=\hat{F}/T$, noting that the relation
\begin{equation}
(j_k)_{\alpha\beta}G_{\alpha\beta}=\int_s\Delta_{ab}(s)g_{ab}(s)+~\textup{const}
\end{equation}
is independent of $k$ and therefore
\begin{equation}\label{160}
\frac{\hat{F}_k}{T}=\Gamma^\prime_k[g,\Delta]-\frac{1}{2}\int_s\Delta_{ab}(s)
g_{ab}(s)-\frac{1}{2} \Tr \ln (-j_k+\Delta)+c
\end{equation}
In particular, in an approximation where $\Gamma^\prime$ is independent of $\Delta$ the
variation eq. (\ref{160}) with respect to $g$ yields a simple relation analogous to
eq. (\ref{82})
\begin{equation}\label{160AA}
\Delta_{ab}(s)=\frac{\partial\Gamma^\prime_k}{\partial g_{ab}(s)}
\end{equation}
This expresses $\Delta$ as a functional of $g$ and allows us to study the flow equation
(\ref{158}) for $\Gamma^\prime$ being a functional only of the local bosonic fields
$g_{ab}(s)$. Nevertheless, the term $\Gamma^{(2)}_{\det}$ still involves the source
$j_k$ according to eqs. (\ref{eq:matrix}), (\ref{156}).

The situation becomes particularly simple for the study of critical phenomena in a context
where all fermions acquire a nonzero mass (gap). Then for low enough $k$ we may neglect
the fermion fluctuations altogether and omit the term $\Gamma^{(2)}_{\det}$ in the flow
equation (\ref{158}). What remains is the well studied flow equation for a bosonic
system that has already given a very satisfactory description of critical phenomena
\cite{ERGE,DE}. This demonstrates how critical phenomena can be understood in the context of
the BEA. If massless fermions are present at the phase transition (as, e.g., the
antiferromagnetic transition in the Hubbard model \cite{BBWN}), the universal
critical exponents will be affected by the fermionic fluctuations and $\Gamma^{(2)}_{\det}$
cannot be neglected even near the phase transition.

Let us summarize sects. \ref{exactfermionic}-\ref{bosonicrenormalizationgroup}
from the viewpoint of practical applications. Employing suitable fermionic and bosonic
cutoffs and approximate truncations we have recovered the ``standard'' well
investigated approximations for non-perturbative flow equations. This holds both for
the flow of fermionic and bosonic systems - both are derived from a common exact
renormalization group equation for the BEA. Furthermore, our 2PI setting points to
extensions of the simplest versions of these ``standard'' flow equations. For
the fermionic flow equation the leading effect of the effective bosons can be
incorporated via the inclusion of the gap in eq. (\ref{140}), using at every $k$ an
approximative solution of eq.  (\ref{198QA}). For the bosonic flow equation the
effect of the fermionic fluctuations is incorporated via the inclusion of
$\Gamma^{(2)}_{\det}[\Delta]$ in eq. (\ref{158}), with $\Delta$ determined by eq.
(\ref{160AA}). However, even for $j_k=\bar{j}$ and using eqs. (\ref{eq:matrix})
(\ref{189BX}) for $\Gamma^{(2)}_{\det}$, the fermionic contributions still appear in a
rather complicated form. At the present stage the bosonic renormalization
group equation may therefore be mainly considered as demonstration how the well known
bosonic critical behaviour arises in a fermionic model. For practical investigations of the
phase transition the functional renormalization in the partially bosonized formulation
(next section) may be easier to handle.

\section{Functional renormalization for partial bosonization}
\label{functionalrenormalization}
In the formulation of partial bosonization (PB) the simultaneous implementation
of cutoffs for fermions and bosons is particularly simple. One adds to the action in eq.
(\ref{PB18}) a piece $\sim R^{(F)}_k$ quadratic in the fermions $\tilde{\psi}$ and
another one $\sim R^{(B)}_k$ quadratic in the bosons $\tilde{g}$. Within PB the
derivation of the exact renormalization group equation is straightforward
\cite{ERGE}
\begin{equation}\label{W1}
\partial_k\Gamma_k=\frac{1}{2}Str\left\{\partial_kR_k\left(\Gamma^{(2)}_k+R_k\right)^{-1}
\right\}
\end{equation}
Here $\Gamma_k$ stands for $\Gamma_{FB,k}$ as defined by the Legendre transform in
presence of the cutoff. Again, the cutoff term is subtracted in generalization of
eq. (\ref{159}) (or eq. (\ref{eq:becomesin})). The supertrace $Str$ contains
a minus sign for the fermions (cf. eq. (\ref{eq:cutoff})) and involves now only one
momentum integration according to a one loop expression for local fields. The matrix
$\Gamma^{(2)}$ acts in the combined space for fermions and bosons, just as $R_k$.
Truncations to eq. (\ref{W1}) have been successfully investigated for Nambu-Jona-Lasino
type models \cite{JuW}, the Gross-Neveu model \cite{21A} and the Hubbard model
\cite{BBWN}.

We may compare eq. (\ref{W1}) with the fermionic flow equation (sects.
\ref{exactfermionic}-\ref{runningcoupling}) and the bosonic flow equation
(sect. \ref{bosonicrenormalizationgroup}). For the fermionic flow equation we have
argued that a major improvement for the treatment of the low temperature phase may
arise from the inclusion of the gap (cf. eqs. (\ref{140}) (\ref{189AX})) and the
concentration on the coupling $\lambda_k$ instead of the four fermion vertex
$\tilde{\lambda}_k$. Both features are naturally implemented by eq. (\ref{W1}).
The gap is present by the Yukawa coupling $\sim \psi\psi g$. Typically, the flow
follows the effective average potential $U_k(g)$. In case of spontaneous symmetry
breaking a minimum of $U_k$ for $g\neq 0$ appears at some value of $k$. Just as
discussed for the gap improved renormalization group equation in sect.
\ref{runningcoupling} we can easily follow the minimum of $\Gamma_{J,FB}$ in the
regime with spontaneous symmetry breaking and thus account for a nonvanishing gap.

The flow in the bosonic sector typically monitors the inverse bosonic propagator
$P$ for a few local fields $g_i\widehat{=}g_{ab}$. Let us assume a basis where $P$
is diagonal and denote the eigenvalues by $P_i(q)$ (in momentum space). Furthermore,
one may compute the flow of the corresponding Yukawa couplings
\begin{equation}\label{W2}
{\cal L}_y=-\int_{q,q'}
H_{iab}(q,q')\psi_a(q)\psi_b(q')g_i(-q-q')
\end{equation}
In the simplest approximation one may keep a fixed momentum dependence such that only the
constants $h_{iab}$ depend on $k$
\begin{equation}\label{W3}
H_{iab}(q,q')=h_{iab}F_{iab}(q,q')
\end{equation}
whereas the momentum dependence in $F_{iab}(q,q')$ describes a given channel, like
antiferromagnetism in sect. \ref{antiferromagnetism}. We can now reconstruct the
part of the effective four fermion coupling that is induced by the exchange of the
bosons. For this purpose one solves the field equation for
$g_i$ as a functional of $\psi$
\begin{eqnarray}\label{W4}
&&g_i(q)=\langle g_i(q)\rangle \nonumber\\
&&+P^{-1}_i(q)\int_{q'}H_{iab}(q-q',q')
\psi_a(q-q')\psi_b(q')\nonumber\\
&&+\cdots
\end{eqnarray}
Inserting this solution into $\Gamma_{j,FB}$ and extracting the term involving
four fermion fields yields the contribution to the four fermion vertex. For
vanishing $\langle g_i\rangle=0, J_i=0$ this reads
\begin{eqnarray}\label{W5}
{\cal L}_{F,4}&=&-\frac{1}{2}\sum_i\int_{q,q',q''}
P^{-1}_i(q)\nonumber\\
&&H_{iab}(q-q',q')
\bar{H}_{icd}(-q-q'',q'')\nonumber\\
&&\cdot \psi_a(q-q')\psi_b(q')\psi_c(-q-q'')\psi_d(q'')
\end{eqnarray}
We recognize the close analogy to the ``boson exchange contribution''
$\sim(\Gamma^{(2)}_B)^{-1}$ in eq. (\ref{177YA}) and recall that $\lambda_k$ is related
to $\Gamma^{(2)}_B$ by eq. ({\ref{177YB}). The information contained in $P_i(q)$ and
$H_{iab}(q,q')$ contains an essential part of the information related to
$\lambda_k$!

An important difference of principle, remains however. This arises from the restriction
to a finite number of local boson fields $g_i(x)$ instead of the bilocal bosonic
quantities $G_{ab}(x,y)$. In consequence, the most general form of the four fermion
vertex cannot be obtained by eq. (\ref{W5}). The contribution (\ref{W5}) is a sum of terms
that factorize in the form
$f(q_1,q_2)\tilde{f}(q_3,-q_1-q_2-q_3)h(q_1+q_2)$ whereas the four fermion vertex
$\tilde{\lambda}_k(q_1,q_2,q_3)$ can be an arbitrary function of three independent momenta.
Of course, the difference can be accounted for by the remaining piece of the four
fermion interaction which is 1PI in the fermion-boson-model. Despite this limitation,
a truncation of the effective four fermion vertex to the sum of factorizable terms
(\ref{W5})  is sufficient for many purposes. Simple approximations can still describe a rich
momentum structure of $\tilde{\lambda}_k$ by a finite number of parameters consisting of
$h_{iab}$ as well as masses and wave function renormalizations for the momentum dependence
in $P_i(q)$. Obviously, increasing the number of boson fields $g_i(x)$ retained
in PB improves the resolution of the momentum dependence of $\tilde{\lambda}_k$.

At this point we may recall the redundancy within the fermion-boson model: a given
piece in $\tilde{\lambda}_k$ with the structure (\ref{W5}) can both be obtained from
the proper vertex (1PI in the fermion-boson-model) and from the exchange of the bosons $g_i$.
Redundancies of this type should be avoided in order to keep the number of running couplings
manageable. After all, the split into different pieces contains no relevant information.
It is indeed possible to shift all contributions of the type (\ref{W5}) uniquely
to the exchange of bosons at every scale $k$. For this purpose the corresponding contribution
from the proper vertex which is generated by the flow can be transferred to the bosonic
sector by the method of ``rebosonization'' \cite{GW}. This uses a $k$-dependent
definition of the bosonic variables. Rebosonization is a crucial ingredient for
truncations that omit the proper four fermion vertex!

Even for simple truncations the functional renormalization in PB is a powerful
method to include the bosonic fluctuations beyond mean field theory. This has
been tested by studies of the mean field ambiguity \cite{IW,BBWN}: For phase transitions the
dependence of the critical temperature on the choice of partial bosonization
$(S_{abcd})$ is often very substantial in mean field theory. It turns out to be greatly
reduced once the effect of the bosonic fluctuations is included via a solution of the flow
equations. For the two dimensional Hubbard model with next neighbour interactions the
partially bosonized flow equations yield already a rather satisfactory picture at small
doping \cite{BBWN}. One finds a pseudo-critical temperature $T_{pc}$ which is
considerably below the value of fig. \ref{fig:rho_T} due to the effect of the bosonic
fluctuations. Furthermore, $T_{pc}$ is not the critical temperature for the transition
to antiferromagnetic order. For a temperature range $T_c<T<T_{pc}$ the system is in the
symmetric phase (vanishing antiferromagetic order parameter) and the correlation
length increases $\sim t^{-1}\exp (-b T_c/T),b\approx 20.7$. This describes the universal
critical behaviour of the classical $O(3)$-Heisenberg model \cite{chak} and reflects
that the evolution for small $k$ is dominated by the Goldstone bosons. For
$k\ll T<T_{pc}$ the flow in the fermion-boson-model reduces to a simple flow of
a bosonic linear $\sigma$-model \cite{ERGE}, \cite{RGKT}. Finally, for $T\approx T_c$
the correlation length reaches the macroscopic sample size $L\approx 1 cm$. For
$T<T_c$ long range antiferromagnetic order occurs and the temperature dependence of the
order parameter has been computed quantitatively. The order becomes possible due to the
absence of fluctuations with wave length larger than $L$ in a finite sample.
The critical temperature shows a weak logarithmic dependence on the sample size
and vanishes for $L\rightarrow\infty$, in accordance with the Mermin-Wagner theorem
\cite{MW}. All this demonstrates the practical viability of functional renormalization
in a partially bosonized setting. Similar achievements for the flow equations discussed
in sects. \ref{exactfermionic}-\ref{bosonicrenormalizationgroup} have still to be
demonstrated.

In the present paper we have developed the formal implementation of the functional
renormalization group for the BEA. Concrete examples have to be investigated in order to
assert the practical use of the proposed approximate flow equations.
Nevertheless, we have already established a rather close contact with the functional
renormalization group in an approach with partial bosonization. The success of the latter
is a strong argument in favor  of the conclusion that the main features of renormalization
can indeed be obtained from rather simple approximations to the exact flow equations.

\section{Conclusions}
\label{conclusions}
We have discussed a bosonic effective action (BEA) for interacting fermion
systems. While the general form depends on bilocal fields or, equivalently, infinitely
many local fields, the structure becomes much simpler for suitable approximations. The
lowest order of a systematic loop expansion is equivalent to the lowest order
Schwinger-Dyson (SD) equation. As an important achievement, the BEA offers a simple and direct
way for the computation of the free energy corresponding to different solutions of the
SD-equation. This is crucial in order to establish the ground state in such a situation.
Furthermore, for local four fermion interactions it is sufficient to consider local gaps
in the lowest order of the loop expansion. Therefore the BEA only needs a finite number of
local fields in this approximation. Already at this level useful computations can be
performed as we have demonstrated for the phase diagram for the Hubbard model
with next-neighbour interactions in two dimensions. We emphasize that the one loop BEA goes
far beyond the perturbative loop result! In particular, it can describe spontaneous symmetry
breaking.

Going beyond the leading order gets more involved. For the BEA
it is a priori not obvious how a
restriction to a finite number of bosonic fields can be implemented in a systematic way.
We propose here to use a truncation of an exact renormalization group equation.
For this purpose we generalize the exact renormalization group equation for the (1PI)
effective average action \cite{ERGE} to a 2PI -formulation. It describes the dependence of
the BEA on a characteristic length scale $k^{-1}$. The price to pay for two particle
irreducibility is the appearance of bilocal fields $G_{\alpha\beta}$. As long as only
a fermionic cutoff is introduced for the description of the renormalization group
running and the truncation is kept to low orders of $G$ one can cope with the problem
of bilocality. We propose to use a
simple renormalized gap equation (\ref{107AA}). The renormalized couplings appearing in
this equation obey rather simple approximative flow equations (\ref{107BB}) which can
avoid the previously observed divergence of the couplings.

A reliable description of the flow for critical phenomena or other situations with a large
bosonic correlation length calls for the additional introduction of a bosonic infrared cutoff.
By a suitable
choice of the cutoff $R_k$ we can reduce the right hand side of
the bosonic flow equation to a
sum of single one-loop momentum integrals over the propagators of a finite number of bosonic
fields. The infinite number of bosonic fields contained in $G_{\alpha\beta}$ appears now
only in the relation between the propagators and the functional derivatives of the BEA. We
propose a truncation that reduces this problem to a finite number of bosonic degrees of
freedom as well. In summary, the flow equation interpolates from the solution of the lowest
order gap equation at short distances $(k=\Lambda)$ to the full 2PI quantum effective action
as the cutoff is removed for $k\rightarrow 0$. The practical use of this particular
approach has, however, to be demonstrated in the future.

The discussion of the functional renormalization in the 2PI-context of the BEA is also
useful for a direct comparison with formulations based on partial bosonization. We find
that already simple truncations for the partially bosonized functional renormalization
flow reproduce the main improvements of the 2PI-flow in a natural way. The simple
formulation of the partially bosonized flow in terms of a finite number of
local field appears to be an important advantage for practical computations.

At this point we may compare the merits and disadvantages of the BEA to the other competing
method to go beyond mean field theory, namely partial bosonization (PB). Let us first look
at the criteria formulated in the introduction. (a) Both BEA and PB permit a reasonable
simple computation of the free energy. (b) Whereas BEA is one loop exact in lowest order by
construction, this property is realized for PB only after inclusion of the bosonic
fluctuations, as demonstrated in a RG approach \cite{IW}. (c) Both BEA and PB can include the
effective bosonic fluctuations in a more or less systematic way by the use of renormalization
group equations. The BEA offers the additional advantage of a simple estimate of the
importance of the bosonic fluctuations from a computation of (parts of) the three loop
effective action (see sect. \ref{freeenergy}). On the other hand, the flow equations for
PB are both conceptually and practically much simpler.

The BEA has the important merit of avoiding redundancy in the description from the outset.
For PB this redundancy leads in lowest order to the problem of the Fierz ambiguity. It can,
however, be (partially) absorbed by rebosonization \cite{GW}, \cite{IW} in the flow
equations. The Fierz ambiguity can even be turned into an advantage by providing an error
estimate for approximations.
Another asset of the BEA is the high degree of symmetry which allows one to
connect different channels. Indeed, the BEA is a very flexible approach since the various
channels where condensates or other interesting physics may occur are all described at once.
This feature is very helpful for the exact mapping between the repulsive and attractive
Hubbard model presented in this paper.
In contrast, PB needs a selection of the interesting channels from the beginning of its
formulation, thereby introducing a certain rigidity and perhaps sometimes a certain bias.
On the other hand, this restriction of PB to a few degrees of freedom is an important
advantage for practical computations. Comparing this last aspect with BEA we find no
practical problem in lowest order. In higher orders, however, an effective reduction to
a finite number of bosons requires some thought as discussed in
sects. \ref{exactfermionic}-\ref{bosonicrenormalizationgroup}.

Finally, PB has the advantage that both bosons and fermions appear explicitely. This
permits a simple description of universality classes for (near) critical behavior for which
only the fermionic and bosonic low-mass fluctuations are relevant. The use of nonperturbative
flow equations for fermionic systems with composite bosons has been well tested in this
context and has reached a highly sophisticated level. For BEA the fermions appear not
explicitely anymore. Nevertheless, the fermion fluctuations
are included and the running of the
``fermionic part of the effective action'' is reflected by the running of $j_k$
(sects. \ref{exactfermionic}, \ref{renormalizedgapequation}).
It may be advantageous to cast the fermionic fluctuation effects into a more explicit form.
This, as well as a practical demonstration of the use of the flow equation remains to be
done in the setting of BEA.

Before a concrete computation of the flow of the BEA for a simple
fermionic system has been performed it is perhaps premature to judge the relative merits of
the two approaches for systems with a long correlation length.
It seems well conceivable to us
that both the bosonic effective action and partial bosonization should be used
simultaneously in order to
assert how reliable are proposed solutions for strongly correlated fermionic systems.
Such an approach can use the presented detailed relations between the quantities
computed by both methods.

\section* {Appendix A: Gross-Neveu-model}
\renewcommand{\theequation}{A.\arabic{equation}}
\setcounter{equation}{0}

The action for the Gross-Neveu model (GN-model) \cite{GN} can be written in the form
\begin{equation}\label{eq:gnmodel}
S=\int\limits^\beta_0 d\tau\int d^Dx
\Big\{i\overline{\psi} \gamma^\mu \partial_\mu \psi
+\frac{G}{2}(\overline{\psi}\psi)^2\Big\}
\end{equation}
For $D \le 3$ and $\beta \rightarrow \infty$ this action is invariant under the rotation
(``Lorentz'') group $SO (D+1)$ and therefore describes relativistic fermions.
This holds provided the matrices $\gamma^\mu$ are appropriately chosen, where $\mu=0$,
$1\dots D$
runs over $d=D+1$ values. For $D=2$ we take $\gamma^0=\tau_3$, $\gamma^1=\tau_1$,
$\gamma^2=\tau_2$. An appropriate choice for $D=3$ is $\gamma^0=1$, $\gamma^1=i\tau_1$,
$\gamma^2=i\tau_2,$ $\gamma^3=i\tau_3$ and we note that $\psi$ accounts only for one
chirality, i. e. $\psi\widehat{=}\psi_L$, $\overline{\psi}\widehat{=}\overline{\psi}_L$.
The $SO(4)$-invariance of this somewhat unusual ``Chiral Gross-Neveu model'' can be
seen by noting the ``Fierz identity'' for two component spinors
\begin{equation}
(\overline{\psi} \psi)^2=-\frac{1}{3}(\overline{\psi}\vec{\tau}\psi)
(\overline{\psi}\vec{\tau}\psi)
=\frac{1}{4}(\overline{\psi}\gamma^\mu\psi)(\overline{\psi}\gamma_\mu\psi)
\end{equation}
Finally, for $D=1$ we may take $\gamma^0=\tau_2,\gamma^1=\tau_1$.

\medskip

The continuous GN-model (\ref{eq:gnmodel}) needs some type of an ultraviolet regularization.
A cutoff preserving the $SO(d)$ symmetry can easily be implemented in momentum space
by restricting $q^\mu q_\mu\leq\Lambda^2$. Another approach provides for a suitable
smooth momentum dependent cutoff by modifying the fermion kinetic term in the action
(\ref{eq:GN}). As an alternative, we may discretize the model on a cubic lattice with
lattice distance $a$ such that $\psi(\vec{x}, \tau)=a^{-D/2} \psi (m_1, m_2, \dots m_D, \tau)$,
with integers $(m_1,m_2, \dots m_D)$ denoting the lattice sites. Depending on the precise
implementation of the discretized derivative the continuum limit $a\rightarrow 0$
of the lattice GN-model may correspond to a continuous GN-model with several species
of fermions (``fermion doubler problem'').

In the following we concentrate on $D=2$. On a cubic lattice with sites $(m, n)$ the
action of the lattice-Gross-Neveu model (LGN-model)
\begin{eqnarray} \label{eq:GN}
S&=&\int d\tau \sum_{(m,n)} \Big\{{\cal L}_{kin}
+\frac{1}{2} U(\overline{\psi} \psi)^2\Big\}~,\nonumber\\
{\cal L}_{kin}&=&i\overline{\psi}\gamma^\mu \partial_\mu \psi
\end{eqnarray}
is formulated in terms of the lattice derivatives
\begin{eqnarray}
\partial_1\psi(m,n)=\frac{1}{2a}\Big(\psi(m+1,n)-\psi(m-1,n)\Big) \nonumber \\
\partial_2\psi(m,n)=\frac{1}{2a}\Big(\psi(m,n+1)-\psi(m,n-1)\Big)
\end{eqnarray}
with $\partial_0=\partial_\tau$ and $\gamma^0=\tau_3,~\gamma^{1,2}=\tau_{1,2}$.
The coupling strength is related to the one of the continuum formulation by
$U=G/a^2$.

\medskip
Without changing the structure of the interaction term $\sim U$ we may bring
the discretized action (\ref{eq:GN}) of the GN model closer to the one for the Hubbard model
(\ref{eq:Hubbard}) by appropriate redefinitions of the spinor fields.
This transformation is an example for the extended symmetry transformations which
leave $\Gamma_B$ invariant, as discussed in sect. \ref{symmetries}. With
\begin{eqnarray} \label{eq:inversemap}
\psi(m,n,\tau)&=&-i\tau_3~\exp(-\frac{i\pi n}{2}\tau_1)~\nonumber\\
&&\exp(\frac{i\pi m}{2}\tau_2) \psi^{\prime}(m,n,\tau) \nonumber \\
\overline{\psi}(m,n,\tau)&=&\overline{\psi}~^\prime(m,n,\tau)~\nonumber\\
&&\exp(-\frac{i\pi m}{2}\tau_2)~\exp(\frac{i \pi n}{2}\tau_1)
\end{eqnarray}
the kinetic term equals the one for the Hubbard model up to an important factor
$(-1)^n$, i. e.
\begin{eqnarray} \label{eq:kineticterm}
{\cal L}_{kin}&=&\overline{\psi}~^\prime(m,n,\tau)\partial_\tau\psi^\prime(m,n,\tau)
-t\overline{\psi}~^\prime (m,n,\tau) \nonumber\\
&&\Big\{(-1)^n\Big(\psi^\prime(m+1,n,\tau)
+\psi^\prime(m-1,n,\tau)\Big) \nonumber \\
&&+\psi^\prime(m,n+1,\tau)
+\psi^\prime(m,n-1,\tau)\Big\}~
\end{eqnarray}
We identify $t=-1$/(2$a$) and note that a change of sign of $t$ can be achieved by further
multiplying both $\psi^\prime(m,n)$ and $\overline{\psi}~^\prime (m,n)$ by a factor
$(-1)^{m+n}$.

The alternating factor $(-1)^n$ multiplying the next neighbour interaction in the
1-direction cannot be absorbed by further local transformations of the fields.
(It may be shifted to a factor $(-1)^m$ in the 2-direction.)
It expresses the difference in the structure
of the Fermi surface between the LGN- and the Hubbard model. We may call the model
with the kinetic term (\ref{eq:kineticterm}) the ``layered Hubbard model'' (LH-model)
since the next neighbour interaction in the 1-direction switches sign between different
layers in $n$. The hopping of the electrons in the 1-direction differs between the layers
with $n$ even or odd. In physical terms, such a situation may be realized
by an appropriate atomic
lattice structure. Results on phase transitions in the Gross-Neveu model \cite{HNW},
\cite{21A} can be
carried over to the LH-model.\footnote{We expect the LGN-model to belong to the universality
class with four independent two component spinors in a continuum formulation $(N=4)$. This is
connected to ``fermion doubling'' on the lattice.}
Similarly, by use of the inverse map (\ref{eq:inversemap}) the Hubbard
model (\ref{eq:Hubbard}) can be mapped onto a ``layered Gross-Neveu'' model where
the factor $(-1)^n$ now multiplies the derivative term $\sim\tau_1\partial_1$.

The exact map between the LGN and the LH models opens new perspectives for the
mutual understanding of both models. For example, the (pseudo-)scalar order parameter
in the LGN-model appears as the third component of an antiferromagnetic order parameter
in the LH model
\begin{eqnarray}
i\overline{\psi}(m,n\tau)\psi(m,n,\tau)&=&\nonumber\\
(-1)^{m+n}\overline{\psi}~^\prime (m,n,\tau)
\tau_3\psi^\prime(m,n,\tau)&=&\tilde{a}_3
\end{eqnarray}
The LH model is invariant under independent global spin rotations
$\psi^\prime \rightarrow \exp (i\vec\alpha\vec\tau)\psi^\prime$, $
\overline{\psi}~^\prime\rightarrow \overline{\psi}~^\prime \exp (-i\vec{\alpha}\vec{\tau})$.
By the mapping (\ref{eq:inversemap}) these symmetries appear as $(m,n)$-dependent
transformtations in the LGN-model. In particular, the other two components of the antiferromagnetic
spin vector in the LH model appear in the LGN model as
\begin{eqnarray}
\tilde{a}_1&=&(-1)^{m+n}\overline{\psi}~^\prime(m,n,\tau)
\tau_1\psi^\prime(m,n,\tau)\nonumber\\
&=&(-1)^n\overline{\psi}(m,n,\tau)\tau_2\psi(m,n,\tau) \nonumber \\
\tilde{a}_2&=&(-1)^{m+n}\overline{\psi}~^\prime(m,n,\tau)
\tau_2\psi^\prime(m,n,\tau)\nonumber\\
&=&(-1)^{m+1}\overline{\psi}(m,n,\tau)\tau_1\psi(m,n,\tau)\nonumber \\
\end{eqnarray}
From the symmetry of the LH model we infer that the free energy in the lattice Gross-Neveu
model is degenerate in the direction of the vector $(\tilde{a}_1,\tilde{a}_2,\tilde{a}_3)$.
This would probably not have been suspected from a direct inspection of the action
(\ref{eq:GN}). In particular, we predict for the ordered low temperature state of the
lattice GN-model the existence of two massless Goldstone bosons.

\vspace{3,5cm}
\noindent
\underline{Acknowledgment}

\medskip
\noindent
The author would like to thank T. Baier and E. Bick for collaboration on the Hubbard model and
J. Berges, H. Gies and J. Jaeckel for discussions and collaboration on various other aspects
of this work.

\bigskip
\noindent
\underline{Note added}

\medskip
\noindent
The formalism presented here has been generalized to higher multi-fermion interactions in
\cite{JW}. In this paper chiral symmetry breaking in strong interactions (QCD) has been
studied for a six fermion interaction based on instantons. The functional renormalization
group in the partially bosonized version has been
applied to an investigation of the antiferromagnetic order parameter in the Hubbard model
in \cite{BBWN}. The material presented in sects. \ref{symmetrieshubbard}-
\ref{partialbosonization} and \ref{functionalrenormalization}
was not included in a first version of this article.

\end{document}